\documentclass[preprint,12pt,authoryear]{elsarticle}

\usepackage{amssymb}
\usepackage{times}

\usepackage{amsmath}
\usepackage{version}
\usepackage{graphicx}
\usepackage{pstricks}
\usepackage{amsthm}
\usepackage{natbib}

\newtheorem{lem}{Lemma}[section]

\newtheorem{theor}{Theorem}[section]

\theoremstyle{remark}
\newtheorem{remark}{Remark}

\newcommand{\n}{^{(n)}}

\newcommand{\cqfd}{\hfill $\square$}
\newcommand{\R}{\mathbb R}

\newcommand{\N}{\mathbb N}

\newcommand{\Z}{\mathbb{Z}}

\newcommand{\varthetab}{{\pmb \vartheta}}

\newcommand{\Deltab}{{\pmb \Delta}}

\newcommand{\taub}{{\pmb \tau}}

\newcommand{\Gammab}{{\pmb \Gamma}}

\newcommand{\zerob}{{\pmb 0}}

\newcommand{\pr}{^{\prime}}

\newcommand{\ny}{n\rightarrow\infty}

\makeatletter
\def\ps@pprintTitle{%
 \let\@oddhead\@empty
 \let\@evenhead\@empty
 \def\@oddfoot{}%
 \let\@evenfoot\@oddfoot}
\makeatother

\begin{document}

\begin{frontmatter}

\title{Optimal tests for circular reflective symmetry about an unknown central direction}

\author[Ameijeiras]{Jose Ameijeiras-Alonso}
\ead{jose.ameijeiras@usc.es}

\author[Ley]{Christophe Ley}
\ead{christophe.ley@ugent.be}

\author[Pewsey]{Arthur Pewsey}
\ead{apewsey@unex.es}

\author[Verdebout]{Thomas Verdebout}
\ead{tverdebo@ulb.ac.be}

\address[Ameijeiras]{Department of Statistics, Mathematical Analysis and Optimization, Universidade de Santiago de Compostela, Spain}

\address[Ley]{Department of Applied Mathematics, Computer Science and Statistics, Ghent University, Belgium}

\address[Pewsey]{Department of Mathematics, University of Extremadura, Spain}

\address[Verdebout]{Mathematics Department, Brussels Free University, Belgium}

\begin{abstract}

\noindent Parametric and semiparametric tests of circular
reflective symmetry about an unknown central direction are
developed that are locally and asymptotically optimal in the Le
Cam sense against asymmetric $k$-sine-skewed alternatives. The
results from Monte Carlo studies comparing the rejection rates of
tests with those of previously proposed tests lead to
recommendations regarding the use of the various tests with small-
to medium-sized samples. Analyses of data on the directions of
cracks in cemented femoral components and the times of gun crimes
in Pittsburgh illustrate the proposed methodology and its
bootstrap extension.

\end{abstract}

\end{frontmatter}

\section{Introduction}\label{sec:intro}

\noindent Symmetry, or more precisely \textit{reflective
symmetry}, is one of the most frequently encountered simplifying
assumptions, the rejection of which generally leads to the
subsequent exploration of models with more parameters than their
symmetric counterparts. Its rejection also raises important issues
as to precisely which of a distribution's characteristics are of
primary and secondary interest.

For data observed on the real line, or \textit{linear data} for
short, numerous procedures have been proposed for testing
symmetry. Such tests divide into two main groups: those for which
the centre of the distribution is assumed known, or specified, and
those for which it is not. \cite{P04} provides references for
tests in the first category, and \cite{P02} for tests in the
second. The latter is the one most directly relevant to the
testing scenario considered here.

For data whose natural support is the unit circle, things are
somewhat more involved because, due to the circle's compactness
and isometries of rotation and reflection, ``symmetry'' is not
uniquely defined. There are thus at least four symmetry testing
set-ups that might be of interest when analyzing circular data.
The first, that of testing for cyclic, or $l$-fold, symmetry has
no equivalent for linear data. Permutation-based procedures for
this scenario were proposed by \cite{JS83}. The second set-up,
testing for symmetry about a specified axis against rotation
alternatives, was considered by \cite{Sch69}. He obtained results
for locally most powerful linear rank tests. The third scenario
involves testing for reflective symmetry about some known or
specified median direction. Tests for this set-up were proposed by
{\cite{P04}} and \cite{LV14b}. Finally, the fourth testing
scenario, and the one that we consider here, is that of reflective
symmetry about some unknown central direction. \cite{P02} proposed
a simple omnibus test for this set-up based on the sample second
sine moment about the mean direction, $\bar{b}_2$.

In this paper we develop optimal tests of the null hypothesis that
the distribution from which a random sample of circular data was
drawn is  reflectively symmetric about an unknown central
direction against the alternative hypothesis that the distribution
is $k$-sine-skewed. The definition and basic properties of the
$k$-sine-skewed family are given in Section \ref{sec:density}, and
a \textit{uniform local asymptotic normality} (ULAN) property
established for the family in Section \ref{sec:ULAN}. In Section
\ref{unkdirknden}, optimal parametric tests for circular
reflective symmetry about an unknown central direction are
developed which assume that the form of the base symmetric
unimodal circular density is known. This last assumption is
relaxed in Section \ref{unkdirunkden} where optimal
semi-parametric tests for circular reflective symmetry about an
unknown central direction are developed which assume that the form
of the base symmetric unimodal circular density is unknown but
posited to be of a specified kind. Results from simulation
experiments designed to explore and compare the size and power
characteristics of the tests proposed here with those of
\cite{P02,P04} and \cite{LV14b} are reported in Section
\ref{sec:simus}. On the basis of those results, recommendations
are made concerning the application of the various tests. In
Section \ref{sec:data} various tests of reflective symmetry are
applied in the analysis of circular data on the cracks in cemented
femoral components and the times of gun crimes. The paper ends
with Section \ref{sec:discussion} in which our findings, related
issues and extensions are discussed. Proofs of
Lemma~\ref{consistency}, Lemma \ref{lemunk1} and Theorem
\ref{thunk} are presented in \ref{sec:prooflem1}, \ref{sec:proof}
and \ref{sec:proofth}, respectively. Additional results
from the Monte Carlo studies reported in Section~\ref{sec:simus} are provided in \ref{sec:addsm}.

\section{The $k$-sine-skewed family of distributions and its ULAN property}\label{sec:density}

\noindent In this section we review the definition of the
$k$-sine-skewed family of distributions and its properties,
including its crucial ULAN property established in \cite{LV14b}.

\subsection{The $k$-sine-skewed family}\label{sec:sine-skewed}

\noindent Let
\begin{eqnarray*}
\mathcal{F}&:=&\bigg\{f_0:f_0(\theta)>0, f_0(\theta+2\pi k)=f_0(\theta)\forall k\in\Z, f_0(-\theta)=f_0(\theta),\\
&&\quad\quad f_0\ \mbox{unimodal in}\ \theta\in[-\pi,\pi)\
\mbox{with mode at}\ 0,\int_{-\pi}^\pi f_0(\theta)d\theta=1\bigg\}
\end{eqnarray*}
denote the family of unimodal circular densities that are
reflectively symmetric about the zero direction. Some of the
best-known members of $\mathcal{F}$ are the von Mises, cardioid,
wrapped Cauchy and wrapped normal densities, given, respectively,
by: $f_{\textrm{VM}_\kappa}(\theta)=\frac{1}{2\pi I_0(\kappa)}
\exp(\kappa$ $\cos \theta)$, for $\kappa>0$, where $I_k$ denotes the
modified Bessel function of the first kind and order $k$;
$f_{\textrm{C}_\rho}(\theta)=\frac{1}{2\pi}(1+2\rho\cos\theta)$,
for $0\leq\rho< 1/2$;
$f_{\textrm{WC}_\rho}(\theta)=\frac{1}{2\pi}\left(\frac{1-\rho^2}{1+\rho^2-2\rho\cos\theta}\right)$,
for $0\leq\rho< 1$;
$f_{\textrm{WN}_\rho}(\theta)=\frac{1}{2\pi}(1+2\sum_{p=1}^\infty\rho^{p^2}\cos(p\theta))$,
for $0\leq\rho< 1$. In these densities, $\kappa$ and $\rho$ are
concentration parameters, with $\rho$ denoting the mean resultant
length. A location parameter $\mu\in[-\pi,\pi)$ can readily be
introduced to change the centre of symmetry, leading to densities
of the form $f_0(\theta-\mu)$ with modal direction $\mu$.

\begin{figure}
 \vspace{-120pt}
 \hspace{-16pt}\includegraphics[angle=90, width=1.1\textwidth]{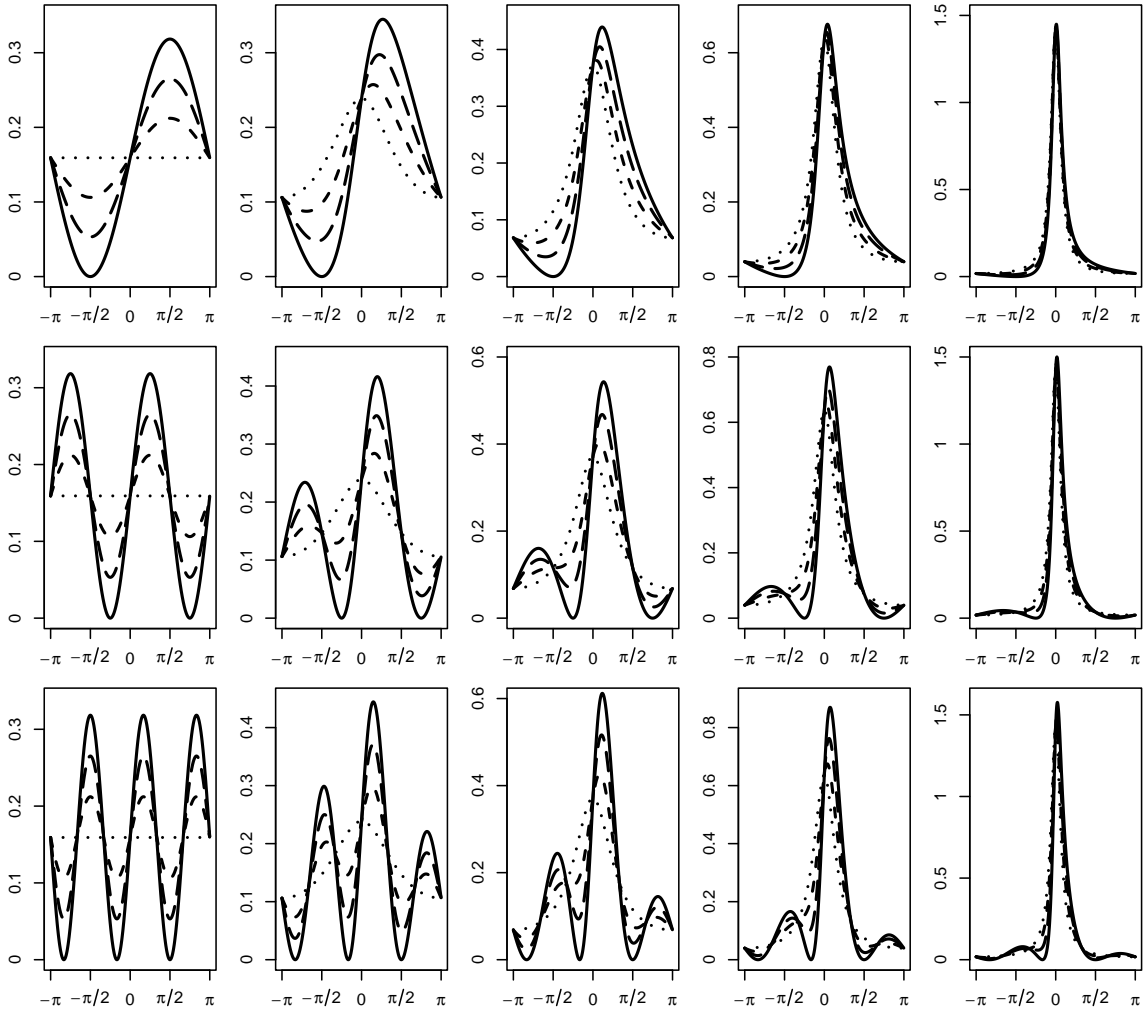}\vspace{-160pt}\\
 \caption{$k$-sine-skewed wrapped Cauchy densities with $\mu=0$ and $k=1$ (top row), $k=2$ (middle row), $k=3$ (bottom row). The five columns correspond,
 from left to right, to $\rho=0,0.2,0.4,0.6,0.8$. In each panel, $\lambda=0$
 (dotted), $\lambda=1/3$ (dashed), $\lambda=2/3$ (long-dashed), $\lambda=1$ (solid).}
 \label{fig:kSSWCDens}
\end{figure}

Inspired by the perturbation approach of \cite{AC03}, \cite{UJ09}
proposed circular densities of the form
$$
2f_0(\theta-\mu)G(\omega(\theta-\mu)),
$$
where $G$ is the cdf of some reflectively symmetric circular
distribution and $\omega$ is a weighting function satisfying: (i)
$\omega(-\theta)=-\omega(\theta)$; (ii) $\omega(\theta+2\pi
k)=\omega(\theta)\forall k\in\Z$; (iii) $|\omega(\theta)|\leq\pi$.
For reasons of mathematical tractability, \cite{UJ09} focused on
the case when $G(\theta)=(\pi+\theta)/(2\pi)$, the cdf of the
circular uniform distribution, and
$\omega(\theta)=\lambda\pi\sin(k\theta), k\in\N_0,$
$\lambda\in[-1,1]$. These choices yield the \emph{$k$-sine-skewed}
family of distributions with density
\begin{equation}\label{densitygen}
f^k_{\mu,\lambda}(\theta):=f_0(\theta-\mu)[1+\lambda\sin(k(\theta-\mu))].
\end{equation}
Appealing properties of such densities include: (i)
$f^k_{\mu,\lambda}(\mu-\theta)=f^k_{\mu,-\lambda}(\mu+\theta)$;
(ii) $f^k_{\mu,\lambda}(\mu)=f_0(0)$ independently of the value of
$\lambda$; (iii) $f^k_{\mu,\lambda}(\mu-\pi)$ and
$f^k_{\mu,\lambda}(\mu+\pi)$ coincide. The base reflectively
symmetric unimodal circular density, $f_0$, is unperturbed if
$\lambda=0$. When $k=1$, $\lambda$ mainly acts as a skewness
parameter, with (\ref{densitygen}) being skew to the left if
$\lambda>0$ or to the right if $\lambda<0$, and the density is
generally unimodal. However, for certain base density and
parameter combinations, (\ref{densitygen}) can be bimodal
\citep{AP11}. For $k\geq 2$ and $\lambda\neq 0$,
(\ref{densitygen}) is generally multimodal, $\lambda$ determining
the number of modes as well as their heights and skewness. Being
interested in unimodal models, \cite{AP11} restricted their
attention to the $k=1$ case, with densities
\begin{equation}\label{density}
f_{\mu,\lambda}(\theta):=f_0(\theta-\mu)[1+\lambda\sin(\theta-\mu)],
\end{equation}
referring to them as \textit{sine-skewed circular} densities. The
reference to \emph{$k$-sine-skewed} distributions extends their
terminology. Figure \ref{fig:kSSWCDens} portrays examples of $k$-sine-skewed densities when $f_0$ is wrapped Cauchy.

In applications, $k$-sine-skewed distributions have been used as
models for ant orientation data and the times of thunder storms,
in \cite{AP11}, the $\mbox{CO}_2$ daily cycle at a rural site, in
\cite{PSGP12}, and forest disturbance regimes, in \cite{AKSAK12}.

The $k$-sine-skewed family is an appealing one in the sense that
it provides a dense family of distributions capable of describing
varied forms of departure from the reflectively symmetric unimodal
circular densities in $\mathcal{F}$. This is the motivation for
considering its cases with $\lambda\neq 0$ as the alternatives in
our tests.

\subsection{The ULAN property of $k$-sine-skewed
densities}\label{sec:ULAN}

\noindent Let $\Theta_1,\ldots,\Theta_n$ be \mbox{i.i.d.} circular
observations with common density~(\ref{densitygen}). For any
reflectively symmetric unimodal base density $f_0\in\mathcal{F}$
and any $k\in\N_0$, denote the joint distribution of the $n$-tuple
$\Theta_1,\ldots,\Theta_n$ by ${\rm P}^{(n)}_{\varthetab;f_0,k}$,
where $\varthetab:=(\mu,\lambda)\pr\in[-\pi,\pi)\times[-1,1]$.
Since $f^k_{\mu,\lambda}=f_0$ when $\lambda=0$, and hence does not
depend on $k$, we drop the index $k$ and simply write ${\rm
P}^{(n)}_{\varthetab;f_0}$ at
$\varthetab=\varthetab_0:=(\mu,0)\pr$. Any pair $(f_0,k)$ induces
the \textit{parametric} model
$$\mathcal{P}^{(n)}_{f_0,k}:=\left\{{\rm P}^{(n)}_{\varthetab;f_0,k}:\varthetab\in[-\pi,\pi)\times[-1,1]\right\},$$
whereas any $k\in\N_0$ induces the \textit{semi-parametric} model
$\mathcal{P}^{(n)}_k:=\cup_{f_0\in\mathcal{F}}\mathcal{P}^{(n)}_{f_0,k}$.

The ULAN property of the parametric model
$\mathcal{P}^{(n)}_{f_0,k}$ in the vicinity of unimodal reflective
symmetry, i.e.\ around $\lambda=0$, was established by
\cite{LV14b} and is crucial to the development of our tests. Its
derivation requires the following mild regularity condition on the
base density $f_0$ to hold.\vspace{4mm}

\noindent {\sc Assumption A}: \textit{The base density
$f_0(\theta)$ is $\mathcal{C}^1$ almost everywhere over
$[-\pi,\pi)$, or equivalently over $\R$ by periodicity, with
derivative $\dot{f_0}$ {almost everywhere}}.\vspace{4mm}

\noindent Most classical reflectively symmetric unimodal densities
satisfy this requirement. Note that the continuously
differentiable condition over a compact manifold, combined with
the fact that $f_0>0$, implies that the Fisher information
quantity for location,
$I_{f_0}:=\int_{-\pi}^\pi\varphi^2_{f_0}(\theta)f_0(\theta)d\theta$,
where $\varphi_{f_0}=-\dot{f_0}/f_0$, is finite. The ULAN property
of the parametric model $\mathcal{P}^{(n)}_{f_0,k}$ with respect
to $\varthetab=(\mu,\lambda)^\prime$, in the vicinity of unimodal
reflective symmetry, then takes the following form.

\begin{theor}\label{ULAN}
Suppose $f_0\in\mathcal{F}$, $k\in\N_0$ and that
Assumption A holds. Then, for any $\mu\in[-\pi,\pi)$, the
parametric family of densities $\mathcal{P}^{(n)}_{f_0,k}$ is ULAN
at $\varthetab_0=(\mu,0)^\prime$ with central sequence
\begin{eqnarray*}
\Deltab^{(n)}_{f_0,k}(\mu) &:=& \left(
\begin{array}{c}
\Delta^{(n)}_{f_0,k;1}(\mu)\\[1mm]
\Delta^{(n)}_{k;2}(\mu)
\end{array}
\right)\\
&:=&
\frac{1}{\sqrt{n}}\, \sum_{i=1}^n
\left(
\begin{array}{c}
\varphi_{f_0}(\Theta_i-\mu)\\
\sin(k(\Theta_i-\mu))
\end{array}
\right),
\end{eqnarray*}
and corresponding Fisher information matrix
$${\Gammab}_{f_0,k}:=\left(
\begin{array}{cc}
\Gamma_{f_0,k;11}&\Gamma_{f_0,k;12}\\
\Gamma_{f_0,k;12}&\Gamma_{f_0,k;22}
\end{array}\right),$$
where $\Gamma_{f_0,k;11}:=I_{f_0}$,
$\Gamma_{f_0,k;12}:=-\int_{-\pi}^\pi\sin(k\theta)\dot{f_0}(\theta)d\theta$
and
$\Gamma_{f_0,k;22}:=\int_{-\pi}^\pi\sin^2(k\theta)f_0(\theta)d\theta$.

More precisely, for any $\mu^{(n)}=\mu+O(n^{-1/2})$ and for any
bounded sequence $\taub^{(n)}=(\tau_1^{(n)},\tau_2^{(n)})'\in\R^2$
such that $\mu^{(n)}+n^{-1/2}\tau_1^{(n)}$ remains in
$[-\pi,\pi)$ and $n^{-1/2}\tau_2^{(n)}$ in $[-1,1]$, we have
{\footnotesize
\begin{eqnarray}
\Lambda^{(n)}_{(\mu^{(n)}+n^{-1/2}\tau_1^{(n)},n^{-1/2}\tau_2^{(n)})'/(\mu^{(n)},0)';f_0,k}&:=&\log\left(d{\rm P}^{(n)}_{(\mu^{(n)}+n^{-1/2}\tau_1^{(n)},n^{-1/2}\tau_2^{(n)})';f_0,k}/d{\rm P}^{(n)}_{(\mu^{(n)},0)';f_0}\right)\nonumber\\
&=&\taub^{(n)'}\Deltab^{(n)}_{f_0,k}(\mu^{(n)})-(1/2)\taub^{(n)'}{\Gammab}_{f_0,k}\taub^{(n)}+o_{\rm P}(1)\nonumber\\
&&\label{Taylor}
\end{eqnarray}
}
and
$\Deltab^{(n)}_{f_0,k}(\mu^{(n)})\stackrel{\mathcal{D}}{\rightarrow}\mathcal{N}_2(\zerob,{\Gammab}_{f_0,k})$,
both under ${\rm P}^{(n)}_{(\mu^{(n)},0)\pr;f_0}$ as
$n\rightarrow\infty$.
\end{theor}

The proof of Theorem~\ref{ULAN} is given in \cite{LV14b}, where a
brief discussion of the minimal conditions required for the ULAN
property to hold is also provided. The Fisher information for
departures from unimodal reflective symmetry, $\Gamma_{f_0,k;22}$,
and hence the cross-information quantity $\Gamma_{f_0,k;12}$, can
easily be shown to be finite by bounding $\sin^2$ by $1$ under the
integral sign. Note that the constant $k$ has no effect on the
validity of Theorem~\ref{ULAN} and that $\Delta^{(n)}_{k;2}(\mu)$
does not depend on $f_0$.

\begin{remark}\label{remsin}
For the ULAN property to hold, the Fisher information matrix
${\Gammab}_{f_0,k}$ must be non-singular. Proposition 1 of
\cite{LV14b} states that this is always the case, except for when
$f_0$ is von Mises and $k=1$. As we shall see in the sequel, a
singular information matrix is of no relevance when building tests
for reflective symmetry about a \textit{known} central direction
but precludes the construction of a powerful test for reflective
symmetry against von-Mises-based sine-skewed alternatives when the
central direction is \textit{unknown}.
\end{remark}

\section{Optimal tests for reflective symmetry about an unknown central direction}

\noindent \cite{LV14b} proposed locally and asymptotically optimal
tests, in the Le Cam sense, for reflective symmetry within the
$k$-sine-skewed family when $\mu$ is \textit{known}. In this
section, we first consider the \emph{parametric} testing problem
\begin{equation} \label{paramprob}
\left\{\begin{array}{l}
\mathcal{H}_{0;f_0}:=\cup_{\mu\in[-\pi,\pi)}{\rm
P}^{(n)}_{(\mu,0)\pr;f_0}, \\
\mathcal{H}_{1;f_0,k}:=\cup_{\lambda\neq0\in[-1,1]}\cup_{\mu\in[-\pi,\pi)}{\rm
P}^{(n)}_{(\mu,\lambda)\pr;f_0,k}, \end{array}\right.
\end{equation}
where $f_0$ is a specified density belonging to ${\cal F}$ and the
unknown central direction under $\mathcal{H}_{0;f_0}$ is
estimated.

A drawback of the above tests is that they are only valid under
the parametric null hypothesis $\mathcal{H}_{0;f_0}$ with $f_0$
specified. In order to address the more general null hypothesis of
reflective symmetry, we need a test statistic whose asymptotic
distribution is valid under any symmetric density
$g_0\in\mathcal{F}$. Thus, we subsequently consider the more
demanding testing problem
\begin{equation} \label{semiprob}
\left\{\begin{array}{l}
\mathcal{H}_0:=\cup_{\mu\in[-\pi,\pi)}\cup_{g_0\in\mathcal{F}}{\rm P}^{(n)}_{(\mu,0)\pr;g_0},
\\
\mathcal{H}_{1;k}:=\cup_{\lambda\neq0\in[-1,1]}\cup_{\mu\in[-\pi,\pi)}\cup_{g_0\in\mathcal{F}}{\rm P}^{(n)}_{(\mu,\lambda)\pr;g_0,k},
\end{array}\right.
\end{equation}
in which the location parameter $\mu$ and the density $g_0$ both
take on nuisance roles.

For both problems, we make use of the ULAN property of Theorem
\ref{ULAN} to derive tests that (i) are valid under the null
hypotheses considered and (ii) achieve local and asymptotic
parametric optimality against a $k$-sine-skewed alternative
characterized by the fixed couple $(f_0,k) \in ({\cal F} \times
{\mathbb N}_0)$. In the {semi-parametric} testing problem \eqref{semiprob}, $f_0$ and
$k$ are chosen \textit{a priori} by the practitioner and we derive
{tests $\phi^{(n);f_0}_{k}$} that are asymptotically optimal against
the $(f_0,k)$-sine-skewed alternative and are such that
$$\lim_{\ny} {\rm E}[\phi^{(n);f_0}_{k}] \leq \alpha,$$
where the expectation is taken under any possible
${\rm P}^{(n)}_{(\mu,0)\pr;g_0}$ belonging to $\mathcal{H}_0$:
i.e., they are valid under \emph{any} density $g_0\in\mathcal{F}$.

\subsection{Optimal tests: parametric scenario}\label{unkdirknden}

\noindent For the testing problem \eqref{paramprob}, our tests are
constructed using a root-$n$ consistent and discretized (see
Assumption B below) estimator
$\hat{\mu}\n$. 
The main reason why this testing problem is more demanding than
the fixed-$\mu$ problem considered in \cite{LV14b} is because the
Fisher information matrix ${\Gammab}_{f_0,k}$ is not, in general,
diagonal. If the information matrix ${\Gammab}_{f_0,k}$ were
diagonal, the substitution of $\hat{\mu}\n$ for $\mu$ would,
asymptotically, have no influence on the behavior of the central
sequence for departures from unimodal reflective symmetry
$\Delta^{(n)}_{k;2}(\mu)$.

\

\begin{remark}\label{remg12}
The information matrix ${\Gammab}_{f_0,k}$ is never diagonal if
$k=1$. This can be seen by noting that
$\sin(\theta)\varphi_{f_0}(\theta)f_0(\theta)>0$ over
$(-\pi,\pi)$. On the other hand, when $k>1$ we can find densities
for which $\Gamma_{f_0,k;12}=0, \forall k\in \{2,3,\ldots\}$. If
the density function is square integrable on $[-\pi,\pi)$ this
happens when $\alpha_k={\rm E} [\cos k\Theta]=0$ {for $\Theta\sim
f_0$}, which can be proved using the Fourier expansion
\citep[see][Section 2.1]{JS01} of density (\ref{densitygen}). A
well-known example where this occurs is the cardioid density, for
which $\alpha_1=\rho$ and $\alpha_k=0$ for $k>1$.
\end{remark}

From Remark \ref{remg12}, the covariance $\Gamma_{f_0,k;12}$ is
only rarely null. Hence, a local perturbation of $\mu$ has the
same asymptotic impact on $\Delta^{(n)}_{k;2}(\mu)$ as a local
perturbation of $\lambda=0$. It follows that the cost of not
knowing the value of $\mu$ is strictly positive when performing
inference on $\lambda$: the stronger the correlation between $\mu$
and $\lambda$, the larger that cost. The worst case occurs when
the information matrix is singular (see Remark~\ref{remsin}),
which leads to asymptotic local powers equal to the nominal level
$\alpha$. For this scenario, the best possible test is that which
ignores the data and simply rejects the null hypothesis with
probability $\alpha$. Henceforth we refer to such a test as the
``trivial test''.

We address the cost of estimating $\mu$ by removing the effect of
the location central sequence  $\Delta^{(n)}_{f_0,k;1}(\mu)$ from
the skewness central sequence $\Delta^{(n)}_{k;2}(\mu)$. To
achieve this we use a Gram-Schmidt orthogonalization approach. We
project $\Delta^{(n)}_{k;2}(\mu)$ onto the subspace orthogonal to
$\Delta^{(n)}_{f_0,k;1}(\mu)$, which ensures that the resulting
$f_0$-\emph{efficient central sequence for skewness}
$\Delta^{(n)eff}_{f_0,k;2}(\mu)$ and $\Delta^{(n)}_{f_0,k;1}(\mu)$
are asymptotically uncorrelated. This new central sequence is of
the form
\begin{eqnarray}
\Delta^{(n)eff}_{f_0,k;2}(\mu)&:=&\Delta^{(n)}_{k;2}(\mu)-\frac{\Gamma_{f_0,k;12}}{\Gamma_{f_0,k;11}}\Delta^{(n)}_{f_0,k;1}(\mu)\nonumber\\
&=&n^{-1/2}\sum_{i=1}^n\left(\sin(k(\Theta_i-\mu))-\frac{\Gamma_{f_0,k;12}}{\Gamma_{f_0,k;11}}\varphi_{f_0}(\Theta_i-\mu)\right).\label{eff}
\end{eqnarray}
Now we make use of another important consequence of the ULAN
property, namely the \emph{asymptotic linearity property}:
\begin{equation}\label{aslingen}
\Delta^{(n)}_{f_0,k}(\mu+n^{-1/2}\tau_1\n)-\Delta^{(n)}_{f_0,k}({\mu})=-{\Gamma}_{f_0,k}(\tau_1\n,0)\pr+o_{\rm
P}(1)
\end{equation}
under ${\rm P}^{(n)}_{(\mu,0)\pr;f_0}$ as $n\rightarrow\infty$,
with $\tau_1\n\in\R$ as in Theorem~\ref{ULAN}. We refer the reader
to Sections 2 and 3 of \cite{KL14} for in-depth discussions of
these issues. It is not difficult to derive the asymptotic
linearity property of $\Delta^{(n)eff}_{f_0,k;2}(\mu)$ from
(\ref{aslingen}), namely:
\begin{equation}\label{aslineff}
\Delta^{(n)eff}_{f_0,k;2}(\mu+n^{-1/2}\tau_1\n)-\Delta^{(n)eff}_{f_0,k;2}(\mu)=o_{\rm
P}(1)
\end{equation}
under ${\rm P}^{(n)}_{(\mu,0)\pr;f_0}$ as $n\rightarrow\infty$.

Now consider replacing the non-random bounded sequence $\tau_1\n$
with $n^{1/2}(\hat{\mu}\n-\mu)$  for some root-$n$ consistent
estimator $\hat{\mu}\n$. The latter is bounded in probability and,
via Lemma~4.4 of \cite{K87}, serves as an ideal candidate for
$\tau_1\n$, provided the following assumption holds.

\vspace{4mm}

\noindent {\sc  Assumption B}: \textit{The sequence of estimators
$\hat\mu\n$ is \textrm{(i)} root-$n$ consistent, i.e.
$n^{1/2}(\hat\mu\n-\mu)=O_{\rm P}(1)$ as $n\rightarrow\infty$,
under  $ {\rm P}\n_{(\mu,0)\pr;f_0}$, and \textrm{(ii)} locally
asymptotically discrete, meaning that, for all $\mu\in[-\pi,\pi)$
and all $c>0$, there exists an $M=M(c)>0$ such that the number of
possible values of $\hat\mu\n$ in intervals of the form $\{t\in\R
: n^{1/2}|t-\mu| \leq c\}$ is bounded by $M$, uniformly as
$n\rightarrow \infty$}.

\vspace{4mm}

Note that Assumption B(ii) is a purely technical requirement, with
little practical implication. Indeed, for fixed sample size, any
estimator can be considered part of a locally asymptotically
discrete sequence. However, it is this assumption that enables us
to replace $\tau_1\n$ by $n^{1/2}(\hat\mu\n-\mu)$
in~(\ref{aslineff}) {thanks to the aforementioned Lemma 4.4 of \cite{K87}}, yielding

\begin{equation}\label{asympeff}
\Delta^{(n)eff}_{f_0,k;2}(\hat\mu\n)-\Delta^{(n)eff}_{f_0,k;2}(\mu)=o_{\rm
P}(1)
\end{equation}
under ${\rm P}^{(n)}_{(\mu,0)\pr;f_0}$ as $n\rightarrow\infty$.

Our locally and asymptotically maximin $f_0$-parametric test,
$\phi^{(n);f_0}_{k}$, rejects $\mathcal{H}_{0;f_0}$ at asymptotic
level $\alpha$ whenever the statistic
$$Q^{(n);f_0}_{k}:=\frac{|\Delta^{(n)eff}_{f_0,k;2}(\hat\mu\n)|}{\Gamma^{1/2}_{f_0,k;22.1}}$$
exceeds the upper $\alpha/2$ quantile of the standard normal
distribution, $z_{1-\alpha/2}$, where
$\Gamma_{f_0,k;22.1}\linebreak
:=\Gamma_{f_0,k;22}-\frac{\Gamma^2_{f_0,k;12}}{\Gamma_{f_0,k;11}}$
is the asymptotic variance of $\Delta^{(n)eff}_{f_0,k;2}(\mu)$
under ${\rm P}^{(n)}_{(\mu,0)\pr;f_0}$. Optimal properties of this
test statistic are described in Section \ref{unkdirunkden}.

Different constructions of the test statistic $Q^{(n);f_0}_{k}$
are available depending on the choice of $f_0$ and $k$. Among the
possible candidate base symmetric densities, here we describe the
test statistic for three well-known models: the von Mises, the
cardioid and the wrapped Cauchy. The sine-skewed extensions of
these models were studied by \cite{AP11}.

\subsubsection{Von Mises distribution}\label{Sec:vMCase}

\noindent For the von Mises distribution,
$\varphi_{f_{\textrm{VM}_\kappa}}(\theta)=\kappa \sin(\theta)$,
$\Gamma_{f_{\textrm{VM}_\kappa},k;11}=\kappa A_1(\kappa)$,
$\Gamma_{f_{\textrm{VM}_\kappa},k;12}=k A_k(\kappa)$ and
$\Gamma_{f_{\textrm{VM}_\kappa},k;22}=(1-A_{2k}(\kappa))/2$, where
$A_k(\kappa)=I_k(\kappa)/I_0(\kappa)$. As mentioned previously,
when $k=1$ the Fisher information matrix is singular and the
resulting test reduces to the trivial test. For $k>1$, the test
statistic is

\begin{equation*}
Q^{(n);f_{\textrm{VM}_\kappa}}_{k}:=\frac{n^{-1/2}\sum_{i=1}^n\left(\sin(k(\Theta_i-\hat\mu\n))-\frac{k I_k(\kappa)}{\kappa I_1(\kappa)}\kappa\sin(\Theta_i-\hat\mu\n)\right)}{\sqrt{\frac{1}{2}\left(1-\frac{I_{2k}(\kappa)}{I_0(\kappa)}\right)-\frac{\left(k I_k(\kappa)\right)^2}{\kappa I_1(\kappa)I_0(\kappa)}}}.
\end{equation*}

\subsubsection{Cardioid distribution}\label{Sec:CardioidDist}

\noindent Here, and in Section \ref{Sec:WCDist}, we exclude the
case when $\rho=0$ as it corresponds to the circular uniform
distribution. Since, when $k>1$,
$\Gamma_{f_{\textrm{C}_\rho},k;12}=0$ (see Remark 2) and
$\Gamma_{f_{\textrm{C}_\rho},k;22}=1/2$ for the cardioid
distribution, the parametric test statistic takes the form
\begin{equation*}
Q^{(n);f_{\textrm{C}_\rho}}_{k}:={\sqrt{2}n^{-1/2}\sum_{i=1}^n\sin(k(\Theta_i-\hat\mu\n))}.
\end{equation*}
When $k=1$,
straightforward calculations yield
$\varphi_{f_{\textrm{C}_\rho}}(\theta)=2\rho
\sin(\theta)/(1+2\rho\cos(\theta))$,
$\Gamma_{f_{\textrm{C}_\rho},1;11}=1-\sqrt{1-4\rho^2}$ and
$\Gamma_{f_{\textrm{C}_\rho},1;12}=\rho$. The test statistic then
becomes
\begin{equation*}
Q^{(n);f_{\textrm{C}_\rho}}_{1}:=\frac{n^{-1/2}\sum_{i=1}^n\left(\sin(\Theta_i-\hat\mu\n)-\frac{2\rho^2
\sin(\Theta_i-\hat\mu\n)}{\left(1-\sqrt{1-4\rho^2}\right)(1+2\rho\cos(\Theta_i-\hat\mu\n))}
\right)}{\sqrt{\frac{1}{2}-\frac{\rho^2}{1-\sqrt{1-4\rho^2}}}}.
\end{equation*}

\subsubsection{Wrapped Cauchy distribution}\label{Sec:WCDist}

\noindent For the wrapped Cauchy model we obtain
$\varphi_{f_{\textrm{WC}_\rho}}(\theta)=2\rho
\sin(\theta)/(1+\rho^2-2\rho\cos(\theta))$,
$\Gamma_{f_{\textrm{WC}_\rho},k;11}=2\rho^2/(1-\rho^2)^2$,
$\Gamma_{f_{\textrm{WC}_\rho},k;12}=k\rho^k$ and
$\Gamma_{f_{\textrm{WC}_\rho},k;22}=(1-\rho^2)(\sum_{l=1}^{k}
\rho^{2(l-1)})/2$. The test statistic is then
\begin{equation*}
Q^{(n);f_{\textrm{WC}_\rho}}_{k}:=\frac{n^{-1/2}\sum_{i=1}^n\left(\sin(k(\Theta_i-\hat\mu\n))-\left(k\rho^{k-1}(1-\rho^2)^2\right) \frac{\sin(\Theta_i-\hat\mu\n)}{1+\rho^2-2\rho\cos(\Theta_i-\hat\mu\n)} \right)}{\sqrt{\frac{1-\rho^2}{2}\left((\sum_{l=1}^{k} \rho^{2(l-1)})-k^2\rho^{2(k-1)}(1-\rho^2)\right)}}.
\end{equation*}

\vspace{6pt}

\noindent Note that all of the test statistics in Sections
\ref{Sec:vMCase}--\ref{Sec:WCDist}, apart from
$Q^{(n);f_{\textrm{C}_\rho}}_{k}$ with $k>1$, assume that the
value of the concentration parameter, $\kappa$ or $\rho$, is
known.

\subsection{Optimal tests: semi-parametric scenario}\label{unkdirunkden}

\noindent Consider now the testing problem in \eqref{semiprob}.
Our objective is still to construct a test that is locally and
asymptotically maximin for detecting an alternative characterized
by a specified couple $(f_0,k) \in ({\cal F} \times {\mathbb
N}_0)$. The main difference between the semi-parametric scenario
addressed here and the parametric one considered in Section
\ref{unkdirknden} is that here we aim to build a test that is
asymptotically valid under
$\mathcal{H}_0:=\cup_{\mu\in[-\pi,\pi)}\cup_{g_0\in\mathcal{F}}{\rm
P}^{(n)}_{(\mu,0)\pr;g_0}$. Specifically, we need to allow for the
substitution of $\mu$ by $\hat{\mu}\n$ in $
\Delta_{k;2}^{(n)}\left({\mu}\right)$ under ${\rm
P}^{(n)}_{(\mu,0)\pr;g_0}$ and the distinct possibility that $g_0
\neq f_0$. The ULAN property combined with Lemma~4.4 of \cite{K87}
leads to
\begin{equation}\label{unk1}
\Delta_{k;2}^{(n)}\left(\hat{\mu}^{(n)}\right)-\Delta_{k;2}^{(n)}(\mu)=-\Gamma_{g_0,k;12}
\sqrt{n} \left(\hat{\mu}^{(n)}-\mu \right)+o_{ \rm P}(1),
\end{equation}
{under ${\rm P}^{(n)}_{(\mu,0)\pr;g_0}$ as $\ny$}, provided that
$\hat{\mu}^{(n)}$ satisfies Assumption B. Then, the substitution
of $\mu$ by $\hat{\mu}^{(n)}$ under ${\rm P}^{(n)}_{(\mu,0)';g_0}$
has no asymptotic cost only if $\Gamma_{g_0,k;12}=0.$ As we saw in
Remark \ref{remg12}, this will rarely be the case. In order to
circumvent this problem and eliminate the asymptotic covariance
$\Gamma_{g_0,k;12}$ while keeping the $(f_0,k)$ target in mind, we
consider an \textit{efficient central sequence}
\begin{equation*}
\Delta_{f_0,g_0,k;2}^{(n); ecd}(\mu):=n^{-1/2} \sum_{i=1}^n \left( \sin
(k (\Theta_i-\mu)) -\eta \varphi_{f_0}(\Theta_i-\mu) \right),
\end{equation*}
where $\eta:=\Gamma_{g_0,k;12}/\Gamma_{f_0,g_0,k;11}$ with
\begin{equation*}
\Gamma_{f_0,g_0,k;11}:=\int_{-\pi}^{\pi}
\varphi_{f_0}(\theta)\varphi_{g_0}(\theta)g_0(\theta) d\theta.
\end{equation*}
Since $f_0$ and $g_0$ are both periodic $\mathcal{C}^1$ functions
over a bounded set and $f_0,g_0>0$, the cross-information quantity
$\Gamma_{f_0,g_0,k;11}$ is finite. When $g_0=f_0$,
$\Gamma_{f_0,f_0,k;11}=\Gamma_{f_0,k;11}$, so that
$\Delta_{f_0,f_0,k;2}^{(n);ecd}(\mu)$ will coincide with
$\Delta_{f_0,k;2}^{(n)eff}(\mu)$ under ${\rm
P}^{(n)}_{(\mu,0)',f_0}$, which, as we will see in the sequel, is
key to maintaining the asymptotic optimality against $(f_0,k)$
alternatives. Integrating by parts, we obtain
 \begin{eqnarray} \label{bypart1}
 \Gamma_{g_0,k;12}&=& \int_{-\pi}^\pi \sin(k\theta) \varphi_{g_0} (\theta) g_0(\theta) d\theta \nonumber \\
 &=& k \int_{-\pi}^\pi \cos(k\theta) g_0(\theta) d\theta \nonumber \\
 &=&k {\rm E}_{g_0}[\cos(k (\Theta_i- \mu))]
 \end{eqnarray}
and
   \begin{eqnarray} \label{bypart2}
\Gamma_{f_0,g_0,k;11} &=& \int_{-\pi}^{\pi}
\varphi_{f_0}(\theta)\varphi_{g_0}(\theta)g_0(\theta) d\theta \nonumber \\
&=& [-\varphi_{f_0}(\theta) g_0(\theta)]^{\pi}_{-\pi}+ \int_{-\pi}^{\pi}
\dot{\varphi}_{f_0}(\theta)g_0(\theta) d\theta \nonumber \\
&=& \int_{-\pi}^{\pi} \dot{\varphi}_{f_0}(\theta)g_0(\theta)
d\theta= { {\rm E}_{g_0}[{\dot{\varphi}_{f_0}}(\Theta_i- \mu)]},
 \end{eqnarray}
provided that the following assumption holds. \vspace{4mm}

\noindent {\sc  Assumption C}. {The mapping $\theta \mapsto
\varphi_{f_0} (\theta)$ is ${\cal C}^1$ almost everywhere over
$[-\pi, \pi)$ with derivative ${\dot{\varphi}_{f_0}(\theta)}$
almost everywhere, where $f_0\in \mathcal{F}$}. \vspace{4mm}

\noindent In the following lemma, we establish that
\begin{equation*}
\hat{\Gamma}_{g_0,k;12}=n^{-1} \sum_{i=1}^n
k\cos\left(k\left(\Theta_i-\hat{\mu}^{(n)}\right)\right)
\end{equation*}
and
\begin{equation*}
\hat{\Gamma}_{f_0,g_0,k;11}=n^{-1} \sum_{i=1}^n
\dot{\varphi}_{f_0}\left(\Theta_i-\hat{\mu}^{(n)}\right)
\end{equation*}
are consistent estimators of $\Gamma_{g_0,k;12}$ and
$\Gamma_{f_0,g_0,k;11}$ in \eqref{bypart1} and \eqref{bypart2},
respectively.
\begin{lem}
\label{consistency}
Suppose $k\in \mathbb{N}_0$, $f_0,g_0\in
\mathcal{F}$ and Assumptions A, B and C hold. Then $\hat{\Gamma}_{g_0,k;12}-{\Gamma}_{g_0,k;12}=o_{\rm P}(1)$ and $\hat{\Gamma}_{f_0,g_0,k;11}-{\Gamma}_{f_0,g_0,k;11}=o_{\rm P}(1)$ as
$n\rightarrow\infty$ under ${\rm P}^{(n)}_{(\mu,0);g_0}$.
\end{lem}
\noindent The proof is provided in~\ref{sec:prooflem1}.

Using these estimators, our test is based on the estimated version
of the efficient central sequence
\begin{equation} \label{effsem}
\Delta_{f_0,k;2}^{*(n); ecd}(\mu):=n^{-1/2} \sum_{i=1}^n \left( \sin
(k (\Theta_i-\mu)) -\hat{\eta} \varphi_{f_0}(\Theta_i-\mu) \right),
\end{equation}
where $\hat{\eta}:=\hat{\Gamma}_{g_0,k;12}/
\hat{\Gamma}_{f_0,g_0,k;11}$. The test $\phi_{f_0,k}^{*(n)}$
rejects $\mathcal{H}_0$ at asymptotic level $\alpha$ whenever the
test statistic $|Q^{*(n)}_{f_0,k}| > z_{1-\alpha/2}$, where
\begin{eqnarray}\label{unk7}
Q^{*(n)}_{f_0,k} &:= & \frac{\Delta_{f_0,k;2}^{*(n);ecd}\left(\hat{\mu}^{(n)}\right)}{(C^{*(n)}_{f_0,k}\left(\hat{\mu}^{(n)}\right))^{1/2}} \nonumber \\
&:= &  \frac{n^{-1/2} \sum_{i=1}^n \left( \sin (k
(\Theta_i-\hat{\mu}^{(n)})) -\hat{\eta}
\varphi_{f_0}\left(\Theta_i-\hat{\mu}^{(n)}\right)\right)}{\left(n^{-1}
\sum_{i=1}^n \left( \sin (k (\Theta_i-\hat{\mu}^{(n)}))
-\hat{\eta}
\varphi_{f_0}\left(\Theta_i-\hat{\mu}^{(n)}\right)\right)^2\right)^{1/2}}.
\end{eqnarray}

The asymptotic distribution of $Q^{*(n)}_{f_0,k}$ is formally
established in Theorem~\ref{thunk}, where we also prove the
optimality properties of $\phi_{f_0,k}^{*(n)}$. Before doing so,
however, we first need the following result on the efficient
central sequence in (\ref{effsem}), whose proof is given in
\ref{sec:proof}.

\begin{lem}\label{lemunk1}
Suppose $k\in \mathbb{N}_0$, $f_0,g_0\in
\mathcal{F}$ and Assumptions A, B and C hold. Then, as
$n\rightarrow\infty$ under ${\rm P}^{(n)}_{(\mu,0);g_0}$: (i)
$\Delta_{f_0,k;2}^{*(n);ecd}\left(\hat{\mu}^{(n)}\right)-\Delta_{f_0,g_0,k;2}^{(n);ecd}(\mu)=o_{\rm P}(1)$;
(ii) $C^{*(n)}_{f_0,k}\left(\hat{\mu}^{(n)}\right)
-C^{(n)}_{f_0,g_0,k}\left(\mu\right) =o_{\rm P}(1)$, where
$$C^{(n)}_{f_0,g_0,k}\left(\mu\right):=n^{-1}\sum_{i=1}^n \left(
\sin (k (\Theta_i-\mu))
-\frac{\Gamma_{g_0,k;12}}{\Gamma_{f_0,g_0,k;11}}
\varphi_{f_0}(\Theta_i-\mu)\right)^2.$$
\end{lem}

Using Lemma \ref{lemunk1}, we can establish the optimality
properties of the semi-parametric test $\phi_{f_0,k}^{*(n)}$.
Given a posited base density $f_0\in\mathcal{F}$ and value of $k$,
in Theorem \ref{thunk} we provide the asymptotic properties of the
test statistic $Q^{*(n)}_{f_0,k}$ both under $\mathcal{H}_0$ and a
sequence of contiguous alternatives. Theorem \ref{thunk} is the
main result of the paper and resolves the complicated issue of the
non-null behavior of semi-parametrically efficient test procedures
for circular reflective symmetry about an unknown central
direction against $k$-sine-skewed alternatives.

\begin{theor}\label{thunk}
Suppose $k\in \mathbb{N}_0$, the posited base
density $f_0\in\mathcal{F}$ and Assumptions (A), (B) and (C)
hold. Then:
\begin{enumerate}[(i)]
\item under $\mathcal{H}_0$, $Q^{*(n)}_{f_0,k}
\overset{\mathcal{D}}\rightarrow \mathcal{N} (0,1)$ as
$n\rightarrow \infty$, so that the test $\phi_{f_0,k}^{*(n)}$ has
asymptotic level $\alpha$ under $\mathcal{H}_0$; \item under
$\cup_{\mu\in[-\pi,\pi)}
{\rm P}^{(n)}_{(\mu,n^{-1/2}\tau_2^{(n)})';g_0,k'}$ with
$g_0\in\mathcal{F}$, $k'\in \mathbb{N}_0$ {and $\tau_2\n$ a bounded sequence as in Theorem~\ref{ULAN}}, $Q^{*(n)}_{f_0,k}$
is asymptotically normal with mean
$V^{g_0}_{f_0}(k)^{-1/2}C^{g_0}_{f_0}(k,k')\tau_2$ and variance 1,
where $\tau_2=\lim_{n\rightarrow\infty} \tau_2^{(n)}$,
\begin{equation*}
V^{g_0}_{f_0}(k)=\int_{-\pi}^{\pi}\left( \sin (k\theta) -
\frac{\Gamma_{g_0,k;12}}{\Gamma_{f_0,g_0,k;11}}
\varphi_{f_0}(\theta) \right)^2 g_0(\theta) d\theta,
\end{equation*}
and
\begin{equation*}
C^{g_0}_{f_0}(k,k')=\int_{-\pi}^{\pi}\left( \sin (k\theta) -
\frac{\Gamma_{g_0,k;12}}{\Gamma_{f_0,g_0,k;11}}
\varphi_{f_0}(\theta) \right) \sin(k'\theta) g_0(\theta) d\theta,
\end{equation*}
(both finite) ;
\item {
the test $\phi_{f_0,k}^{*(n)}$ is locally and asymptotically maximin at
asymptotic level $\alpha$ when testing $\mathcal{H}_0$ against $\mathcal{H}_{1; f_0, k}$.}
\end{enumerate}
\end{theor}
\noindent The proof is given in \ref{sec:proofth}.

Theorem \ref{thunk} states that $\phi_{f_0,k}^{*(n)}$ is valid
under the entire null hypothesis $\mathcal{H}_0$, and so is
asymptotically distribution and location free. Theorem
\ref{thunk}(ii) provides an important result which can be used to
calculate the asymptotic power of $\phi_{f_0,k}^{*(n)}$ against
local alternatives of the form $\cup_{\mu\in[-\pi,\pi)} {\rm
P}^{(n)}_{(\mu,n^{-1/2}\tau_2^{(n)});g_0,k'}$ as a function of the
posited density $f_0$.

As in Section \ref{unkdirknden}, here we focus on details of the
test statistic when the posited density is von Mises, cardioid or
wrapped Cauchy. To save on space, we only provide formulae for the
numerator $\Delta_{f_0,k;2}^{*(n); ecd}(\hat{\mu}\n)$ rather than
the full test statistic $Q^{*(n)}_{f_0,k}$.

\subsubsection{Von Mises distribution}\label{Sec:vMSPCase}

\noindent For the von Mises distribution,
$\dot{\varphi}_{f_{\textrm{VM}_\kappa}}(\theta)= \kappa
\sin(\theta)$. As in the parametric case, the trivial test is
obtained when $k=1$. When $k>1$, the numerator of the test
statistic,
$\Delta_{\textrm{VM}_\kappa,k;2}^{*(n);ecd}(\hat{\mu}^{(n)})$, is
\begin{equation*}
n^{-1/2} \sum_{i=1}^n \left( \sin (k(\Theta_i-\hat{\mu}^{(n)}))
-\left[\frac{\sum_{l=1}^n
k\cos\left(k\left(\Theta_l-\hat{\mu}^{(n)}\right)\right)}{\sum_{m=1}^n
\cos(\Theta_m-\hat\mu\n)}\right]
  \sin(\Theta_i-\hat{\mu}^{(n)}) \right),
\end{equation*}
which, importantly, does not depend on $\kappa$. When $k=2$ and
the method of moments estimator of $\mu$ is used, straightforward
calculations lead to $Q^{*(n)}_{f_{\textrm{VM}_\kappa},2}$ being
asymptotically equivalent (in the sense that the difference is
$o_{\rm P}(1)$ as $\ny$) to the $\bar{b}_2$ based test statistic
of \cite{P02}. The latter is of the form
$$\frac{n^{-1/2}\sum_{i=1}^n \sin (2(\Theta_i-\hat{\mu}^{(n)}))}{\sqrt{M_n}},$$
where $M_n-n^{-1} \sum_{i=1}^n
\left(\sin(2(\Theta_i-\hat{\mu}^{(n)})) \right)^2$ is $o_{\rm
P}(1)$ as $\ny$ under ${\rm P}^{(n)}_{(\mu,0)';g_0}$ for any
$g_0$. It follows therefore that the $\bar{b}_2$ based test is
locally and asymptotically maximin against any 2-sine-skewed von
Mises alternative, irrespective of the value of $\kappa$, when
$\mu$ is unknown.

\subsubsection{Cardioid distribution}\label{Sec:CardioidSPDist}

\noindent Taking the cardioid density with $\rho \neq 0$ as the
posited density $f_0$, the derivative of
$\varphi_{f_{\textrm{C}_\rho}}$ with respect to $\theta$ is
$\dot{\varphi}_{f_{\textrm{C}_\rho}}(\theta)=2\rho
(2\rho+\cos(\theta))/(1+2\rho\cos(\theta))^2$ and the numerator of
the test statistic, $\Delta_{\textrm{C}_\rho,k;2}^{*(n);
ecd}(\hat{\mu}^{(n)})$, becomes
\begin{equation*}
n^{-1/2} \sum_{i=1}^n \left( \sin (k(\Theta_i-\hat{\mu}^{(n)}))
-\left[\frac{\sum_{l=1}^n
k\cos\left(k\left(\Theta_l-\hat{\mu}^{(n)}\right)\right)}{\sum_{m=1}^n
\frac{
(2\rho+\cos(\Theta_m-\hat\mu\n))}{(1+2\rho\cos(\Theta_m-\hat\mu\n))^2}}\right]
 \frac{ \sin(\Theta_i-\hat{\mu}^{(n)})}{1+2\rho\cos(\Theta_i-\hat{\mu}^{(n)})} \right).
\end{equation*}

\vspace{4pt}

\begin{remark}\label{remg23}
When the true underlying density, $g_0$, is cardioid with $\rho
\neq 0$, it follows from Remark \ref{remg12} that, for $k>1$,
$\Gamma_{g_0,k;12}=0$ and hence $\eta=0$. Then
$$Q^{*(n)}_{f_0,k}= \frac{n^{-1/2}\sum_{i=1}^n \sin (k(\Theta_i-\hat{\mu}^{(n)}))}{\left(n^{-1}
\sum_{i=1}^n \left( \sin (k (\Theta_i-\hat{\mu}^{(n)}))
\right)^2\right)^{1/2}}+o_{\rm P}(1)$$ as $\ny$ under ${\rm
P}^{(n)}_{(\mu,0)';g_0}$, irrespective of the posited density
$f_0$. This implies that, for fixed $k>1$, the choice of the
posited density $f_0$ has no effect on the local power of the test
based on $Q^{*(n)}_{f_0,k}$ when the true underlying distribution
is cardioid. More specifically, when $k=2$ the local asymptotic
power of tests based on $Q^{*(n)}_{f_0,k}$ is the same as that of
the $\phi^{*(n)}_{f_{\textrm{VM}_\kappa},2}$ test (see Section
\ref{Sec:vMSPCase}), and hence that of the $\bar{b}_2$ based test
of Pewsey (2002), irrespective of the posited density $f_0$.
\end{remark}

\subsubsection{Wrapped Cauchy distribution}\label{Sec:WCSPDist}

\noindent When the posited density is wrapped Cauchy,
$\dot{\varphi}_{f_{\textrm{WC}_\rho}}(\theta)= 2 \rho (-2 \rho +
(1 + \rho^2) \cos(\theta))/(1 + \rho^2 - 2 \rho \cos(\theta))^2$
and the numerator of the test statistic,
$\Delta_{\textrm{WC}_\rho,k;2}^{*(n); ecd}(\hat{\mu}^{(n)})$, is
{\small
\begin{equation*}
n^{-1/2} \sum_{i=1}^n \left( \sin (k(\Theta_i-\hat{\mu}^{(n)}))
-\left[\frac{\sum_{l=1}^n
k\cos\left(k\left(\Theta_l-\hat{\mu}^{(n)}\right)\right)}{\sum_{m=1}^n
\frac{
2 \rho (-2 \rho + (1 + \rho^2) \cos(\Theta_m-\hat\mu\n))
}{(1 + \rho^2 - 2 \rho \cos(\Theta_m-\hat\mu\n))^2}}\right]
 \frac{ \sin(\Theta_i-\hat{\mu}^{(n)})}{1+2\rho\cos(\Theta_i-\hat{\mu}^{(n)})} \right).
\end{equation*}
}

\vspace{4pt}

\noindent Note that, unlike
$\Delta_{\textrm{VM}_\kappa,k;2}^{*(n);ecd}(\hat{\mu}^{(n)})$,
$\Delta_{\textrm{C}_\rho,k;2}^{*(n); ecd}(\hat{\mu}^{(n)})$ and
$\Delta_{\textrm{WC}_\rho,k;2}^{*(n); ecd}(\hat{\mu}^{(n)})$
assume that the value of the concentration parameter is known.
However, as showed in Theorem \ref{thunk}, they are asymptotically
well calibrated irrespective of the underlying density.

\section{Monte Carlo results}\label{sec:simus}

\subsection{Simulation experiment}\label{sec:simexp}

\noindent In an extensive simulation experiment we compared the
size and power characteristics of the parametric and
semi-parametric tests proposed in Sections \ref{unkdirknden} and
\ref{unkdirunkden} with those of their counterparts, which we
denote by $\phi^{(n);\mu;g_0}_k$ and $\phi^{*(n);\mu}_k$,
respectively, proposed by \cite{LV14b} for when $\mu$ is
specified. We also compared them with those of the $b^*_2$ based
and $\bar{b}_2$ based tests proposed by \cite{P04, P02} for when
$\mu$ is specified and estimated, respectively. As the
$\phi^{*(n);\mu}_2$ and $b^*_2$ based tests are identical
\citep[Section 3]{LV14b}, henceforth we present the results for
the common test as being those for the $b^*_2$ based test. Recall
that, from Section \ref{Sec:vMSPCase} and Remark \ref{remg23}, the
$\bar{b}_2$ based test is asymptotically equivalent to the
semi-parametric tests proposed here when $k=2$ and $f_0$ is von
Mises or $g_0$ is cardioid.

In our Monte Carlo study we simulated samples of size
$n=30,100,500$ from $k'$-sine-skewed distributions with $\mu=0$,
$\lambda=0,0.2,0.4,0.6$, $k'=1,2,3$ and $f_{\textrm{VM}_1}$,
$f_{\textrm{VM}_{10}}$, $f_{\textrm{WN}_{0.5}}$,
$f_{\textrm{WN}_{0.9}}$, $f_{\textrm{C}_{0.45}}$,
$f_{\textrm{WC}_{0.5}}$ densities (see Section
\ref{sec:sine-skewed}) for the base symmetric density $g_0$. For
each $(n, \lambda, k', g_0)$ combination we simulated 1000 samples
and performed the different tests at a nominal significance level
of $\alpha=0.05$ with $k=1,2,3$. The rejection rates obtained for
$k=2$ are reproduced in Tables \ref{estsim1}--\ref{estsim3}. Their
counterparts for $k=1,3$ are presented in Tables \ref{estsim4}--\ref{estsim6} and
\ref{estsim7}--\ref{estsim9}, respectively, of the \ref{sec:addsm}. In all nine
tables, the null hypothesis of reflective symmetry corresponds to
$\lambda=0$. For that value of $\lambda$, the results are
invariant to the value of $k'$ because of the form of the density
\eqref{densitygen}. Clearly, the scenario in which we would expect
the tests to perform best is when the true value of $k'$ and the
value posited for $k$ are the same. In Table S1, no results are
given for $\phi^{(n);g_0}_{1}$ because, as explained in Section
\ref{Sec:vMCase}, it reduces to the trivial test when $g_0$ is von
Mises. In Tables S1--S6, the rejection rates for the $b^*_2$ based
and $\bar{b}_2$ based tests have been included to aid comparisons.

\begin{table}[!h]
\caption{Rejection rates, for a nominal significance level of
$\alpha=0.05$, of the $\phi^{(n);\mu;g_0}_{2}$, $b^*_2$ based,
$\phi^{(n);g_0}_{2}$ and
$\bar{b}_2$ based tests calculated using
$1000$ samples of size $n$ simulated from the $k'$-sine-skewed
distribution with the specified base von Mises density $g_0$ and
values of $\lambda$ and $k'$.}\label{estsim1}
\vspace{3pt}\centering \scriptsize{
\begin{tabular}{|ll||ccc|ccc|ccc|ccc|}
  \hline
   & $\lambda$ & \multicolumn{3}{c|}{0} & \multicolumn{3}{c|}{0.2} & \multicolumn{3}{c|}{0.4} & \multicolumn{3}{c|}{0.6}\\
   & $n$ & 30 & 100 & 500 & 30 & 100 & 500 & 30 & 100 & 500 & 30 & 100 & 500 \\
   \hline \hline  Test & $k'$ & \multicolumn{12}{c|}{$g_0=f_{\textrm{VM}_1}$}\\
  \hline
  $\phi^{(n);\mu;g_0}_{2}$ & 1 & 0.050 & 0.046 & 0.050 & 0.057 & 0.087 & 0.306 & 0.097 & 0.232 & 0.797 & 0.145 & 0.463 & 0.990 \\
  $b^*_2$ &  & 0.053 & 0.043 & 0.048 & 0.051 & 0.088 & 0.298 & 0.095 & 0.232 & 0.802 & 0.148 & 0.473 & 0.990 \\
  $\phi^{(n);g_0}_{2}$ &  & 0.057 & 0.043 & 0.056 & 0.046 & 0.046 & 0.054 & 0.046 & 0.044 & 0.068 & 0.036 & 0.064 & 0.219 \\
  $\bar{b}_2$ &  & 0.033 & 0.036 & 0.050 & 0.029 & 0.043 & 0.056 & 0.038 & 0.038 & 0.068 & 0.044 & 0.079 & 0.248 \\
 \hline
  $\phi^{(n);\mu;g_0}_{2}$ & 2 & 0.050 & 0.046 & 0.050 & 0.100 & 0.295 & 0.895 & 0.349 & 0.825 & 1 & 0.648 & 0.994 & 1 \\
  $b^*_2$ &  & 0.053 & 0.043 & 0.048 & 0.111 & 0.293 & 0.897 & 0.347 & 0.826 & 1 & 0.646 & 0.995 & 1 \\
  $\phi^{(n);g_0}_{2}$ &  & 0.057 & 0.043 & 0.056 & 0.069 & 0.211 & 0.783 & 0.195 & 0.605 & 0.999 & 0.351 & 0.826 & 1 \\
$\bar{b}_2$ &  & 0.033 & 0.036 & 0.050 & 0.039 & 0.183 & 0.777 & 0.139 & 0.550 & 0.999 & 0.235 & 0.747 & 1 \\
\hline
  $\phi^{(n);\mu;g_0}_{2}$ & 3 & 0.050 & 0.046 & 0.050 & 0.062 & 0.100 & 0.321 & 0.103 & 0.256 & 0.823 & 0.164 & 0.475 & 0.991 \\
  $b^*_2$ &  & 0.053 & 0.043 & 0.048 & 0.061 & 0.105 & 0.319 & 0.107 & 0.259 & 0.823 & 0.172 & 0.482 & 0.991 \\
  $\phi^{(n);g_0}_{2}$ &  & 0.057 & 0.043 & 0.056 & 0.053 & 0.077 & 0.293 & 0.086 & 0.218 & 0.794 & 0.122 & 0.405 & 0.982 \\
 $\bar{b}_2$ &  & 0.033 & 0.036 & 0.050 & 0.032 & 0.065 & 0.288 & 0.057 & 0.198 & 0.777 & 0.086 & 0.343 & 0.975 \\
   \hline \hline  Test & $k'$ & \multicolumn{12}{c|}{$g_0=f_{\textrm{VM}_{10}}$}\\
  \hline
  $\phi^{(n);\mu;g_0}_{2}$ & 1 & 0.049 & 0.046 & 0.044 & 0.060 & 0.085 & 0.282 & 0.097 & 0.233 & 0.786 & 0.163 & 0.476 & 0.983 \\
  $b^*_2$ &  & 0.051 & 0.042 & 0.044 & 0.058 & 0.090 & 0.282 & 0.095 & 0.234 & 0.791 & 0.162 & 0.464 & 0.986 \\
  $\phi^{(n);g_0}_{2}$ &  & 0.050 & 0.047 & 0.040 & 0.045 & 0.045 & 0.034 & 0.045 & 0.040 & 0.039 & 0.047 & 0.036 & 0.044 \\
 $\bar{b}_2$ &  & 0.028 & 0.040 & 0.040 & 0.029 & 0.037 & 0.034 & 0.029 & 0.030 & 0.041 & 0.027 & 0.038 & 0.053 \\
\hline
  $\phi^{(n);\mu;g_0}_{2}$ & 2 & 0.049 & 0.046 & 0.044 & 0.083 & 0.183 & 0.694 & 0.222 & 0.586 & 0.998 & 0.417 & 0.911 & 1 \\
  $b^*_2$ &  & 0.051 & 0.042 & 0.044 & 0.083 & 0.190 & 0.699 & 0.230 & 0.586 & 0.998 & 0.431 & 0.918 & 1 \\
  $\phi^{(n);g_0}_{2}$ &  & 0.050 & 0.047 & 0.040 & 0.052 & 0.053 & 0.044 & 0.045 & 0.051 & 0.070 & 0.032 & 0.040 & 0.053 \\
 $\bar{b}_2$ &  & 0.028 & 0.040 & 0.040 & 0.029 & 0.044 & 0.046 & 0.038 & 0.042 & 0.068 & 0.037 & 0.047 & 0.068 \\
\hline
  $\phi^{(n);\mu;g_0}_{2}$ & 3 & 0.049 & 0.046 & 0.044 & 0.107 & 0.238 & 0.843 & 0.273 & 0.728 & 1 & 0.575 & 0.977 & 1 \\
  $b^*_2$ &  & 0.051 & 0.042 & 0.044 & 0.104 & 0.243 & 0.842 & 0.280 & 0.743 & 1 & 0.581 & 0.982 & 1 \\
  $\phi^{(n);g_0}_{2}$ &  & 0.050 & 0.047 & 0.040 & 0.061 & 0.058 & 0.138 & 0.050 & 0.098 & 0.412 & 0.059 & 0.138 & 0.543 \\
$\bar{b}_2$ &  & 0.028 & 0.040 & 0.040 & 0.034 & 0.057 & 0.146 & 0.033 & 0.087 & 0.397 & 0.031 & 0.091 & 0.524 \\
\hline
\end{tabular}
}
\end{table}

\begin{table}[!h]
\caption{Rejection rates, for a nominal significance level of
$\alpha=0.05$, of the $\phi^{(n);\mu;g_0}_{2}$,
$b^*_2$ based, $\phi^{(n);g_0}_{2}$ and 
$\bar{b}_2$ based tests calculated using
$1000$ samples of size $n$ simulated from the $k'$-sine-skewed
distribution with the specified base wrapped normal density $g_0$
and values of $\lambda$ and $k'$.}\label{estsim2}
\vspace{3pt}\centering \scriptsize{
\begin{tabular}{|ll||ccc|ccc|ccc|ccc|}
  \hline
   & $\lambda$ & \multicolumn{3}{c|}{0} & \multicolumn{3}{c|}{0.2} & \multicolumn{3}{c|}{0.4} & \multicolumn{3}{c|}{0.6}\\
   & $n$ & 30 & 100 & 500 & 30 & 100 & 500 & 30 & 100 & 500 & 30 & 100 & 500 \\
   \hline \hline  Test & $k'$ & \multicolumn{12}{c|}{$g_0=f_{\textrm{WN}_{0.5}}$}\\
  \hline
  $\phi^{(n);\mu;g_0}_{2}$ & 1 & 0.054 & 0.048 & 0.052 & 0.066 & 0.113 & 0.354 & 0.117 & 0.292 & 0.894 & 0.202 & 0.578 & 0.997 \\
  $b^*_2$ &  & 0.052 & 0.053 & 0.049 & 0.065 & 0.112 & 0.351 & 0.111 & 0.294 & 0.895 & 0.212 & 0.579 & 0.998 \\
  $\phi^{(n);g_0}_{2}$ &  & 0.036 & 0.047 & 0.052 & 0.040 & 0.059 & 0.145 & 0.033 & 0.066 & 0.223 & 0.021 & 0.040 & 0.112 \\
 $\bar{b}_2$ &  & 0.030 & 0.041 & 0.050 & 0.039 & 0.059 & 0.148 & 0.037 & 0.077 & 0.257 & 0.041 & 0.058 & 0.167 \\
\hline
  $\phi^{(n);\mu;g_0}_{2}$ & 2 & 0.054 & 0.048 & 0.052 & 0.116 & 0.289 & 0.894 & 0.327 & 0.826 & 1 & 0.656 & 0.995 & 1 \\
  $b^*_2$ &  & 0.052 & 0.053 & 0.049 & 0.112 & 0.289 & 0.896 & 0.329 & 0.825 & 1 & 0.669 & 0.996 & 1 \\
  $\phi^{(n);g_0}_{2}$ &  & 0.036 & 0.047 & 0.052 & 0.090 & 0.243 & 0.864 & 0.210 & 0.717 & 1 & 0.399 & 0.941 & 1 \\
  $\bar{b}_2$ &  & 0.030 & 0.041 & 0.050 & 0.084 & 0.249 & 0.866 & 0.186 & 0.728 & 1 & 0.355 & 0.930 & 1 \\
\hline
  $\phi^{(n);\mu;g_0}_{2}$ & 3 & 0.054 & 0.048 & 0.052 & 0.062 & 0.118 & 0.364 & 0.114 & 0.284 & 0.889 & 0.187 & 0.553 & 0.999 \\
  $b^*_2$ &  & 0.052 & 0.053 & 0.049 & 0.061 & 0.118 & 0.363 & 0.114 & 0.277 & 0.889 & 0.189 & 0.558 & 0.999 \\
  $\phi^{(n);g_0}_{2}$ &  & 0.036 & 0.047 & 0.052 & 0.050 & 0.112 & 0.354 & 0.084 & 0.265 & 0.893 & 0.167 & 0.531 & 0.999 \\
$\bar{b}_2$ &  & 0.030 & 0.041 & 0.050 & 0.062 & 0.111 & 0.387 & 0.086 & 0.269 & 0.912 & 0.165 & 0.534 & 1 \\
   \hline \hline  Test & $k'$ & \multicolumn{12}{c|}{$g_0=f_{\textrm{WN}_{0.9}}$}\\
  \hline
  $\phi^{(n);\mu;g_0}_{2}$ & 1 & 0.043 & 0.047 & 0.053 & 0.059 & 0.109 & 0.437 & 0.109 & 0.339 & 0.954 & 0.227 & 0.671 & 1 \\
  $b^*_2$ &  & 0.050 & 0.041 & 0.055 & 0.069 & 0.112 & 0.434 & 0.100 & 0.340 & 0.953 & 0.230 & 0.679 & 1 \\
  $\phi^{(n);g_0}_{2}$ &  & 0.052 & 0.050 & 0.059 & 0.044 & 0.047 & 0.051 & 0.047 & 0.041 & 0.058 & 0.030 & 0.031 & 0.051 \\
 $\bar{b}_2$ &  & 0.024 & 0.044 & 0.060 & 0.033 & 0.044 & 0.055 & 0.032 & 0.039 & 0.063 & 0.033 & 0.036 & 0.065 \\
\hline
  $\phi^{(n);\mu;g_0}_{2}$ & 2 & 0.043 & 0.047 & 0.053 & 0.089 & 0.219 & 0.824 & 0.263 & 0.717 & 1 & 0.564 & 0.982 & 1 \\
  $b^*_2$ &  & 0.050 & 0.041 & 0.055 & 0.091 & 0.227 & 0.825 & 0.263 & 0.731 & 1 & 0.571 & 0.983 & 1 \\
  $\phi^{(n);g_0}_{2}$ &  & 0.052 & 0.050 & 0.059 & 0.050 & 0.068 & 0.128 & 0.055 & 0.088 & 0.354 & 0.051 & 0.124 & 0.446 \\
 $\bar{b}_2$ &  & 0.024 & 0.044 & 0.060 & 0.043 & 0.051 & 0.133 & 0.031 & 0.066 & 0.357 & 0.024 & 0.104 & 0.462 \\
\hline
  $\phi^{(n);\mu;g_0}_{2}$ & 3 & 0.043 & 0.047 & 0.053 & 0.097 & 0.233 & 0.829 & 0.275 & 0.737 & 1 & 0.599 & 0.986 & 1 \\
  $b^*_2$ &  & 0.050 & 0.041 & 0.055 & 0.096 & 0.232 & 0.824 & 0.280 & 0.741 & 1 & 0.608 & 0.988 & 1 \\
  $\phi^{(n);g_0}_{2}$ &  & 0.052 & 0.050 & 0.059 & 0.073 & 0.113 & 0.483 & 0.131 & 0.326 & 0.951 & 0.194 & 0.546 & 0.998 \\
 $\bar{b}_2$ &  & 0.024 & 0.044 & 0.060 & 0.049 & 0.123 & 0.486 & 0.077 & 0.297 & 0.958 & 0.123 & 0.497 & 0.997 \\
\hline
\end{tabular}
}
\end{table}

\begin{table}[!h]
\caption{Rejection rates, for a nominal significance level of
$\alpha=0.05$, of the $\phi^{(n);\mu;g_0}_{2}$,
$b^*_2$ based, $\phi^{(n);g_0}_{2}$ and 
$\bar{b}_2$ based tests calculated using
$1000$ samples of size $n$ simulated from the $k'$-sine-skewed
distribution with the specified base density $g_0$ and values of
$\lambda$ and $k'$.}\label{estsim3} \vspace{3pt}\centering
\scriptsize{
\begin{tabular}{|ll||ccc|ccc|ccc|ccc|}
  \hline
   & $\lambda$ & \multicolumn{3}{c|}{0} & \multicolumn{3}{c|}{0.2} & \multicolumn{3}{c|}{0.4} & \multicolumn{3}{c|}{0.6}\\
   & $n$ & 30 & 100 & 500 & 30 & 100 & 500 & 30 & 100 & 500 & 30 & 100 & 500 \\
  \hline \hline  Test & $k'$ & \multicolumn{12}{c|}{$g_0=f_{\textrm{C}_{0.45}}$}\\
  \hline
  $\phi^{(n);\mu;g_0}_{2}$ & 1 & 0.056 & 0.057 & 0.042 & 0.068 & 0.090 & 0.285 & 0.106 & 0.260 & 0.814 & 0.175 & 0.489 & 0.993 \\
  $b^*_2$ &  & 0.061 & 0.057 & 0.041 & 0.070 & 0.093 & 0.287 & 0.103 & 0.252 & 0.811 & 0.171 & 0.484 & 0.992 \\
  $\phi^{(n);g_0}_{2}$ &  & 0.054 & 0.061 & 0.044 & 0.041 & 0.064 & 0.246 & 0.042 & 0.101 & 0.485 & 0.025 & 0.055 & 0.338 \\
 $\bar{b}_2$ &  & 0.049 & 0.057 & 0.038 & 0.045 & 0.067 & 0.269 & 0.055 & 0.148 & 0.577 & 0.048 & 0.118 & 0.488 \\
\hline
  $\phi^{(n);\mu;g_0}_{2}$ & 2 & 0.056 & 0.057 & 0.042 & 0.120 & 0.289 & 0.905 & 0.336 & 0.830 & 1 & 0.661 & 0.991 & 1 \\
  $b^*_2$ &  & 0.061 & 0.057 & 0.041 & 0.119 & 0.292 & 0.905 & 0.342 & 0.834 & 1 & 0.667 & 0.991 & 1 \\
  $\phi^{(n);g_0}_{2}$ &  & 0.054 & 0.061 & 0.044 & 0.073 & 0.261 & 0.890 & 0.200 & 0.729 & 1 & 0.402 & 0.948 & 1 \\
 $\bar{b}_2$ &  & 0.049 & 0.057 & 0.038 & 0.083 & 0.274 & 0.901 & 0.205 & 0.763 & 1 & 0.373 & 0.950 & 1 \\
\hline
  $\phi^{(n);\mu;g_0}_{2}$ & 3 & 0.056 & 0.057 & 0.042 & 0.057 & 0.094 & 0.296 & 0.113 & 0.237 & 0.829 & 0.157 & 0.493 & 0.989 \\
  $b^*_2$ &  & 0.061 & 0.057 & 0.041 & 0.058 & 0.100 & 0.299 & 0.112 & 0.237 & 0.832 & 0.157 & 0.496 & 0.988 \\
  $\phi^{(n);g_0}_{2}$ &  & 0.054 & 0.061 & 0.044 & 0.051 & 0.092 & 0.287 & 0.070 & 0.217 & 0.816 & 0.105 & 0.435 & 0.987 \\
 $\bar{b}_2$ &  & 0.049 & 0.057 & 0.038 & 0.050 & 0.085 & 0.302 & 0.074 & 0.215 & 0.802 & 0.106 & 0.408 & 0.981 \\
 \hline \hline  Test & $k'$ & \multicolumn{12}{c|}{$g_0=f_{\textrm{WC}_{0.5}}$}\\
  \hline
  $\phi^{(n);\mu;g_0}_{2}$ & 1 & 0.058 & 0.041 & 0.045 & 0.061 & 0.098 & 0.241 & 0.092 & 0.222 & 0.682 & 0.132 & 0.389 & 0.959 \\
  $b^*_2$ &  & 0.054 & 0.043 & 0.047 & 0.062 & 0.097 & 0.241 & 0.084 & 0.213 & 0.691 & 0.133 & 0.387 & 0.962 \\
  $\phi^{(n);g_0}_{2}$ &  & 0.045 & 0.054 & 0.043 & 0.046 & 0.064 & 0.162 & 0.059 & 0.122 & 0.507 & 0.083 & 0.264 & 0.894 \\
  $\bar{b}_2$ &  & 0.038 & 0.053 & 0.053 & 0.051 & 0.069 & 0.227 & 0.077 & 0.187 & 0.719 & 0.120 & 0.427 & 0.979 \\
\hline
  $\phi^{(n);\mu;g_0}_{2}$ & 2 & 0.058 & 0.041 & 0.045 & 0.114 & 0.294 & 0.852 & 0.322 & 0.807 & 1 & 0.629 & 0.985 & 1 \\
  $b^*_2$ &  & 0.054 & 0.043 & 0.047 & 0.110 & 0.292 & 0.852 & 0.325 & 0.808 & 1 & 0.624 & 0.987 & 1 \\
  $\phi^{(n);g_0}_{2}$ &  & 0.045 & 0.054 & 0.043 & 0.069 & 0.144 & 0.478 & 0.113 & 0.372 & 0.957 & 0.219 & 0.628 & 0.997 \\
 $\bar{b}_2$ &  & 0.038 & 0.053 & 0.053 & 0.060 & 0.123 & 0.417 & 0.106 & 0.310 & 0.910 & 0.160 & 0.467 & 0.990 \\
\hline
  $\phi^{(n);\mu;g_0}_{2}$ & 3 & 0.058 & 0.041 & 0.045 & 0.067 & 0.113 & 0.334 & 0.110 & 0.290 & 0.857 & 0.186 & 0.531 & 0.997 \\
  $b^*_2$ &  & 0.054 & 0.043 & 0.047 & 0.067 & 0.114 & 0.336 & 0.110 & 0.289 & 0.859 & 0.190 & 0.543 & 0.996 \\
  $\phi^{(n);g_0}_{2}$ &  & 0.045 & 0.054 & 0.043 & 0.046 & 0.061 & 0.057 & 0.056 & 0.060 & 0.088 & 0.047 & 0.060 & 0.127 \\
 $\bar{b}_2$ &  & 0.038 & 0.053 & 0.053 & 0.051 & 0.074 & 0.152 & 0.074 & 0.128 & 0.413 & 0.097 & 0.206 & 0.679 \\
    \hline
\end{tabular}
}
\end{table}

From a consideration of the results in Tables
\ref{estsim1}--\ref{estsim3}, where $k$ is posited to be 2, it
would appear that the various tests are correctly calibrated apart
from the $\bar{b}_2$ based 
test which tends to be somewhat conservative when $n=30$ and $g_0$
is von Mises or wrapped normal. The results from another
simulation study, not presented here, indicate that the bootstrap
analogue of the test \citep{P02} maintains the nominal
significance better for
samples of size 30. 

As expected, the rejection rates for the different tests generally
increase with the sample size $n$ and the value of $\lambda$, and
are generally highest when $k=k'$. Exceptions to these general
patterns are, in Tables 1 and 2, the $\phi^{(n);\mu;g_0}_{2}$ and
$b^*_2$ tests which perform better for $k'=3$, rather than for
$k'=2$, when $g_0$ is $f_{\text{VM}_{10}}$ or
$f_{\text{WN}_{0.9}}$. In Tables S1 and S2, for the same two $g_0$
densities, the $\phi^{(n);\mu;g_0}_{1}$ and $\phi^{*(n);\mu}_{1}$
tests perform better for $k'=2,3$ than for $k'=1$. The base
$f_{\text{VM}_{10}}$ and $f_{\text{WN}_{0.9}}$ densities are both
highly concentrated and their $k'$-sine-skewed densities, like
their counterparts in the right-hand column of Figure
\ref{fig:kSSWCDens}, are close to unimodal.

When $k\neq k'$, some of the tests perform, at best, like the
trivial test. This is the case, in Tables 1 and 2, for the
$\bar{b}_2$ based and $\phi^{(n);g_0}_2$ 
tests when $k'=1$ and, again, $g_0$ is the
highly concentrated $f_{\text{VM}_{10}}$ or $f_{\text{WN}_{0.9}}$
density. See also the results for the:  $\bar{b}_2$ based 
test when $g_0=f_{\text{VM}_{10}}$ and $k'=1$, in Table S1;
$\phi^{(n);g_0}_1$ and $\bar{b}_2$ based 
tests when $g_0=f_{\text{WN}_{0.9}}$ and $k'=1$, in Table S2;
$\phi^{(n);\mu;g_0}_{1}$ and $\phi^{*(n);\mu}_{1}$ tests when
$g_0=f_{\text{C}_{0.45}}$ and $k'=3$, in Table S3;
$\phi^{(n);g_0}_3$ and $\bar{b}_2$ based 
tests when $g_0=f_{\text{VM}_{10}}$ and $k'=1,2$, in Table S4,
$g_0=f_{\text{WN}_{0.5}}$ and $k'=1$, in Table S5,
$g_0=f_{\text{WN}_{0.9}}$ and $k'=1$, in Table S5;
$\phi^{(n);\mu;g_0}_3$, $\phi^{*(n);\mu}_3$ and $\phi^{(n);g_0}_3$
tests when
$g_0=f_{\text{C}_{0.45}}$ and $k'=1$, in Table S6. So, the problem
is not exclusive to when the base $g_0$ density is highly
concentrated.

The rejection rates for the $\phi^{(n);\mu;g_0}_k$ and
$\phi^{*(n);\mu}_k$ tests in Tables 1-3 and S1--S6 are very
similar. Thus, when $\mu$ is correctly specified, there is little
or no benefit gained from knowing the form of the underlying
density, $g_0$. To aid comparisons, results for the $b^*_2$ based
test have been included in Tables S1--S6. Comparing the results
for the three tests, we conclude that when $\mu$ and $k'$ are
known, the $\phi^{*(n);\mu}_{k'}$ test should be used. Otherwise,
if $k'$ is unknown, the $b^*_2$ based test should be used. The
results in Tables S1--S6 provide an indication of the power loss
or gain associated with this testing strategy.

Again as might be expected, the rejection rates for the
$\phi^{(n);g_0}_{k'}$ 
tests, for which $\mu$ is assumed unknown, are lower than those
for their counterparts $\phi^{(n);\mu;g_0}_{k'}$ and
$\phi^{*(n);\mu}_{k'}$ for which $\mu$ is specified. The same does
not always hold when $k\neq k'$: see, for example, the results for
$k'=1$ and $g_0=f_{\text{WC}_{0.5}}$ in Table 3. Comparing the
results in Tables 1--3 and S1--S6 for the $\phi^{(n);g_0}_{k}$
and $\bar{b}_2$ based 
tests, we conclude that when $\mu$ is unknown but $g_0$ and $k'$
are known, the $\phi^{(n);g_0}_{k'}$ test
should be employed, except, of course, when $g_0$ is von Mises and $k'=1$. 
When $\mu$, $g_0$ and $k'$ are all unknown, we recommend the use
of the $\bar{b}_2$ based test as an omnibus test.

Tables S7--S9 illustrate what can happen to the rejection rates of
the $\phi^{(n);\mu;f_0}_{2}$ and $\phi^{(n);f_0}_{2}$ tests when
the posited density $f_0$ is misspecified. When $f_0$ is more
concentrated than $g_0$, the tests tend to be liberal or very
liberal, respectively: see the results for the
$\phi^{(n);\mu;f_{\textrm{VM}_{10}}}_{2}$ and
$\phi^{(n);f_{\textrm{VM}_{10}}}_{2}$ tests in Tables S8 and S9
and the top half of Table S7. On the other hand, when $f_0$ is
less concentrated than $g_0$, the tests tend to be conservative or
very conservative, respectively: see the results for the
$\phi^{(n);\mu;f_{\textrm{C}_{\rho}}}_{2}$,
$\phi^{(n);\mu;f_{\textrm{WC}_{0.5}}}_{2}$,
$\phi^{(n);f_{\textrm{C}_{\rho}}}_{2}$ and
$\phi^{(n);f_{\textrm{WC}_{0.5}}}_{2}$ tests in the bottom halves
of Tables S7 and S8.

In Tables S10--S12, we observe that the rejection rate of the
$\phi_{f_0;2}^{*(n)}$ test is little affected by the choice of
$f_0$. However, at least for a sample size of $n=30$, the test
tends to be somewhat conservative. When the $\phi^{*(n);\mu}_{2}$
test of \cite{LV14b} is used with $\mu$ estimated from the data,
we obtain the test denoted as $\phi^{*(n);\hat{\mu}^{(n)}}_{2}$.
From the rejection rates for it, we conclude that the test is even
more conservative than its $\phi_{f_0;2}^{*(n)}$ counterpart, the
true size being 0 for all three sample sizes considered when $g_0$
is highly concentrated. For less concentrated $g_0$, its power can
be lower or higher than that of its $\phi_{f_0;2}^{*(n)}$
counterpart, depending on whether $k'$ is less than or greater
than $k$, respectively.

Finally, we also considered the performance of the $b^*_2$ based
and $\bar{b}_2$ based tests for data drawn from distributions
outside the $k$-sine-skewed family. Specifically we simulated data
from: (i) the distribution proposed by \cite{KJ10}
($\text{KJ}_{10}$) with $\mu=0$, $r=0.5$, $\kappa=0.5,0.9$ and
values of the skewness parameter of $\nu=0,0.2,0.4,0.6$; (ii) the
three-parameter asymmetric submodel given in Equation (7) of
\cite{KJ15} ($\text{KJ}_{15}$) with $\mu=0$, $\gamma=0.5,0.9$ and
$\bar{\beta}_2=\nu\gamma(1-\gamma)$ for values of the skewness
parameter of $\nu=0,0.2,0.4,0.6$. The rejection rates obtained are
presented in Table S13. For both choices of $\kappa$ for the
$\text{KJ}_{10}$ distribution, the power of the $b^*_2$ based test
is far higher than that of the $\bar{b}_2$ based test, the latter
being very low. For the $\text{KJ}_{15}$ distribution, the
rejection rates of the two tests are all very similar. For
$\gamma=0.9$ and a sample size of $n=500$, the $\bar{b}_2$ based
test can even be more powerful than the $b^*_2$ based test with
$\mu$ specified.

\subsection{Recommendations}\label{sec:simrec}

\noindent On the basis of the conclusions drawn from the
simulation experiment described above, combined with the
theoretical results obtained in Sections \ref{unkdirknden} and
\ref{unkdirunkden}, we make the following recommendations
concerning the use of the various tests for circular reflective
symmetry.
\begin{enumerate}
    \item When $\mu$ and $k'$ are
\textit{known}, use the $\phi^{*(n);\mu}_{k'}$ test of
\cite{LV14b}.
    \item When $\mu$ is \textit{known} but $k'$ is \textit{unknown}, use the $b^*_2$ based omnibus test of \cite{P04}.
    \item When $\mu$ is \textit{unknown} but $g_0$ and $k'$
are \textit{known}, use the parametric $\phi^{(n);g_0}_{k'}$ test
proposed here, except when $g_0$ is von Mises and $k'=1$. \item
When $\mu$ and $g_0$ are both \textit{unknown} but $k'>1$ is
\textit{known}, use the semi-parametric
$\phi^{*(n)}_{f_{\textrm{VM}_\kappa},k'}$ test proposed here.
\item When $\mu$ and $g_0$ are both \textit{unknown} but $k'=1$ is
\textit{known}, or $\mu$ is \textit{unknown} and it is
\textit{known} that $g_0$ is von Mises and $k'= 1$, use any
semiparametric $\phi_{f_0,k}^{*(n)}$, with $f_0\in\mathcal{F}$,
apart from a von Mises $f_0$.
    \item When $\mu$, $k'$ and $g_0$ are all \textit{unknown} use
the $\bar{b}_2$ based omnibus test of \cite{P02}.
\end{enumerate}

\section{Illustrative applications}\label{sec:data}

\noindent In this section we illustrate the application of various
tests of reflective symmetry in analyses of two datasets taken
from the Biomechanical and Political Sciences literature,
respectively.

\subsection{Cracks in cemented femoral components}\label{sec:data1}

\begin{figure}
\vspace{-30pt}
\hspace{-20pt}    \includegraphics[width=0.57\textwidth]{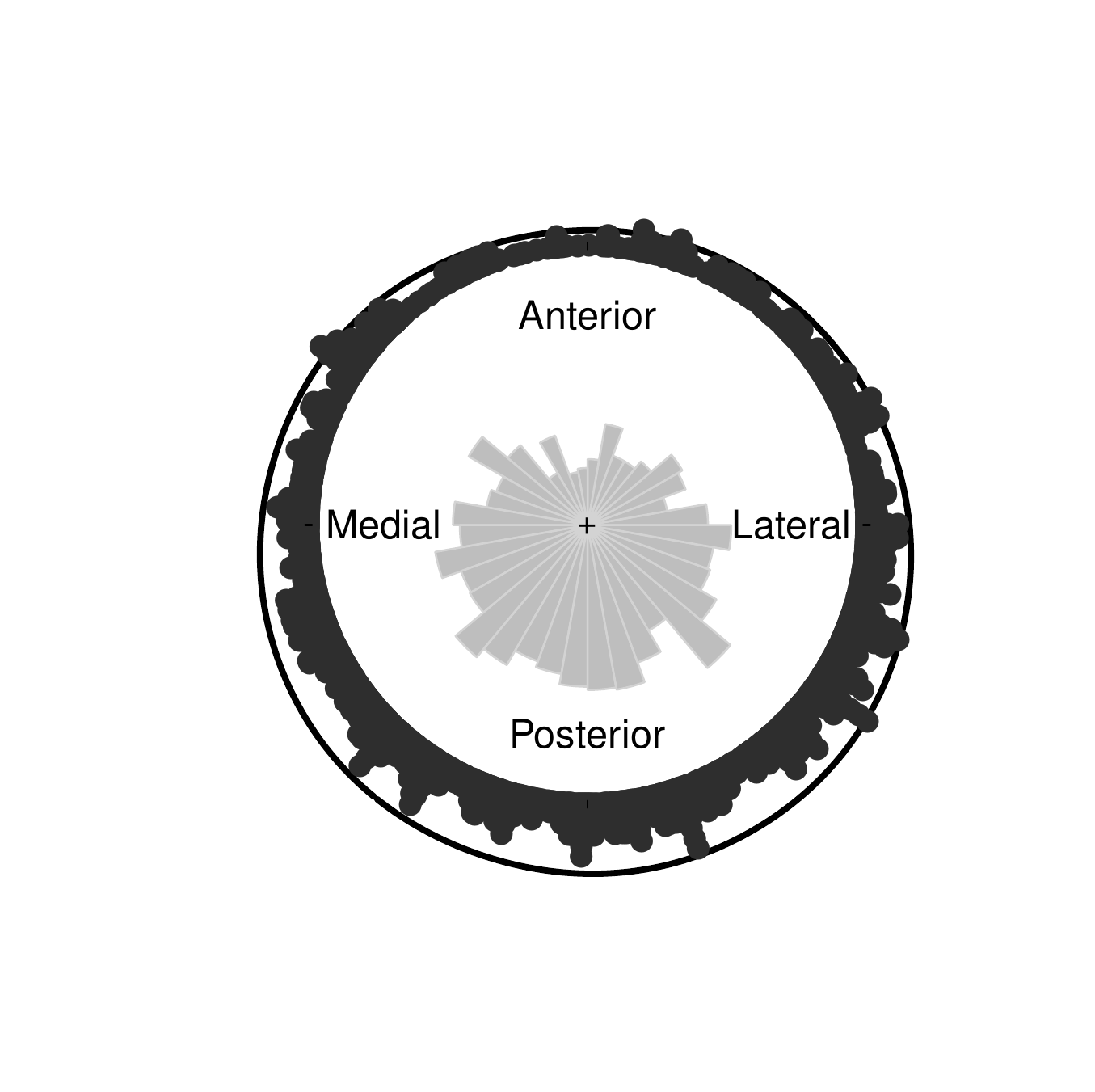}
\hspace{-40pt}
    \includegraphics[width=0.57\textwidth]{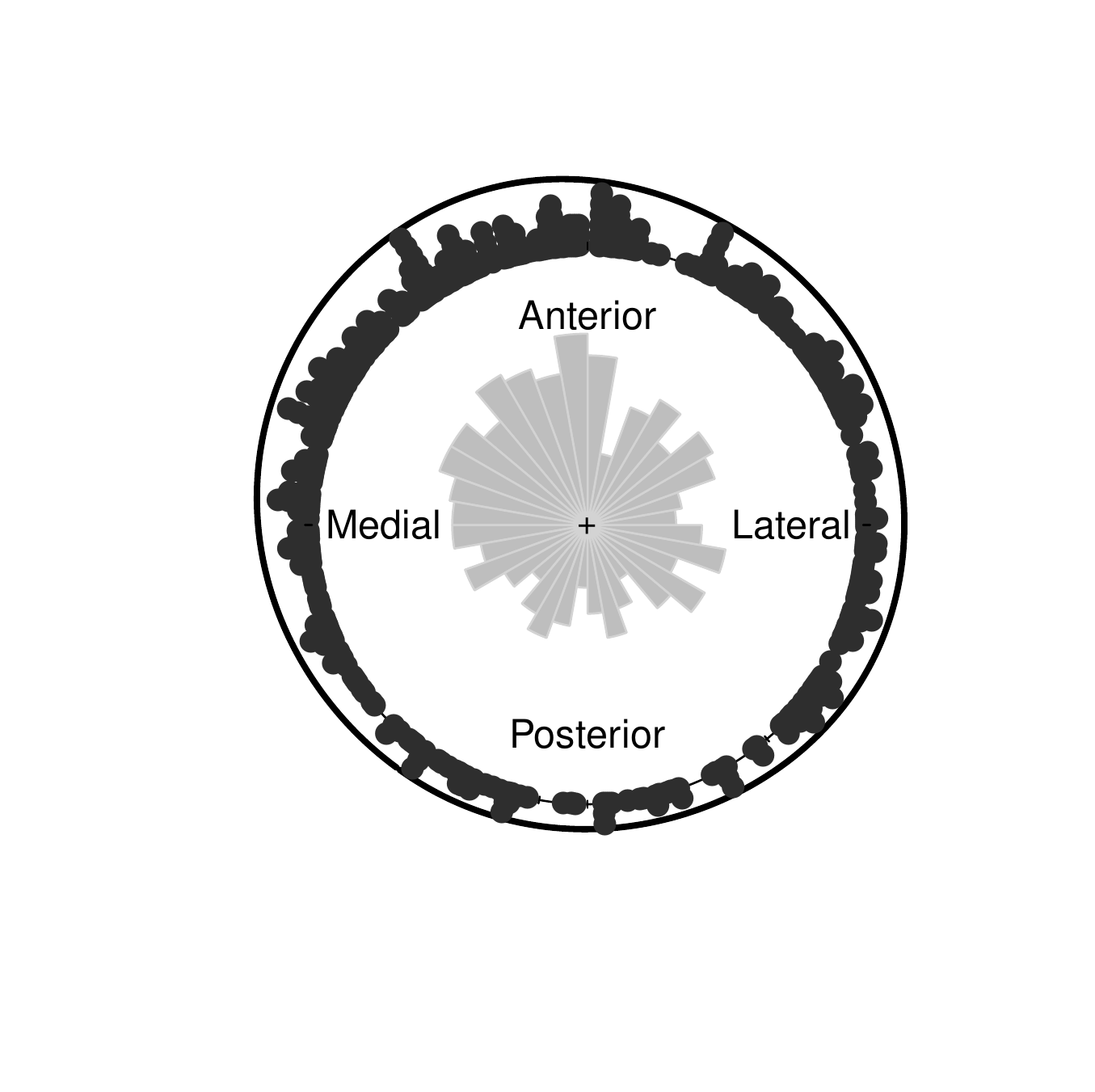} \\ \vspace{-50pt}
 \caption{Raw circular plot, rose diagram and kernel density estimate for the directions of the cracks in the cemented femoral components in the proximal region (left) and
 the distal region (right).}
 \label{figcracks}
\end{figure}

\noindent The first dataset we analyze was collected during an
\textit{in vitro} fatigue study of total hip replacements
described in \cite{MGRMC03}. Here we consider the directions,
measured in angles relative to the centre of the stem, of fatigue
cracks around the cemented femoral components in six hip implants.
After an extended stress cycle had been applied, each femur was
sectioned in 10 mm intervals from the level of the implant collar
to the distal tip of the stem. Measurements at 60 and 70 mm were
not made because of limiting physical constraints imposed by the
experimental setup. As a result, two groups of measurements were
obtained: those in the proximal (10-50 mm) region and those in the
distal (80-110 mm) region. After removing one bone described by
\cite{MGRMC03} as having ``an inferior cement mantle with
substantial stem-cement voids'', the total number of cement cracks
was 2001: 1567 in the proximal region, and 434 in the distal
region. Circular data plots for the two regions, together with
rose diagrams and kernel density estimates obtained using the
plug-in rule of \cite{OCRC12} to select the concentration
parameter, are portrayed in Figure \ref{figcracks}. \cite{MGRMC03}
showed that the directions of the fatigue cracks are not uniformly
distributed and that their distributions in the two regions
differ. Here we investigate whether the cracks in the two regions
are symmetrically distributed about some unknown centre.

The plot on the left-hand side of Figure \ref{figcracks} suggests
that the underlying distribution for the crack directions in the
proximal region is probably unimodal, i.e.\ $k=1$. The emboldened
results in Table \ref{pvalproximal} are the $p$-values for the
parametric tests of Sections \ref{Sec:vMCase}--\ref{Sec:WCDist}
and their semi-parametric counterparts of Sections
\ref{Sec:vMSPCase}--\ref{Sec:WCSPDist} when $k=1,2,3$. The others
are the $p$-values for parametric bootstrap versions of those
tests that assume the concentration parameter to be known, the
mean resultant length having been estimated using the mean
resultant length. The similarity between the non-bootstrapped and
bootstrapped $p$-values in the four pairings where they coincide,
is striking. As in this case $\mu$, $g_0$ and $k'$ are all
unknown, our recommended test in this context is the $\bar{b}_2$
based test. The $p$-value for it is $0.61$, equal to or just
slightly larger than all of the $p$-values  for $k=2,3$ in Table
\ref{pvalproximal}. At least for this dataset and $k=2,3$, then,
the type of test, assumed or posited underlying distribution, and
estimation or not of the concentration parameter, would appear to
have little effect on the $p$-value. For $k=1$, perhaps the more
relevant case for the crack directions under consideration, three
of the $p$-values are larger than $0.61$ and the other is slightly
lower. Clearly, none of the $p$-values in Table \ref{pvalproximal}
provides significant statistical evidence against reflective
symmetry. We note that for these data the sample mean direction is
$-1.644$ radians, just below $-\pi/2 = -1.571$ radians which would
correspond to an estimated mean crack direction in the posterior
region of the
femur. 
The sample skewness, $\bar{b}_2/(1-\bar{R})^{3/2}$, for these data
is $0.017$, corroborating reflective symmetry for the underlying
distribution.

\begin{table}[t]
\caption{P-values for the parametric $\phi^{(n);g_0}_{k}$ test
and, in brackets, the semi-parametric $\phi_{f_0,k}^{*(n)}$ test,
for assumed $g_0$ and $k$ or posited $f_0$ and $k$, respectively,
applied to the $1567$ crack directions in the proximal region. The
emboldened results correspond to tests which do not require the
estimation of the concentration parameter. The others were
obtained using parametric bootstrap versions of the tests with
$\mu$ estimated by the sample mean direction, $\bar{\theta}$,
$\rho$ by the sample mean resultant length, $\bar{R}$, truncated
when necessary to 0.4999, and $B=1000$
bootstrap replications.}\label{pvalproximal} \vspace{5pt}\centering 
\begin{tabular}{l|ccc}
 \hline
$g_0$ or $f_0$ &        $k=1$     &    $k=2$    &     $k=3$\\
\hline
von Mises &  & 0.564 (\textbf{0.610}) &  0.571 (\textbf{0.567})\\
cardioid & 0.886 (0.758) & \textbf{0.590} (0.598) & \textbf{0.567} (0.566)\\
wrapped Cauchy & 0.528 (0.763) & 0.591 (0.577) & 0.582 (0.585)\\
\hline
\end{tabular}
\end{table}

Table \ref{pvaldistal} contains analogous results to those in
Table \ref{pvalproximal} for the crack directions in the distal
region. For these data, the $p$-value of the recommended
$\bar{b}_2$ based test is $0.048$. From a consideration of the
right-hand panel of Figure \ref{figcracks}, there appears to be no
reason to assume that the underlying distribution has any more
than two modes, and so we ignore the results for $k=3$. For
$k=1,2$, nine of the ten $p$-values in Table \ref{pvaldistal} are
equal to or marginally less than that for the $\bar{b}_2$ based
test. The one discordant $p$-value of $0.088$ corresponds to the
bootstrapped version of the parametric test for an assumed wrapped
Cauchy distribution and $k'=1$. Reflective symmetry for the
underlying crack directions in the distal region is thus rejected
at the 5\% significance level, sometimes marginally, by 10 of the
11 tests. For these data the sample mean direction is $1.921$
radians, marginally less than $2\pi/3 = 2.094$ radians which would
correspond to an estimated mean crack direction midway between the
anterior and medial regions of the femur. The sample skewness is
$0.106$, somewhat larger than it was for the crack directions in
the proximal region.

\begin{table}[t]
\caption{P-values for the parametric $\phi^{(n);g_0}_{k}$ test
and, in brackets, the semi-parametric $\phi_{f_0,k}^{*(n)}$ test,
for assumed $g_0$ and $k$ or posited $f_0$ and $k$, respectively,
applied to the $434$ crack directions in the distal region. The
emboldened results correspond to tests which do not require the
estimation of the concentration parameter. The others were
obtained using parametric bootstrap versions of the tests with
$\mu$ estimated by the sample mean direction, $\bar{\theta}$,
$\rho$ by the sample mean resultant length, $\bar{R}$, truncated
when necessary to 0.4999, and $B=1000$
bootstrap replications.}\label{pvaldistal} \vspace{5pt}\centering 
\begin{tabular}{l|ccc}
 \hline
$g_0$ or $f_0$ &        $k=1$     &    $k=2$    &     $k=3$\\
\hline
von Mises &  & 0.034 (\textbf{0.048}) &  0.800 (\textbf{0.796})\\
cardioid & 0.036 (0.025) & \textbf{0.042} (0.040) & \textbf{0.778} (0.736)\\
wrapped Cauchy & 0.088 (0.025) & 0.039 (0.030) & 0.702 (0.757)\\
\hline
\end{tabular}
\end{table}

These results shed further light on the data, and complement the
findings in \cite{MGRMC03} regarding the different distributions
of the crack directions in the two regions.

\subsection{Times of gun crimes}\label{sec:data2}

\begin{figure}
 \centering
    \includegraphics[width=0.4\textwidth]{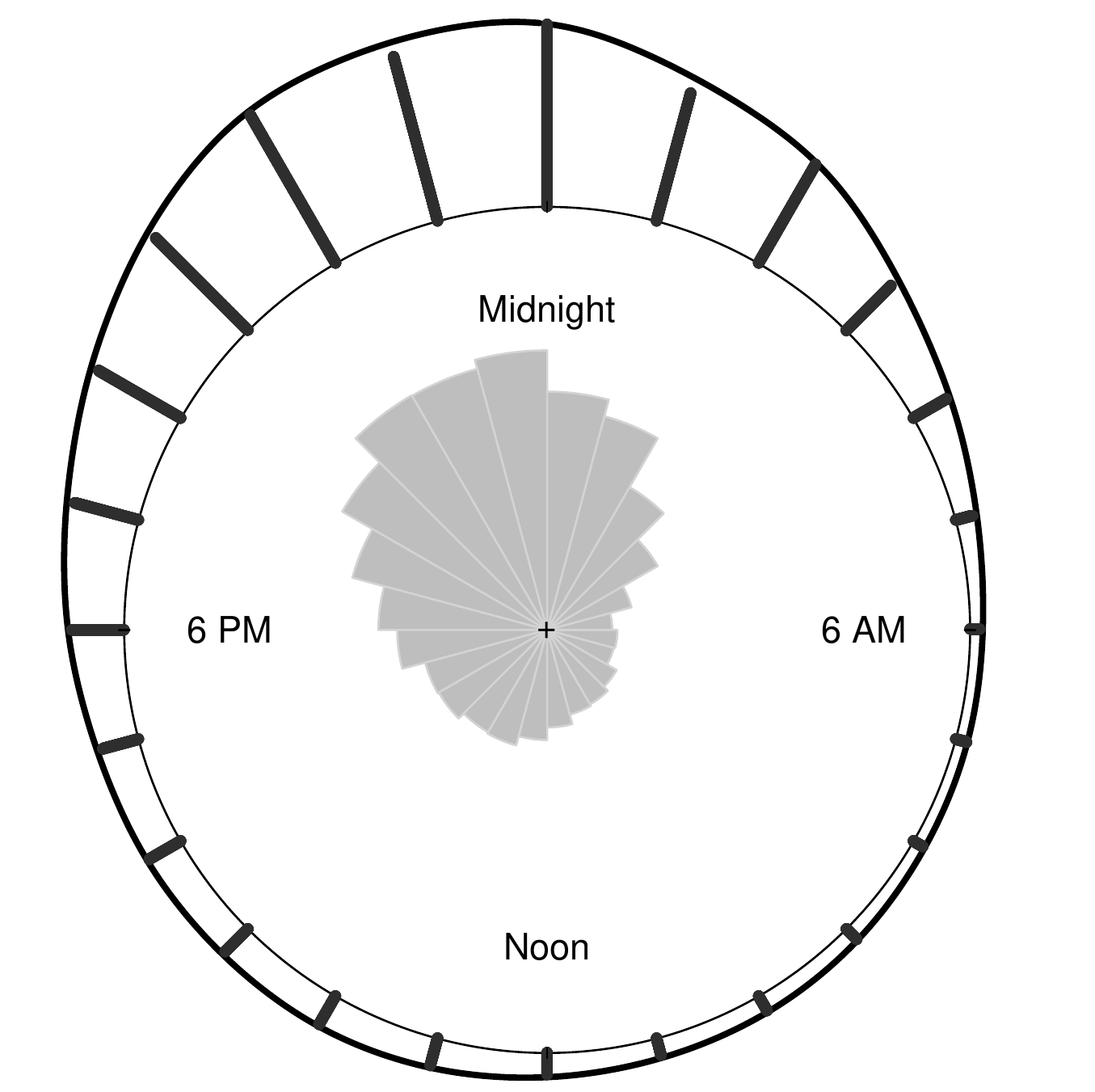}
 \caption{Raw circular plot, rose diagram and kernel density estimate for the times of gun crimes committed in Pittsburgh.}
 \label{figgun}
\end{figure}

\noindent Our second illustrative example involves data on the
times of gun crimes committed in Pittsburgh, Pennsylvania, during
the period from 1st January 1992 to 31st May 1996. The time of
each crime was taken to be the nearest hour to the time it was
reported to the emergency telephone number 911. During the period
in question, there was a total of 15831 registered gun crimes. A
circular plot of the data, together with a rose diagram and a
kernel density estimate calculated using the rule of thumb of
\cite{T08}, are provided in Figure~\ref{figgun}. The data were
first presented in \cite{CG01} and were previously analyzed by
\cite{GH10} to explore their distribution and establish whether
gun crimes were more frequent in the afternoon than in the
morning. The combined plot in Figure \ref{figgun} suggests the
underlying distribution is unimodal and skew.

Suppose we were interested in testing whether the underlying
distribution of the gun crime times was reflectively symmetric
about midnight. Ignoring the fact that the data have been
discretized, converting them to radians and assuming that $k'=1$,
the $p$-value of the semi-parametric $\phi^{*(n);\mu}_{1}$ of
\cite{LV14b} is 0. If, instead, $k'$ is assumed to be unknown and
the recommended $b^*_2$ based test of \cite{P04} applied, the
$p$-value obtained is also 0. Hence, whichever of the two
scenarios is thought to apply, reflective symmetry about midnight
is emphatically rejected.

As there is no obvious reason why the centre of the distribution
should be taken as midnight, we next consider results for tests
which assume that $\mu$ is unknown. Applying the bootstrapped
versions of the parametric and semi-parametric tests proposed here
with $g_0$ ($f_0$) assumed (posited) to be cardioid or wrapped
Cauchy and $k'$ ($k$) assumed (posited) to be 1, all four
$p$-values obtained are also 0. And so is the $p$-value of the
$\bar{b}_2$ based omnibus test. Hence, there is overwhelming
evidence that the distribution underlying the gun crimes is not
reflectively symmetric: neither about midnight, nor about any
other central time. The sample mean direction is $-0.367$ radians,
corresponding to a mean time of 22:40. The sample skewness is
$0.368$, supporting the findings from the tests that the
underlying distribution is not reflectively symmetric.

\section{Discussion}\label{sec:discussion}

\noindent In this paper we have developed tests for circular reflective
symmetry about an unknown centre that are optimal against $k$-sine-skewed alternatives. Recommendations for their
use, as well as other tests that have been proposed in the
literature, were established in the light of the simulation based
results reported in Section \ref{sec:simus}. As mentioned there,
the proposed tests are generally conservative when the sample size
is of the order of 30. In such circumstances, their bootstrap
analogues tend to maintain the nominal significance level better.

In Section \ref{sec:data} we applied bootstrap versions of the
tests proposed here incorporating estimation of the concentration
parameter of $g_0$ or $f_0$. An in-depth treatment of such tests
will be the focus of a future paper. In addition, theoretical
consideration can be given to the non-bootstrap analogues of the
tests presented here when the concentration parameter is
estimated. This would involve: (i) considering a general
location-concentration-skewness model; (ii) establishing the ULAN
property for this general model; (iii) finding conditions under
which the central sequence for the concentration parameter is
independent of the other parameters; (iv) checking if appealing
models satisfy such conditions; (v) deriving test statistics and
investing their optimality properties.

The tests proposed here, as well as those in \cite{LV14b}, are
locally and asymptotically optimal in the Le Cam sense. Clearly,
there are other methodologies one might adopt to derive powerful
tests for reflective symmetry about an unknown centre. One
possibility would be to explore a data-driven approach, similar to
that used by \cite{BBF02} for testing circular uniformity, to
select the value of $k$. Another, presently being developed by the
first author, is to combine the developed test procedures with a
pre-test for the number of modes of the underlying distribution.

In recent years, numerous families of skew-symmetric circular
distributions have been proposed in the literature. \cite{KJ15}
refer to a number of them. The development of powerful tests of
reflective symmetry for use with such families certainly merits
future attention.

Our second illustrative application involved discretized data,
whereas the methodology we have employed assumes the data to be
continuous. Another line of potential future research would be to
develop test procedures for discretized data based on the
bootstrap and symmetrization approaches described in \cite{P02}.

Circular data are just one class of directional data. Others
include bivariate circular data distributed on the torus,
cylindrical data, spherical data and data distributed on the
surfaces of the extensions of such Riemannian manifolds. The
development of tests for reflective symmetry on such manifolds
would be of considerable interest. Ideas underpinning such tests
are explored in \cite{JS83} and \cite{JRA16}.

\section*{Acknowledgments}

\noindent  The research work underpinning this paper was supported
by {grants MTM2016-76969-P (Spanish State Research
Agency, AEI) and MTM2013-41383-P (Spanish Ministry of Economy,
Industry and Competitiveness), both co--funded by the European
Regional Development Fund (ERDF), grant GR15013 (Junta de
Extremadura and the European Union), and funding from the IAP
network P7/06 StUDyS of the Belgian Government (Belgian Science
Policy) and the National Bank of Belgium. Part of the research was carried out by the first author
during a visit to Ghent University supported by grants
BES-2014-071006 and EEBB-I-16-11503 (Spanish Ministry of Economy
and Competitiveness)}.

\appendix

{
\section{Proof of Lemma \ref{consistency}}\label{sec:prooflem1}
\noindent We show that
$\hat{\Gamma}_{f_0,g_0,k;11}-{\Gamma}_{f_0,g_0,k;11}=o_{\rm P}(1)$
as $\ny$ under ${\rm P}^{(n)}_{(\mu,0);g_0}$. Showing that
$\hat{\Gamma}_{g_0,k;12}-{\Gamma}_{g_0,k;12}=o_{\rm P}(1)$ as
$\ny$ under ${\rm P}^{(n)}_{(\mu,0);g_0}$ proceeds along the same
lines. In this proof, we set $\mu\n:= \mu+n^{-1/2} \tau_1\n$ for
some bounded sequence $\tau_1\n$ as in Theorem~\ref{ULAN}. Due to
the local discreteness of $\hat{\mu}\n$ (Assumption B), it is
sufficient to show that $$\frac{1}{n} \sum_{i=1}^n
\dot{\varphi}_{f_0}\left(\Theta_i-{\mu}^{(n)}\right)-{\rm
E}_{g_0}[\dot{\varphi}_{f_0}\left(\Theta_i-{\mu}\right)]=o_{\rm
P}(1)$$ as $\ny$ under ${\rm P}^{(n)}_{(\mu,0);g_0}$. The law of
large numbers leads to
$$\frac{1}{n}  \sum_{i=1}^n
\dot{\varphi}_{f_0}\left(\Theta_i-{\mu}\right)-{\rm E}_{g_0}[\dot{\varphi}_{f_0}\left(\Theta_i-{\mu}\right)]=o_{\rm P}(1)$$
as $\ny$ under ${\rm P}^{(n)}_{(\mu,0);g_0}$ so that it only remains to show that
$$S_n:=\frac{1}{n}  \sum_{i=1}^n
(\dot{\varphi}_{f_0}\left(\Theta_i-{\mu}\n\right)-\dot{\varphi}_{f_0}\left(\Theta_i-{\mu}\right))$$
is $o_{\rm P}(1)$ as $\ny$ under ${\rm P}^{(n)}_{(\mu,0);g_0}$. As
the $\Theta_i$ are i.i.d.,
\begin{eqnarray*}
{\rm E}[| S_n|] & \leq & \frac{1}{n}  \sum_{i=1}^n
{\rm E}[|(\dot{\varphi}_{f_0}\left(\Theta_i-{\mu}\n\right)-\dot{\varphi}_{f_0}\left(\Theta_i-{\mu}\right))|] \\
&=& {\rm E}[|(\dot{\varphi}_{f_0}\left(\Theta_1-{\mu}\n\right)-\dot{\varphi}_{f_0}\left(\Theta_1-{\mu}\right))|].
\end{eqnarray*}
Since $\dot{\varphi}_{f_0}$ is continuous on a compact support, it
is bounded. The result then follows by applying Lebesgue's
dominated convergence theorem. } \cqfd
\section{Proof of Lemma \ref{lemunk1}}\label{sec:proof}

\noindent We start by showing that
$\Delta_{f_0,g_0,k;2}^{(n);ecd}\left(\hat{\mu}^{(n)}\right)-\Delta_{f_0,g_0,k;2}^{(n);ecd}(\mu)=o_{\rm
P}(1)$ under ${\rm P}^{(n)}_{(\mu,0);g_0}$ as $\ny$. First note
that, due to Assumption C, under ${\rm P}^{(n)}_{(\mu,0);g_0}$ as
$\ny$,
\begin{eqnarray} \label{firststep}
n^{-1/2} \sum_{i=1}^n \varphi_{f_0}(\Theta_i- \hat{\mu}\n) &=& \Delta_{f_0,k;1}^{(n)} -{\rm E}_{g_0} [\dot{\varphi}_{f_0}(\Theta_i- {\mu})] n^{1/2} (\hat{\mu}^{(n)}- \mu)+ o_{\rm P}(1) \nonumber \\
&=& \Delta_{f_0,k;1}^{(n)}-\Gamma_{f_0,g_0,k;11} n^{1/2}
(\hat{\mu}^{(n)}- \mu)+ o_{\rm P}(1).
\end{eqnarray} Therefore, using \eqref{unk1} with~\eqref{firststep}, it follows that
\begin{eqnarray*}
\Delta_{f_0,g_0,k;2}^{(n);ecd}\left(\hat{\mu}^{(n)}\right) &=& \Delta_{k;2}^{(n)}(\mu)- \eta  \Delta_{f_0,k;1}^{(n)} - (\Gamma_{g_0,k;12}- \eta \Gamma_{f_0,g_0,k;11}) n^{1/2} (\hat{\mu}^{(n)}- \mu)+ o_{\rm P}(1) \\
&=& \Delta_{f_0,g_0,k;2}^{(n);ecd}\left({\mu}\right)+ o_{\rm P}(1),
\end{eqnarray*}
since $(\Gamma_{g_0,k;12}- \eta \Gamma_{f_0,g_0,k;11})=0$. It
remains to show that
\begin{eqnarray}\label{unk4}
\Delta_{f_0,k;2}^{*(n);ecd}\left(\hat{\mu}^{(n)}\right)-\Delta_{f_0,g_0,k;2}^{(n);ecd}\left(\hat{\mu}^{(n)}\right)&=&-\left(\frac{\hat{\Gamma}_{g_0,k;12}}{\hat{\Gamma}_{f_0,g_0,k;11}}-\frac{\Gamma_{g_0,k;12}}{\Gamma_{f_0,g_0,k;11}}\right)\Delta_{f_0,k;1}^{(n)}\left(\hat{\mu}^{(n)}\right)\nonumber\\
&=&o_{\rm P}(1)
\end{eqnarray}
under ${\rm P}^{(n)}_{(\mu,0);g_0}$ as $\ny$. To prove
(\ref{unk4}), first note that \eqref{firststep} and the central
limit theorem (CLT) imply that
$\Delta_{f_0,k;1}^{(n)}\left(\hat{\mu}^{(n)}\right)$ is $O_{\rm
P}(1)$ under ${\rm P}^{(n)}_{(\mu,0);g_0}$ as $\ny$. Therefore, we
only need to show that
\begin{equation} \label{little}
\frac{\hat{\Gamma}_{g_0,k;12}}{\hat{\Gamma}_{f_0,g_0,k;11}}-\frac{\Gamma_{g_0,k;12}}{\Gamma_{f_0,g_0,k;11}}=o_{\rm P}(1)
\end{equation}
 as $\ny$ under ${\rm P}^{(n)}_{(\mu,0);g_0}$ as $\ny$. Since
\begin{eqnarray}\label{unk5}
\frac{\hat{\Gamma}_{g_0,k;12}}{\hat{\Gamma}_{f_0,g_0,k;11}}-\frac{\Gamma_{g_0,k;12}}{\Gamma_{f_0,g_0,k;11}}&=&
\frac{\hat{\Gamma}_{g_0,k;12}-\Gamma_{g_0,k;12}}{\hat{\Gamma}_{f_0,g_0,k;11}}-\frac{\Gamma_{g_0,k;12}\left(
\hat{\Gamma}_{f_0,g_0,k;11} - \Gamma_{f_0,g_0,k;11}
\right)}{\hat{\Gamma}_{f_0,g_0,k;11}
\Gamma_{f_0,g_0,k;11}},\nonumber
\end{eqnarray}
the result follows directly from Lemma \ref{consistency}.

Turning to the proof of (ii), and working along the same lines as
those at the end of the proof of Lemma \ref{consistency}, we
easily obtain that
\begin{equation} \label{lln1}
n^{-1} \sum_{i=1}^n \varphi_{f_0}^2(\Theta_i- \hat{\mu}\n)-{\rm E}_{g_0}[\varphi_{f_0}^2(\Theta_i- {\mu})]
\end{equation}
and
\begin{equation} \label{lln2}
n^{-1} \sum_{i=1}^n \sin (k(\Theta_i- \hat{\mu}\n)) \varphi_{f_0}(\Theta_i- \hat{\mu}\n)-{\rm E}_{g_0}[\sin (k(\Theta_i- {\mu})) \varphi_{f_0}(\Theta_i- {\mu})]
\end{equation}
are $o_{\rm P}(1)$ under ${\rm P}^{(n)}_{(\mu,0);g_0}$ as $\ny$.
It follows that
$C^{(n)}_{f_0,g_0,k}\left(\hat{\mu}^{(n)}\right)-C^{(n)}_{f_0,g_0,k}\left(\mu\right)
=o_{\rm P}(1)$ under ${\rm P}^{(n)}_{(\mu,0);g_0}$ as $\ny$.
Therefore it remains to show that
$C^{*(n)}_{f_0,k}\left(\hat{\mu}^{(n)}\right)
-C^{(n)}_{f_0,g_0,k}\left(\hat{\mu}^{(n)}\right)$ is $o_{\rm
P}(1)$ under ${\rm P}^{(n)}_{(\mu,0);g_0}$ as $\ny$ . We readily
obtain that
\begin{eqnarray*}
&&C^{*(n)}_{f_0,k}\left(\hat{\mu}^{(n)}\right) -C^{(n)}_{f_0,g_0,k}\left(\hat{\mu}^{(n)}\right) =  \left(\frac{\hat{\Gamma}_{g_0,k;12}^2}{\hat{\Gamma}_{f_0,g_0,k;11}^2}-\frac{\Gamma_{g_0,k;12}^2}{\Gamma_{f_0,g_0,k;11}^2}\right) n^{-1} \sum_{i=1}^n \varphi_{f_0}^2(\Theta_i- \hat{\mu}\n) \\
&&\quad\quad-2
\left(\frac{\hat{\Gamma}_{g_0,k;12}}{\hat{\Gamma}_{f_0,g_0,k;11}}-\frac{\Gamma_{g_0,k;12}}{\Gamma_{f_0,g_0,k;11}}\right) n^{-1} \sum_{i=1}^n \sin (k(\Theta_i- \hat{\mu}\n)) \varphi_{f_0}(\Theta_i- \hat{\mu}\n)
\end{eqnarray*}
so that \eqref{lln1} and \eqref{lln2} together with \eqref{little}
and the continuous mapping theorem imply that
$C^{*(n)}_{f_0,k}\left(\hat{\mu}^{(n)}\right)
-C^{(n)}_{f_0,g_0,k}\left(\hat{\mu}^{(n)}\right)$ is $o_{\rm
P}(1)$ under ${\rm P}^{(n)}_{(\mu,0);g_0}$ as $\ny$. The result
follows.

 \cqfd

\section{Proof of Theorem \ref{thunk}}\label{sec:proofth}

\noindent Fix $g_0\in \mathcal{F}$ and $\mu\in[-\pi,\pi)$.
{Lemma~\ref{lemunk1} combined with Slutsky's lemma leads to
\begin{equation}\label{41i}
Q^{*(n)}_{f_0,k}=\frac{\Delta_{f_0,g_0,k;2}^{(n);ecd}(\mu)}{C^{(n)}_{f_0,g_0,k}\left(\mu\right)}+o_P(1)=\frac{\Delta_{f_0,g_0,k;2}^{(n);ecd}(\mu)}{V^{g_0}_{f_0}(k)}+o_P(1)
\end{equation}
as $n\rightarrow \infty$ under ${\rm P}^{(n)}_{(\mu,0)';g_0}$.
Part (i) then follows from the CLT.}

Part (ii) is obtained via Le Cam's third lemma. First, it is
necessary to calculate the joint distribution of
$\Delta_{f_0,k;2}^{*(n);ecd}\left(\hat{\mu}^{(n)}\right)$ and
$\log(d{\rm P}^{(n)}_{(\mu,n^{-1/2}\tau_2^{(n)})';g_0,k'}/d{\rm
P}^{(n)}_{(\mu,0)';g_0})$ under ${\rm P}^{(n)}_{(\mu,0)';g_0}$. We
use Lemma \ref{lemunk1} and the fact that

\begin{eqnarray*}
&&  \frac{1}{\sqrt{n}} \sum_{i=1}^n \left(
\begin{array}{c}
 \sin(k( \Theta_i- \mu)) -\eta \varphi_{f_0}(\Theta_i- \mu)\\
 \tau_2^{(n)} \sin(k' (\Theta_i- \mu))
 \end{array}
\right)-\left(\begin{array}{c}
0\\
\frac{1}{2} (\tau_2^{(n)})^2 \Gamma_{g_0,k';22}
\end{array}
\right) \overset{\mathcal{D}}{\rightarrow} \\ && \mathcal{N}_2\left(  \left(
\begin{array}{c}
 0\\
-\frac{1}{2} (\tau_2)^2 \Gamma_{g_0,k';22}
\end{array}
\right),
\left(
\begin{array}{cc}
V^{g_0}_{f_0}(k)& \tau_2 C^{g_0}_{f_0}(k,k') \\
\tau_2 C^{g_0}_{f_0}(k,k') &(\tau_2)^2 \Gamma_{g_0,k';22}
\end{array}
\right)
\right) \\
\end{eqnarray*}
as $n\rightarrow\infty$ under ${\rm P}^{(n)}_{(\mu,0)';g_0}$,
obtained using the multivariate CLT. Now, since ${\rm
P}^{(n)}_{(\mu,0)';g_0}$ and ${\rm
P}^{(n)}_{(\mu,n^{-1/2}\tau_2^{(n)})';g_0,k'}$ are mutually
contiguous, applying Le Cam's third lemma we obtain that
$\Delta_{f_0,k;2}^{*(n);ecd}\left(\hat{\mu}^{(n)}\right)
\overset{\mathcal{D}}{\rightarrow}\mathcal{N} (\tau_2
C^{g_0}_{f_0}(k,k'), V^{g_0}_{f_0}(k))$ under ${\rm
P}^{(n)}_{(\mu,n^{-1/2}\tau_2^{(n)})';g_0,k'}$ as $n\rightarrow
\infty$.

Part (iii) can be shown by combining result (\ref{41i}) under
${\rm P}^{(n)}_{(\mu,0)';f_0}$ with the result from the beginning
of Section \ref{unkdirunkden}, namely that
$\Delta_{f_0,f_0,k;2}^{(n);ecd}(\mu)-\Delta_{f_0,k;2}^{(n)eff}(\mu)=o_{\rm
P}(1)$ as $\ny$ under ${\rm P}^{(n)}_{(\mu,0)',f_0}$, and
therefore under contiguous alternatives, together with the
optimality of the parametric test $\phi_{f_{0},k}^{(n)}$.

 \cqfd

\section{Additional results from the Monte Carlo studies}\label{sec:addsm}

In Tables \ref{estsim4}--\ref{estsim13} we present
additional rejection rates to complement those presented in Tables
1--3 for the Monte Carlo experiments referred to in Section 4 of
the paper.

\begin{table}[!h]
\caption{Rejection rates, for a nominal significance level of
$\alpha=0.05$, of the $\phi^{(n);\mu;g_0}_{1}$,
$\phi^{*(n);\mu}_{1}$, $b^*_2$ based 
and $\bar{b}_2$ based tests calculated using
$1000$ samples of size $n$ simulated from the $k'$-sine-skewed
distribution with the specified base von Mises density $g_0$ and
values of $\lambda$ and $k'$. Results for the $\phi^{(n);g_0}_{1}$
test are not included because when $g_0$ is von Mises it is a
trivial test.}\label{estsim4} \vspace{3pt} \centering \scriptsize{
\begin{tabular}{|ll||ccc|ccc|ccc|ccc|}
  \hline
   & $\lambda$ & \multicolumn{3}{c|}{0} & \multicolumn{3}{c|}{0.2} & \multicolumn{3}{c|}{0.4} & \multicolumn{3}{c|}{0.6}\\
   & $n$ & 30 & 100 & 500 & 30 & 100 & 500 & 30 & 100 & 500 & 30 & 100 & 500 \\
   \hline \hline  Test & $k'$ & \multicolumn{12}{c|}{$g_0=f_{\textrm{VM}_1}$}\\
  \hline
$\phi^{(n);\mu;g_0}_{1}$ & 1& 0.040 & 0.050 & 0.052 & 0.125 & 0.302 & 0.862 & 0.326 & 0.788 & 1 & 0.636 & 0.995 & 1 \\
  $\phi^{*(n);\mu}_{1}$ &  & 0.043 & 0.052 & 0.052 & 0.115 & 0.302 & 0.865 & 0.330 & 0.791 & 1 & 0.650 & 0.996 & 1 \\
  $b^*_2$ & & 0.053 & 0.043 & 0.048 & 0.051 & 0.088 & 0.298 & 0.095 & 0.232 & 0.802 & 0.148 & 0.473 & 0.990 \\
$\bar{b}_2$ &  & 0.033 & 0.036 & 0.050 & 0.029 & 0.043 & 0.056 & 0.038 & 0.038 & 0.068 & 0.044 & 0.079 & 0.248 \\
  \hline $\phi^{(n);\mu;g_0}_{1}$ &2& 0.040 & 0.050 & 0.052 & 0.076 & 0.116 & 0.337 & 0.127 & 0.281 & 0.845 & 0.196 & 0.512 & 0.993 \\
  $\phi^{*(n);\mu}_{1}$ &  & 0.043 & 0.052 & 0.052 & 0.074 & 0.114 & 0.334 & 0.127 & 0.279 & 0.845 & 0.190 & 0.516 & 0.992 \\
  $b^*_2$ & & 0.053 & 0.043 & 0.048 & 0.111 & 0.293 & 0.897 & 0.347 & 0.826 & 1 & 0.646 & 0.995 & 1 \\
$\bar{b}_2$ &  & 0.033 & 0.036 & 0.050 & 0.039 & 0.183 & 0.777 & 0.139 & 0.550 & 0.999 & 0.235 & 0.747 & 1 \\
  \hline $\phi^{(n);\mu;g_0}_{1}$ &3& 0.040 & 0.050 & 0.052 & 0.058 & 0.061 & 0.084 & 0.052 & 0.051 & 0.131 & 0.060 & 0.083 & 0.204 \\
  $\phi^{*(n);\mu}_{1}$ &  & 0.043 & 0.052 & 0.052 & 0.060 & 0.060 & 0.085 & 0.049 & 0.057 & 0.127 & 0.061 & 0.084 & 0.203 \\
  $b^*_2$ & & 0.053 & 0.043 & 0.048 & 0.061 & 0.105 & 0.319 & 0.107 & 0.259 & 0.823 & 0.172 & 0.482 & 0.991 \\
$\bar{b}_2$ &  & 0.033 & 0.036 & 0.050 & 0.032 & 0.065 & 0.288 & 0.057 & 0.198 & 0.777 & 0.086 & 0.343 & 0.975 \\
   \hline \hline  Test & $k'$ & \multicolumn{12}{c|}{$g_0=f_{\textrm{VM}_{10}}$}\\
  \hline
$\phi^{(n);\mu;g_0}_{1}$ & 1& 0.051 & 0.046 & 0.043 & 0.059 & 0.093 & 0.282 & 0.101 & 0.241 & 0.788 & 0.168 & 0.491 & 0.984 \\
  $\phi^{*(n);\mu}_{1}$ &  & 0.045 & 0.043 & 0.044 & 0.057 & 0.095 & 0.285 & 0.101 & 0.235 & 0.793 & 0.167 & 0.482 & 0.986 \\
  $b^*_2$ & & 0.051 & 0.042 & 0.044 & 0.058 & 0.090 & 0.282 & 0.095 & 0.234 & 0.791 & 0.162 & 0.464 & 0.986 \\
$\bar{b}_2$ &  & 0.028 & 0.040 & 0.040 & 0.029 & 0.037 & 0.034 & 0.029 & 0.030 & 0.041 & 0.027 & 0.038 & 0.053 \\
  \hline $\phi^{(n);\mu;g_0}_{1}$ &2& 0.051 & 0.046 & 0.043 & 0.082 & 0.179 & 0.681 & 0.224 & 0.575 & 0.998 & 0.413 & 0.903 & 1 \\
  $\phi^{*(n);\mu}_{1}$ &  & 0.045 & 0.043 & 0.044 & 0.078 & 0.186 & 0.693 & 0.218 & 0.584 & 0.997 & 0.434 & 0.913 & 1 \\
  $b^*_2$ & & 0.051 & 0.042 & 0.044 & 0.083 & 0.190 & 0.699 & 0.230 & 0.586 & 0.998 & 0.431 & 0.918 & 1 \\
$\bar{b}_2$ &  & 0.028 & 0.040 & 0.040 & 0.029 & 0.044 & 0.046 & 0.038 & 0.042 & 0.068 & 0.037 & 0.047 & 0.068 \\
  \hline $\phi^{(n);\mu;g_0}_{1}$ &3& 0.051 & 0.046 & 0.043 & 0.101 & 0.221 & 0.811 & 0.259 & 0.702 & 1 & 0.535 & 0.970 & 1 \\
  $\phi^{*(n);\mu}_{1}$ &  & 0.045 & 0.043 & 0.044 & 0.101 & 0.231 & 0.808 & 0.263 & 0.718 & 1 & 0.563 & 0.975 & 1 \\
  $b^*_2$ & & 0.051 & 0.042 & 0.044 & 0.104 & 0.243 & 0.842 & 0.280 & 0.743 & 1 & 0.581 & 0.982 & 1 \\
$\bar{b}_2$ &  & 0.028 & 0.040 & 0.040 & 0.034 & 0.057 & 0.146 & 0.033 & 0.087 & 0.397 & 0.031 & 0.091 & 0.524 \\
   \hline
\end{tabular}
}

\end{table}

\begin{table}[!h]
\caption{Rejection rates, for a nominal significance level of
$\alpha=0.05$, of the $\phi^{(n);\mu;g_0}_{1}$,
$\phi^{*(n);\mu}_{1}$, $b^*_2$ based 
and $\bar{b}_2$ based tests
calculated using $1000$ samples of size $n$ simulated from the
$k'$-sine-skewed distribution with the specified base wrapped
normal density $g_0$ and values of $\lambda$ and
$k'$.}\label{estsim5} \vspace{3pt} \centering \scriptsize{
\begin{tabular}{|ll||ccc|ccc|ccc|ccc|}
  \hline
   & $\lambda$ & \multicolumn{3}{c|}{0} & \multicolumn{3}{c|}{0.2} & \multicolumn{3}{c|}{0.4} & \multicolumn{3}{c|}{0.6}\\
   & $n$ & 30 & 100 & 500 & 30 & 100 & 500 & 30 & 100 & 500 & 30 & 100 & 500 \\
   \hline \hline  Test & $k'$ & \multicolumn{12}{c|}{$g_0=f_{\textrm{WN}_{0.5}}$}\\
  \hline
$\phi^{(n);\mu;g_0}_{1}$ & 1& 0.045 & 0.048 & 0.046 & 0.114 & 0.238 & 0.879 & 0.301 & 0.792 & 1 & 0.629 & 0.990 & 1 \\
  $\phi^{*(n);\mu}_{1}$ &  & 0.046 & 0.049 & 0.047 & 0.109 & 0.233 & 0.880 & 0.303 & 0.789 & 1 & 0.647 & 0.991 & 1 \\
  $b^*_2$ & & 0.052 & 0.053 & 0.049 & 0.065 & 0.112 & 0.351 & 0.111 & 0.294 & 0.895 & 0.212 & 0.579 & 0.998 \\
  $\phi^{(n);g_0}_{1}$ &  & 0.039 & 0.047 & 0.063 & 0.052 & 0.048 & 0.155 & 0.041 & 0.065 & 0.276 & 0.022 & 0.044 & 0.152 \\
$\bar{b}_2$ &  & 0.030 & 0.041 & 0.050 & 0.039 & 0.059 & 0.148 & 0.037 & 0.077 & 0.257 & 0.041 & 0.058 & 0.167 \\
  \hline $\phi^{(n);\mu;g_0}_{1}$ &2& 0.045 & 0.048 & 0.046 & 0.052 & 0.113 & 0.357 & 0.119 & 0.296 & 0.913 & 0.209 & 0.586 & 0.999 \\
  $\phi^{*(n);\mu}_{1}$ &  & 0.046 & 0.049 & 0.047 & 0.057 & 0.112 & 0.354 & 0.120 & 0.301 & 0.914 & 0.207 & 0.588 & 0.999 \\
  $b^*_2$ & & 0.052 & 0.053 & 0.049 & 0.112 & 0.289 & 0.896 & 0.329 & 0.825 & 1 & 0.669 & 0.996 & 1 \\
  $\phi^{(n);g_0}_{1}$ &  & 0.039 & 0.047 & 0.063 & 0.083 & 0.218 & 0.783 & 0.176 & 0.626 & 0.998 & 0.361 & 0.902 & 1 \\
$\bar{b}_2$ &  & 0.030 & 0.041 & 0.050 & 0.084 & 0.249 & 0.866 & 0.186 & 0.728 & 1 & 0.355 & 0.930 & 1 \\
  \hline $\phi^{(n);\mu;g_0}_{1}$ &3& 0.045 & 0.048 & 0.046 & 0.049 & 0.043 & 0.057 & 0.051 & 0.048 & 0.088 & 0.041 & 0.049 & 0.109 \\
  $\phi^{*(n);\mu}_{1}$ &  & 0.046 & 0.049 & 0.047 & 0.051 & 0.045 & 0.060 & 0.050 & 0.049 & 0.089 & 0.047 & 0.052 & 0.107 \\
  $b^*_2$ & & 0.052 & 0.053 & 0.049 & 0.061 & 0.118 & 0.363 & 0.114 & 0.277 & 0.889 & 0.189 & 0.558 & 0.999 \\
  $\phi^{(n);g_0}_{1}$ &  & 0.039 & 0.047 & 0.063 & 0.050 & 0.064 & 0.088 & 0.047 & 0.070 & 0.186 & 0.044 & 0.100 & 0.353 \\
$\bar{b}_2$ &  & 0.030 & 0.041 & 0.050 & 0.062 & 0.111 & 0.387 & 0.086 & 0.269 & 0.912 & 0.165 & 0.534 & 1 \\
   \hline \hline  Test & $k'$ & \multicolumn{12}{c|}{$g_0=f_{\textrm{WN}_{0.9}}$}\\
  \hline
$\phi^{(n);\mu;g_0}_{1}$ & 1& 0.037 & 0.042 & 0.055 & 0.064 & 0.117 & 0.460 & 0.111 & 0.367 & 0.966 & 0.236 & 0.694 & 1 \\
  $\phi^{*(n);\mu}_{1}$ &  & 0.048 & 0.043 & 0.054 & 0.067 & 0.115 & 0.462 & 0.113 & 0.364 & 0.967 & 0.253 & 0.703 & 1 \\
  $b^*_2$ & & 0.050 & 0.041 & 0.055 & 0.069 & 0.112 & 0.434 & 0.100 & 0.340 & 0.953 & 0.230 & 0.679 & 1 \\
  $\phi^{(n);g_0}_{1}$ &  & 0.051 & 0.056 & 0.052 & 0.045 & 0.049 & 0.052 & 0.043 & 0.041 & 0.062 & 0.032 & 0.030 & 0.045 \\
$\bar{b}_2$ &  & 0.024 & 0.044 & 0.060 & 0.033 & 0.044 & 0.055 & 0.032 & 0.039 & 0.063 & 0.033 & 0.036 & 0.065 \\
  \hline $\phi^{(n);\mu;g_0}_{1}$ &2& 0.037 & 0.042 & 0.055 & 0.098 & 0.213 & 0.787 & 0.239 & 0.685 & 1 & 0.532 & 0.974 & 1 \\
  $\phi^{*(n);\mu}_{1}$ &  & 0.048 & 0.043 & 0.054 & 0.093 & 0.212 & 0.789 & 0.249 & 0.703 & 1 & 0.558 & 0.979 & 1 \\
$b^*_2$ & & 0.050 & 0.041 & 0.055 & 0.091 & 0.227 & 0.825 & 0.263 & 0.731 & 1 & 0.571 & 0.983 & 1 \\
  $\phi^{(n);g_0}_{1}$ &  & 0.051 & 0.056 & 0.052 & 0.048 & 0.068 & 0.133 & 0.054 & 0.087 & 0.349 & 0.055 & 0.140 & 0.460 \\
$\bar{b}_2$ &  & 0.024 & 0.044 & 0.060 & 0.043 & 0.051 & 0.133 & 0.031 & 0.066 & 0.357 & 0.024 & 0.104 & 0.462 \\
  \hline $\phi^{(n);\mu;g_0}_{1}$ &3& 0.037 & 0.042 & 0.055 & 0.085 & 0.183 & 0.705 & 0.223 & 0.637 & 1 & 0.470 & 0.946 & 1 \\
  $\phi^{*(n);\mu}_{1}$ &  & 0.048 & 0.043 & 0.054 & 0.081 & 0.191 & 0.708 & 0.237 & 0.640 & 1 & 0.505 & 0.947 & 1 \\
  $b^*_2$ & & 0.050 & 0.041 & 0.055 & 0.096 & 0.232 & 0.824 & 0.280 & 0.741 & 1 & 0.608 & 0.988 & 1 \\
  $\phi^{(n);g_0}_{1}$ &  & 0.051 & 0.056 & 0.052 & 0.073 & 0.108 & 0.446 & 0.128 & 0.308 & 0.941 & 0.189 & 0.519 & 0.998 \\
$\bar{b}_2$ &  & 0.024 & 0.044 & 0.060 & 0.049 & 0.123 & 0.486 & 0.077 & 0.297 & 0.958 & 0.123 & 0.497 & 0.997 \\
    \hline
\end{tabular}
}
\end{table}

\begin{table}[!h]
\caption{Rejection rates, for a nominal significance level of
$\alpha=0.05$, of the $\phi^{(n);\mu;g_0}_{1}$,
$\phi^{*(n);\mu}_{1}$, $b^*_2$ based, $\phi^{(n);g_0}_{1}$
and $\bar{b}_2$ based tests
calculated using $1000$ samples of size $n$ simulated from the
$k'$-sine-skewed distribution with the specified base density
$g_0$ and values of $\lambda$ and $k'$.}\label{estsim6}
\vspace{3pt} \centering \scriptsize{
\begin{tabular}{|ll||ccc|ccc|ccc|ccc|}
  \hline
   & $\lambda$ & \multicolumn{3}{c|}{0} & \multicolumn{3}{c|}{0.2} & \multicolumn{3}{c|}{0.4} & \multicolumn{3}{c|}{0.6}\\
   & $n$ & 30 & 100 & 500 & 30 & 100 & 500 & 30 & 100 & 500 & 30 & 100 & 500 \\
  \hline \hline  Test & $k'$ & \multicolumn{12}{c|}{$g_0=f_{\textrm{C}_{0.45}}$}\\
 \hline
$\phi^{(n);\mu;g_0}_{1}$ & 1& 0.039 & 0.053 & 0.057 & 0.123 & 0.299 & 0.876 & 0.334 & 0.829 & 1 & 0.673 & 0.996 & 1 \\
  $\phi^{*(n);\mu}_{1}$ &  & 0.036 & 0.050 & 0.056 & 0.121 & 0.306 & 0.880 & 0.332 & 0.834 & 1 & 0.679 & 0.998 & 1 \\
$b^*_2$ & & 0.061 & 0.057 & 0.041 & 0.070 & 0.093 & 0.287 & 0.103 & 0.252 & 0.811 & 0.171 & 0.484 & 0.992 \\
  $\phi^{(n);g_0}_{1}$ &  & 0.059 & 0.052 & 0.052 & 0.046 & 0.087 & 0.311 & 0.048 & 0.133 & 0.630 & 0.035 & 0.069 & 0.499 \\
$\bar{b}_2$ &  & 0.049 & 0.057 & 0.038 & 0.045 & 0.067 & 0.269 & 0.055 & 0.148 & 0.577 & 0.048 & 0.118 & 0.488 \\
  \hline $\phi^{(n);\mu;g_0}_{1}$ &2& 0.039 & 0.053 & 0.057 & 0.063 & 0.088 & 0.311 & 0.105 & 0.239 & 0.812 & 0.179 & 0.461 & 0.988 \\
  $\phi^{*(n);\mu}_{1}$ &  & 0.036 & 0.050 & 0.056 & 0.059 & 0.091 & 0.311 & 0.108 & 0.242 & 0.811 & 0.177 & 0.466 & 0.989 \\
  $b^*_2$ & & 0.061 & 0.057 & 0.041 & 0.119 & 0.292 & 0.905 & 0.342 & 0.834 & 1 & 0.667 & 0.991 & 1 \\
  $\phi^{(n);g_0}_{1}$ &  & 0.059 & 0.052 & 0.052 & 0.083 & 0.193 & 0.764 & 0.169 & 0.588 & 1 & 0.321 & 0.903 & 1 \\
$\bar{b}_2$ &  & 0.049 & 0.057 & 0.038 & 0.083 & 0.274 & 0.901 & 0.205 & 0.763 & 1 & 0.373 & 0.950 & 1 \\
  \hline $\phi^{(n);\mu;g_0}_{1}$ &3& 0.039 & 0.053 & 0.057 & 0.057 & 0.048 & 0.064 & 0.067 & 0.042 & 0.053 & 0.059 & 0.052 & 0.038 \\
  $\phi^{*(n);\mu}_{1}$ &  & 0.036 & 0.050 & 0.056 & 0.055 & 0.048 & 0.063 & 0.068 & 0.045 & 0.052 & 0.064 & 0.048 & 0.036 \\
  $b^*_2$ & & 0.061 & 0.057 & 0.041 & 0.058 & 0.100 & 0.299 & 0.112 & 0.237 & 0.832 & 0.157 & 0.496 & 0.988 \\
  $\phi^{(n);g_0}_{1}$ &  & 0.059 & 0.052 & 0.052 & 0.048 & 0.054 & 0.066 & 0.040 & 0.045 & 0.037 & 0.039 & 0.047 & 0.045 \\
$\bar{b}_2$ &  & 0.049 & 0.057 & 0.038 & 0.050 & 0.085 & 0.302 & 0.074 & 0.215 & 0.802 & 0.106 & 0.408 & 0.981 \\
           \hline \hline  Test & $k'$ & \multicolumn{12}{c|}{$g_0=f_{\textrm{WC}_{0.5}}$}\\
  \hline
$\phi^{(n);\mu;g_0}_{1}$ & 1& 0.051 & 0.056 & 0.046 & 0.098 & 0.241 & 0.763 & 0.268 & 0.708 & 1 & 0.544 & 0.963 & 1 \\
  $\phi^{*(n);\mu}_{1}$ &  & 0.050 & 0.055 & 0.044 & 0.092 & 0.231 & 0.763 & 0.273 & 0.716 & 1 & 0.561 & 0.967 & 1 \\
$b^*_2$ & & 0.054 & 0.043 & 0.047 & 0.062 & 0.097 & 0.241 & 0.084 & 0.213 & 0.691 & 0.133 & 0.387 & 0.962 \\
  $\phi^{(n);g_0}_{1}$ &  & 0.037 & 0.048 & 0.052 & 0.041 & 0.072 & 0.263 & 0.062 & 0.173 & 0.742 & 0.102 & 0.363 & 0.979 \\
$\bar{b}_2$ &  & 0.038 & 0.053 & 0.053 & 0.051 & 0.069 & 0.227 & 0.077 & 0.187 & 0.719 & 0.120 & 0.427 & 0.979 \\
  \hline $\phi^{(n);\mu;g_0}_{1}$ &2& 0.051 & 0.056 & 0.046 & 0.066 & 0.083 & 0.257 & 0.103 & 0.233 & 0.769 & 0.165 & 0.454 & 0.982 \\
  $\phi^{*(n);\mu}_{1}$ &  & 0.050 & 0.055 & 0.044 & 0.061 & 0.090 & 0.259 & 0.114 & 0.227 & 0.768 & 0.171 & 0.456 & 0.981 \\
  $b^*_2$ & & 0.054 & 0.043 & 0.047 & 0.110 & 0.292 & 0.852 & 0.325 & 0.808 & 1 & 0.624 & 0.987 & 1 \\
  $\phi^{(n);g_0}_{1}$ &  & 0.037 & 0.048 & 0.052 & 0.047 & 0.077 & 0.271 & 0.090 & 0.217 & 0.707 & 0.137 & 0.344 & 0.892 \\
$\bar{b}_2$ &  & 0.038 & 0.053 & 0.053 & 0.060 & 0.123 & 0.417 & 0.106 & 0.310 & 0.910 & 0.160 & 0.467 & 0.990 \\
  \hline $\phi^{(n);\mu;g_0}_{1}$ &3& 0.051 & 0.056 & 0.046 & 0.055 & 0.057 & 0.093 & 0.046 & 0.110 & 0.256 & 0.067 & 0.148 & 0.534 \\
  $\phi^{*(n);\mu}_{1}$ &  & 0.050 & 0.055 & 0.044 & 0.055 & 0.051 & 0.097 & 0.054 & 0.110 & 0.262 & 0.076 & 0.152 & 0.533 \\
  $b^*_2$ & & 0.054 & 0.043 & 0.047 & 0.067 & 0.114 & 0.336 & 0.110 & 0.289 & 0.859 & 0.190 & 0.543 & 0.996 \\
  $\phi^{(n);g_0}_{1}$ &  & 0.037 & 0.048 & 0.052 & 0.047 & 0.119 & 0.392 & 0.111 & 0.326 & 0.897 & 0.184 & 0.546 & 0.990 \\
$\bar{b}_2$ &  & 0.038 & 0.053 & 0.053 & 0.051 & 0.074 & 0.152 & 0.074 & 0.128 & 0.413 & 0.097 & 0.206 & 0.679 \\
    \hline
\end{tabular}
}
\end{table}

\begin{table}[!h]
\caption{Rejection rates, for a nominal significance level of
$\alpha=0.05$, of the $\phi^{(n);\mu;g_0}_{3}$,
$\phi^{*(n);\mu}_{3}$, $b^*_2$ based, $\phi^{(n);g_0}_{3}$
and $\bar{b}_2$ based tests
calculated using $1000$ samples of size $n$ simulated from the
$k'$-sine-skewed distribution with the specified base von Mises
density $g_0$ and values of $\lambda$ and $k'$.}\label{estsim7}
\vspace{3pt} \centering \scriptsize{
\begin{tabular}{|ll||ccc|ccc|ccc|ccc|}
  \hline
   & $\lambda$ & \multicolumn{3}{c|}{0} & \multicolumn{3}{c|}{0.2} & \multicolumn{3}{c|}{0.4} & \multicolumn{3}{c|}{0.6}\\
   & $n$ & 30 & 100 & 500 & 30 & 100 & 500 & 30 & 100 & 500 & 30 & 100 & 500 \\
   \hline \hline  Test & $k'$ & \multicolumn{12}{c|}{$g_0=f_{\textrm{VM}_1}$}\\
  \hline
$\phi^{(n);\mu;g_0}_{3}$ & 1& 0.044 & 0.072 & 0.056 & 0.048 & 0.047 & 0.049 & 0.041 & 0.058 & 0.092 & 0.050 & 0.081 & 0.153 \\
  $\phi^{*(n);\mu}_{3}$ &  & 0.043 & 0.070 & 0.056 & 0.043 & 0.048 & 0.049 & 0.046 & 0.061 & 0.091 & 0.053 & 0.083 & 0.150 \\
  $b^*_2$ & & 0.053 & 0.043 & 0.048 & 0.051 & 0.088 & 0.298 & 0.095 & 0.232 & 0.802 & 0.148 & 0.473 & 0.990 \\
  $\phi^{(n);g_0}_{3}$ &  & 0.045 & 0.069 & 0.064 & 0.046 & 0.040 & 0.038 & 0.051 & 0.046 & 0.056 & 0.053 & 0.056 & 0.088 \\
 $\bar{b}_2$ &  & 0.033 & 0.036 & 0.050 & 0.029 & 0.043 & 0.056 & 0.038 & 0.038 & 0.068 & 0.044 & 0.079 & 0.248 \\
  \hline $\phi^{(n);\mu;g_0}_{3}$ &2& 0.044 & 0.072 & 0.056 & 0.056 & 0.096 & 0.298 & 0.088 & 0.240 & 0.818 & 0.148 & 0.480 & 0.994 \\
  $\phi^{*(n);\mu}_{3}$ &  & 0.043 & 0.070 & 0.056 & 0.065 & 0.096 & 0.299 & 0.090 & 0.245 & 0.818 & 0.149 & 0.474 & 0.994 \\
   $b^*_2$ & & 0.053 & 0.043 & 0.048 & 0.111 & 0.293 & 0.897 & 0.347 & 0.826 & 1 & 0.646 & 0.995 & 1 \\
  $\phi^{(n);g_0}_{3}$ &  & 0.045 & 0.069 & 0.064 & 0.054 & 0.074 & 0.220 & 0.059 & 0.146 & 0.530 & 0.080 & 0.221 & 0.672 \\
 $\bar{b}_2$ &  & 0.033 & 0.036 & 0.050 & 0.039 & 0.183 & 0.777 & 0.139 & 0.550 & 0.999 & 0.235 & 0.747 & 1 \\
  \hline $\phi^{(n);\mu;g_0}_{3}$ &3& 0.044 & 0.072 & 0.056 & 0.118 & 0.298 & 0.897 & 0.313 & 0.812 & 1 & 0.651 & 0.990 & 1 \\
  $\phi^{*(n);\mu}_{3}$ &  & 0.043 & 0.070 & 0.056 & 0.124 & 0.299 & 0.896 & 0.318 & 0.810 & 1 & 0.660 & 0.992 & 1 \\
  $b^*_2$ & & 0.053 & 0.043 & 0.048 & 0.061 & 0.105 & 0.319 & 0.107 & 0.259 & 0.823 & 0.172 & 0.482 & 0.991 \\
  $\phi^{(n);g_0}_{3}$ &  & 0.045 & 0.069 & 0.064 & 0.091 & 0.231 & 0.875 & 0.196 & 0.700 & 1 & 0.397 & 0.936 & 1 \\
 $\bar{b}_2$ &  & 0.033 & 0.036 & 0.050 & 0.032 & 0.065 & 0.288 & 0.057 & 0.198 & 0.777 & 0.086 & 0.343 & 0.975 \\
   \hline \hline  Test & $k'$ & \multicolumn{12}{c|}{$g_0=f_{\textrm{VM}_{10}}$}\\
  \hline
$\phi^{(n);\mu;g_0}_{3}$ & 1& 0.054 & 0.046 & 0.040 & 0.055 & 0.084 & 0.255 & 0.081 & 0.209 & 0.746 & 0.138 & 0.440 & 0.971 \\
  $\phi^{*(n);\mu}_{3}$ &  & 0.051 & 0.046 & 0.041 & 0.055 & 0.088 & 0.255 & 0.082 & 0.214 & 0.746 & 0.135 & 0.432 & 0.972 \\
$b^*_2$ & & 0.051 & 0.042 & 0.044 & 0.058 & 0.090 & 0.282 & 0.095 & 0.234 & 0.791 & 0.162 & 0.464 & 0.986 \\
  $\phi^{(n);g_0}_{3}$ &  & 0.044 & 0.051 & 0.033 & 0.040 & 0.048 & 0.035 & 0.043 & 0.043 & 0.037 & 0.043 & 0.038 & 0.044 \\
  $\bar{b}_2$ &  & 0.028 & 0.040 & 0.040 & 0.029 & 0.037 & 0.034 & 0.029 & 0.030 & 0.041 & 0.027 & 0.038 & 0.053 \\
  \hline $\phi^{(n);\mu;g_0}_{3}$ &2& 0.054 & 0.046 & 0.040 & 0.081 & 0.177 & 0.680 & 0.206 & 0.553 & 0.997 & 0.394 & 0.902 & 1 \\
  $\phi^{*(n);\mu}_{3}$ &  & 0.051 & 0.046 & 0.041 & 0.082 & 0.182 & 0.682 & 0.213 & 0.563 & 0.997 & 0.412 & 0.905 & 1 \\
  $b^*_2$ & & 0.051 & 0.042 & 0.044 & 0.083 & 0.190 & 0.699 & 0.230 & 0.586 & 0.998 & 0.431 & 0.918 & 1 \\
  $\phi^{(n);g_0}_{3}$ &  & 0.044 & 0.051 & 0.033 & 0.050 & 0.052 & 0.043 & 0.044 & 0.046 & 0.066 & 0.035 & 0.040 & 0.049 \\
  $\bar{b}_2$ &  & 0.028 & 0.040 & 0.040 & 0.029 & 0.044 & 0.046 & 0.038 & 0.042 & 0.068 & 0.037 & 0.047 & 0.068 \\
  \hline $\phi^{(n);\mu;g_0}_{3}$ &3& 0.054 & 0.046 & 0.040 & 0.102 & 0.247 & 0.858 & 0.287 & 0.758 & 0.999 & 0.568 & 0.985 & 1 \\
  $\phi^{*(n);\mu}_{3}$ &  & 0.051 & 0.046 & 0.041 & 0.101 & 0.250 & 0.859 & 0.289 & 0.758 & 0.999 & 0.590 & 0.985 & 1 \\
  $b^*_2$ & & 0.051 & 0.042 & 0.044 & 0.104 & 0.243 & 0.842 & 0.280 & 0.743 & 1 & 0.581 & 0.982 & 1 \\
  $\phi^{(n);g_0}_{3}$ &  & 0.044 & 0.051 & 0.033 & 0.057 & 0.060 & 0.143 & 0.049 & 0.106 & 0.399 & 0.061 & 0.139 & 0.513 \\
 $\bar{b}_2$ &  & 0.028 & 0.040 & 0.040 & 0.034 & 0.057 & 0.146 & 0.033 & 0.087 & 0.397 & 0.031 & 0.091 & 0.524 \\
\hline
\end{tabular}
}
\end{table}

\begin{table}[!h]
\caption{Rejection rates, for a nominal significance level of
$\alpha=0.05$, of the $\phi^{(n);\mu;g_0}_{3}$,
$\phi^{*(n);\mu}_{3}$, $b^*_2$ based, $\phi^{(n);g_0}_{3}$
and $\bar{b}_2$ based tests
calculated using $1000$ samples of size $n$ simulated from the
$k'$-sine-skewed distribution with the specified base wrapped
normal density $g_0$ and values of $\lambda$ and
$k'$.}\label{estsim8} \vspace{3pt} \centering \scriptsize{
\begin{tabular}{|ll||ccc|ccc|ccc|ccc|}
  \hline
   & $\lambda$ & \multicolumn{3}{c|}{0} & \multicolumn{3}{c|}{0.2} & \multicolumn{3}{c|}{0.4} & \multicolumn{3}{c|}{0.6}\\
   & $n$ & 30 & 100 & 500 & 30 & 100 & 500 & 30 & 100 & 500 & 30 & 100 & 500 \\
   \hline \hline  Test & $k'$ & \multicolumn{12}{c|}{$g_0=f_{\textrm{WN}_{0.5}}$}\\
  \hline
$\phi^{(n);\mu;g_0}_{3}$ & 1& 0.034 & 0.048 & 0.041 & 0.050 & 0.062 & 0.056 & 0.048 & 0.055 & 0.071 & 0.046 & 0.054 & 0.092 \\
  $\phi^{*(n);\mu}_{3}$ &  & 0.031 & 0.047 & 0.041 & 0.051 & 0.063 & 0.053 & 0.051 & 0.060 & 0.069 & 0.044 & 0.057 & 0.092 \\
  $b^*_2$ & & 0.052 & 0.053 & 0.049 & 0.065 & 0.112 & 0.351 & 0.111 & 0.294 & 0.895 & 0.212 & 0.579 & 0.998 \\
  $\phi^{(n);g_0}_{3}$ &  & 0.034 & 0.051 & 0.045 & 0.054 & 0.061 & 0.056 & 0.045 & 0.053 & 0.056 & 0.044 & 0.039 & 0.036 \\
  $\bar{b}_2$ &  & 0.030 & 0.041 & 0.050 & 0.039 & 0.059 & 0.148 & 0.037 & 0.077 & 0.257 & 0.041 & 0.058 & 0.167 \\
  \hline $\phi^{(n);\mu;g_0}_{3}$ &2& 0.034 & 0.048 & 0.041 & 0.072 & 0.116 & 0.369 & 0.122 & 0.316 & 0.880 & 0.217 & 0.586 & 0.995 \\
  $\phi^{*(n);\mu}_{3}$ &  & 0.031 & 0.047 & 0.041 & 0.069 & 0.112 & 0.370 & 0.128 & 0.311 & 0.879 & 0.220 & 0.583 & 0.995 \\
  $b^*_2$ & & 0.052 & 0.053 & 0.049 & 0.112 & 0.289 & 0.896 & 0.329 & 0.825 & 1 & 0.669 & 0.996 & 1 \\
  $\phi^{(n);g_0}_{3}$ &  & 0.034 & 0.051 & 0.045 & 0.076 & 0.105 & 0.329 & 0.088 & 0.227 & 0.747 & 0.127 & 0.294 & 0.861 \\
  $\bar{b}_2$ &  & 0.030 & 0.041 & 0.050 & 0.084 & 0.249 & 0.866 & 0.186 & 0.728 & 1 & 0.355 & 0.930 & 1 \\
  \hline $\phi^{(n);\mu;g_0}_{3}$ &3& 0.034 & 0.048 & 0.041 & 0.121 & 0.334 & 0.884 & 0.347 & 0.807 & 1 & 0.679 & 0.989 & 1 \\
  $\phi^{*(n);\mu}_{3}$ &  & 0.031 & 0.047 & 0.041 & 0.124 & 0.333 & 0.886 & 0.346 & 0.810 & 1 & 0.679 & 0.990 & 1 \\
  $b^*_2$ & & 0.052 & 0.053 & 0.049 & 0.061 & 0.118 & 0.363 & 0.114 & 0.277 & 0.889 & 0.189 & 0.558 & 0.999 \\
  $\phi^{(n);g_0}_{3}$ &  & 0.034 & 0.051 & 0.045 & 0.101 & 0.307 & 0.866 & 0.235 & 0.737 & 1 & 0.472 & 0.971 & 1 \\
  $\bar{b}_2$ &  & 0.030 & 0.041 & 0.050 & 0.062 & 0.111 & 0.387 & 0.086 & 0.269 & 0.912 & 0.165 & 0.534 & 1 \\
      \hline \hline  Test & $k'$ & \multicolumn{12}{c|}{$g_0=f_{\textrm{WN}_{0.9}}$}\\
  \hline
$\phi^{(n);\mu;g_0}_{3}$ & 1& 0.045 & 0.058 & 0.053 & 0.066 & 0.098 & 0.320 & 0.084 & 0.242 & 0.867 & 0.176 & 0.517 & 0.996 \\
  $\phi^{*(n);\mu}_{3}$ &  & 0.048 & 0.058 & 0.054 & 0.061 & 0.097 & 0.320 & 0.083 & 0.237 & 0.872 & 0.172 & 0.517 & 0.996 \\
  $b^*_2$ & & 0.050 & 0.041 & 0.055 & 0.069 & 0.112 & 0.434 & 0.100 & 0.340 & 0.953 & 0.230 & 0.679 & 1 \\
  $\phi^{(n);g_0}_{3}$ &  & 0.052 & 0.057 & 0.060 & 0.040 & 0.051 & 0.060 & 0.049 & 0.037 & 0.061 & 0.033 & 0.031 & 0.053 \\
  $\phi_{f_{{\rm
C}_{0.25}};3}^{*(n)}$ &  & 0.030 & 0.050 & 0.066 & 0.040 & 0.048 & 0.056 & 0.040 & 0.041 & 0.064 & 0.039 & 0.040 & 0.070 \\
  $\bar{b}_2$ &  & 0.024 & 0.044 & 0.060 & 0.033 & 0.044 & 0.055 & 0.032 & 0.039 & 0.063 & 0.033 & 0.036 & 0.065 \\
  \hline $\phi^{(n);\mu;g_0}_{3}$ &2& 0.045 & 0.058 & 0.053 & 0.081 & 0.216 & 0.753 & 0.229 & 0.654 & 1 & 0.496 & 0.956 & 1 \\
  $\phi^{*(n);\mu}_{3}$ &  & 0.048 & 0.058 & 0.054 & 0.078 & 0.214 & 0.751 & 0.219 & 0.656 & 1 & 0.496 & 0.956 & 1 \\
  $b^*_2$ & & 0.050 & 0.041 & 0.055 & 0.091 & 0.227 & 0.825 & 0.263 & 0.731 & 1 & 0.571 & 0.983 & 1 \\
  $\phi^{(n);g_0}_{3}$ &  & 0.052 & 0.057 & 0.060 & 0.047 & 0.064 & 0.121 & 0.048 & 0.078 & 0.326 & 0.045 & 0.098 & 0.380 \\
  $\bar{b}_2$ &  & 0.024 & 0.044 & 0.060 & 0.043 & 0.051 & 0.133 & 0.031 & 0.066 & 0.357 & 0.024 & 0.104 & 0.462 \\
  \hline $\phi^{(n);\mu;g_0}_{3}$ &3& 0.045 & 0.058 & 0.053 & 0.101 & 0.263 & 0.881 & 0.300 & 0.815 & 1 & 0.676 & 0.992 & 1 \\
  $\phi^{*(n);\mu}_{3}$ &  & 0.048 & 0.058 & 0.054 & 0.103 & 0.262 & 0.884 & 0.296 & 0.816 & 1 & 0.675 & 0.993 & 1 \\
  $b^*_2$ & & 0.050 & 0.041 & 0.055 & 0.096 & 0.232 & 0.824 & 0.280 & 0.741 & 1 & 0.608 & 0.988 & 1 \\
  $\phi^{(n);g_0}_{3}$ &  & 0.052 & 0.057 & 0.060 & 0.071 & 0.123 & 0.488 & 0.132 & 0.333 & 0.959 & 0.195 & 0.537 & 0.995 \\
$\bar{b}_2$ &  & 0.024 & 0.044 & 0.060 & 0.049 & 0.123 & 0.486 & 0.077 & 0.297 & 0.958 & 0.123 & 0.497 & 0.997 \\
            \hline
\end{tabular}
}
\end{table}

\begin{table}[!h]
\caption{Rejection rates, for a nominal significance level of
$\alpha=0.05$, of the $\phi^{(n);\mu;g_0}_{3}$,
$\phi^{*(n);\mu}_{3}$, $b^*_2$ based, $\phi^{(n);g_0}_{3}$
and $\bar{b}_2$ based tests
calculated using $1000$ samples of size $n$ simulated from the
$k'$-sine-skewed distribution with the specified base density
$g_0$ and values of $\lambda$ and $k'$.}\label{estsim9}
\vspace{3pt} \centering \scriptsize{
\begin{tabular}{|ll||ccc|ccc|ccc|ccc|}
  \hline
   & $\lambda$ & \multicolumn{3}{c|}{0} & \multicolumn{3}{c|}{0.2} & \multicolumn{3}{c|}{0.4} & \multicolumn{3}{c|}{0.6}\\
   & $n$ & 30 & 100 & 500 & 30 & 100 & 500 & 30 & 100 & 500 & 30 & 100 & 500 \\
  \hline \hline  Test & $k'$ & \multicolumn{12}{c|}{$g_0=f_{\textrm{C}_{0.45}}$}\\
  \hline
$\phi^{(n);\mu;g_0}_{3}$ & 1& 0.056 & 0.052 & 0.046 & 0.059 & 0.044 & 0.050 & 0.053 & 0.040 & 0.044 & 0.049 & 0.043 & 0.041 \\
  $\phi^{*(n);\mu}_{3}$ &  & 0.052 & 0.052 & 0.047 & 0.054 & 0.044 & 0.050 & 0.049 & 0.042 & 0.042 & 0.049 & 0.045 & 0.041 \\
$b^*_2$ & & 0.061 & 0.057 & 0.041 & 0.070 & 0.093 & 0.287 & 0.103 & 0.252 & 0.811 & 0.171 & 0.484 & 0.992 \\
  $\phi^{(n);g_0}_{3}$ &  & 0.047 & 0.053 & 0.043 & 0.050 & 0.041 & 0.057 & 0.046 & 0.035 & 0.047 & 0.048 & 0.032 & 0.035 \\
$\bar{b}_2$ &  & 0.049 & 0.057 & 0.038 & 0.045 & 0.067 & 0.269 & 0.055 & 0.148 & 0.577 & 0.048 & 0.118 & 0.488 \\
  \hline $\phi^{(n);\mu;g_0}_{3}$ &2& 0.056 & 0.052 & 0.046 & 0.061 & 0.092 & 0.314 & 0.108 & 0.260 & 0.819 & 0.186 & 0.473 & 0.992 \\
  $\phi^{*(n);\mu}_{3}$ &  & 0.052 & 0.052 & 0.047 & 0.061 & 0.092 & 0.314 & 0.118 & 0.250 & 0.819 & 0.174 & 0.476 & 0.992 \\
$b^*_2$ & & 0.061 & 0.057 & 0.041 & 0.119 & 0.292 & 0.905 & 0.342 & 0.834 & 1 & 0.667 & 0.991 & 1 \\
  $\phi^{(n);g_0}_{3}$ &  & 0.047 & 0.053 & 0.043 & 0.061 & 0.075 & 0.268 & 0.098 & 0.181 & 0.645 & 0.119 & 0.247 & 0.754 \\
$\bar{b}_2$ &  & 0.049 & 0.057 & 0.038 & 0.083 & 0.274 & 0.901 & 0.205 & 0.763 & 1 & 0.373 & 0.950 & 1 \\
  \hline $\phi^{(n);\mu;g_0}_{3}$ &3& 0.056 & 0.052 & 0.046 & 0.134 & 0.307 & 0.897 & 0.361 & 0.814 & 1 & 0.660 & 0.993 & 1 \\
  $\phi^{*(n);\mu}_{3}$ &  & 0.052 & 0.052 & 0.047 & 0.140 & 0.306 & 0.898 & 0.363 & 0.811 & 1 & 0.674 & 0.993 & 1 \\
$b^*_2$ & & 0.061 & 0.057 & 0.041 & 0.058 & 0.100 & 0.299 & 0.112 & 0.237 & 0.832 & 0.157 & 0.496 & 0.988 \\
  $\phi^{(n);g_0}_{3}$ &  & 0.047 & 0.053 & 0.043 & 0.100 & 0.246 & 0.889 & 0.224 & 0.694 & 1 & 0.429 & 0.938 & 1 \\
$\bar{b}_2$ &  & 0.049 & 0.057 & 0.038 & 0.050 & 0.085 & 0.302 & 0.074 & 0.215 & 0.802 & 0.106 & 0.408 & 0.981 \\
  \hline \hline  Test & $k'$ & \multicolumn{12}{c|}{$g_0=f_{\textrm{WC}_{0.5}}$}\\
  \hline
$\phi^{(n);\mu;g_0}_{3}$ & 1& 0.069 & 0.055 & 0.065 & 0.052 & 0.077 & 0.086 & 0.067 & 0.109 & 0.218 & 0.091 & 0.142 & 0.406 \\
  $\phi^{*(n);\mu}_{3}$ &  & 0.067 & 0.055 & 0.067 & 0.053 & 0.078 & 0.087 & 0.068 & 0.108 & 0.222 & 0.084 & 0.146 & 0.412 \\
  $b^*_2$ & & 0.054 & 0.043 & 0.047 & 0.062 & 0.097 & 0.241 & 0.084 & 0.213 & 0.691 & 0.133 & 0.387 & 0.962 \\
  $\phi^{(n);g_0}_{3}$ &  & 0.051 & 0.058 & 0.061 & 0.044 & 0.057 & 0.121 & 0.054 & 0.100 & 0.380 & 0.069 & 0.151 & 0.683 \\
$\bar{b}_2$ &  & 0.038 & 0.053 & 0.053 & 0.051 & 0.069 & 0.227 & 0.077 & 0.187 & 0.719 & 0.120 & 0.427 & 0.979 \\
  \hline $\phi^{(n);\mu;g_0}_{3}$ &2& 0.069 & 0.055 & 0.065 & 0.066 & 0.131 & 0.301 & 0.126 & 0.271 & 0.848 & 0.205 & 0.546 & 0.996 \\
  $\phi^{*(n);\mu}_{3}$ &  & 0.067 & 0.055 & 0.067 & 0.064 & 0.122 & 0.308 & 0.122 & 0.276 & 0.847 & 0.197 & 0.547 & 0.996 \\
  $b^*_2$ & & 0.054 & 0.043 & 0.047 & 0.110 & 0.292 & 0.852 & 0.325 & 0.808 & 1 & 0.624 & 0.987 & 1 \\
  $\phi^{(n);g_0}_{3}$ &  & 0.051 & 0.058 & 0.061 & 0.035 & 0.061 & 0.049 & 0.051 & 0.051 & 0.051 & 0.042 & 0.056 & 0.057 \\
  $\bar{b}_2$ &  & 0.038 & 0.053 & 0.053 & 0.060 & 0.123 & 0.417 & 0.106 & 0.310 & 0.910 & 0.160 & 0.467 & 0.990 \\
  \hline $\phi^{(n);\mu;g_0}_{3}$ &3& 0.069 & 0.055 & 0.065 & 0.133 & 0.317 & 0.873 & 0.334 & 0.822 & 1 & 0.654 & 0.991 & 1 \\
  $\phi^{*(n);\mu}_{3}$ &  & 0.067 & 0.055 & 0.067 & 0.129 & 0.317 & 0.876 & 0.340 & 0.821 & 1 & 0.657 & 0.994 & 1 \\
  $b^*_2$ & & 0.054 & 0.043 & 0.047 & 0.067 & 0.114 & 0.336 & 0.110 & 0.289 & 0.859 & 0.190 & 0.543 & 0.996 \\
  $\phi^{(n);g_0}_{3}$ &  & 0.051 & 0.058 & 0.061 & 0.082 & 0.203 & 0.726 & 0.179 & 0.585 & 0.997 & 0.343 & 0.826 & 1 \\
  $\bar{b}_2$ &  & 0.038 & 0.053 & 0.053 & 0.051 & 0.074 & 0.152 & 0.074 & 0.128 & 0.413 & 0.097 & 0.206 & 0.679 \\
    \hline
\end{tabular}
}
\end{table}

\begin{table}[!h]
\caption{Rejection rates, for a nominal significance level of
$\alpha=0.05$, of the $\phi^{(n);\mu;f_0}_{2}$ and
$\phi^{(n);f_0}_{2}$ tests when $f_0$ is posited to be
$f_{\textrm{VM}_{10}}$, $f_{\textrm{C}_{\rho}}$ (cardioid with any
valid value of $\rho$) or $f_{\textrm{WC}_{0.5}}$, calculated
using $1000$ samples of size $n$ simulated from the
$k'$-sine-skewed distribution with the specified base von Mises
density $g_0$ and values of $\lambda$ and $k'$.}
\label{estsim7b}\vspace{3pt} \centering \scriptsize{
\begin{tabular}{|ll||ccc|ccc|ccc|ccc|}
  \hline
   & $\lambda$ & \multicolumn{3}{c|}{0} & \multicolumn{3}{c|}{0.2} & \multicolumn{3}{c|}{0.4} & \multicolumn{3}{c|}{0.6}\\
   & $n$ & 30 & 100 & 500 & 30 & 100 & 500 & 30 & 100 & 500 & 30 & 100 & 500 \\
   \hline \hline  Test & $k'$ & \multicolumn{12}{c|}{$g_0=f_{\textrm{VM}_1}$}\\
  \hline

   $\phi^{(n);\mu;f_{\textrm{VM}_{10}}}_{2}$ & 1 & 0.152 & 0.131 & 0.147 & 0.170 & 0.203 & 0.471 & 0.226 & 0.402 & 0.905 & 0.312 & 0.635 & 0.998 \\
   $\phi^{(n);\mu;f_{\textrm{C}_{\rho}}}_{2}$ &  & 0.050 & 0.046 & 0.049 & 0.057 & 0.086 & 0.305 & 0.095 & 0.230 & 0.796 & 0.145 & 0.463 & 0.990 \\
   $\phi^{(n);\mu;f_{\textrm{WC}_{0.5}}}_{2}$ &  & 0.061 & 0.054 & 0.060 & 0.069 & 0.098 & 0.325 & 0.111 & 0.248 & 0.821 & 0.159 & 0.488 & 0.991 \\
   $\phi^{(n);f_{\textrm{VM}_{10}}}_{2}$ &  & 0.820 & 0.820 & 0.817 & 0.802 & 0.833 & 0.821 & 0.810 & 0.834 & 0.835 & 0.829 & 0.833 & 0.914 \\
   $\phi^{(n);f_{\textrm{C}_{\rho}}}_{2}$ &  & 0.027 & 0.026 & 0.028 & 0.024 & 0.027 & 0.030 & 0.025 & 0.025 & 0.040 & 0.022 & 0.040 & 0.154 \\
   $\phi^{(n);f_{\textrm{WC}_{0.5}}}_{2}$ &  & 0.041 & 0.039 & 0.041 & 0.042 & 0.048 & 0.045 & 0.038 & 0.035 & 0.052 & 0.035 & 0.048 & 0.184 \\
\hline 
   $\phi^{(n);\mu;f_{\textrm{VM}_{10}}}_{2}$ & 2 & 0.152 & 0.131 & 0.147 & 0.246 & 0.488 & 0.961 & 0.544 & 0.920 & 1 & 0.825 & 1 & 1 \\
   $\phi^{(n);\mu;f_{\textrm{C}_{\rho}}}_{2}$ &  & 0.050 & 0.046 & 0.049 & 0.100 & 0.293 & 0.894 & 0.347 & 0.825 & 1 & 0.647 & 0.994 & 1 \\
   $\phi^{(n);\mu;f_{\textrm{WC}_{0.5}}}_{2}$ &  & 0.061 & 0.054 & 0.060 & 0.116 & 0.323 & 0.912 & 0.381 & 0.838 & 1 & 0.672 & 0.994 & 1 \\
   $\phi^{(n);f_{\textrm{VM}_{10}}}_{2}$ &  & 0.820 & 0.820 & 0.817 & 0.860 & 0.920 & 0.997 & 0.901 & 0.978 & 1 & 0.946 & 0.992 & 1 \\
   $\phi^{(n);f_{\textrm{C}_{\rho}}}_{2}$ &  & 0.027 & 0.026 & 0.028 & 0.043 & 0.161 & 0.701 & 0.134 & 0.512 & 0.998 & 0.262 & 0.768 & 1 \\
   $\phi^{(n);f_{\textrm{WC}_{0.5}}}_{2}$ &  & 0.041 & 0.039 & 0.041 & 0.073 & 0.196 & 0.712 & 0.163 & 0.550 & 0.999 & 0.319 & 0.817 & 1 \\
\hline
   $\phi^{(n);\mu;f_{\textrm{VM}_{10}}}_{2}$ & 3 & 0.152 & 0.131 & 0.147 & 0.160 & 0.227 & 0.501 & 0.241 & 0.424 & 0.922 & 0.336 & 0.680 & 0.994 \\
   $\phi^{(n);\mu;f_{\textrm{C}_{\rho}}}_{2}$ &  & 0.050 & 0.046 & 0.049 & 0.061 & 0.100 & 0.318 & 0.102 & 0.255 & 0.822 & 0.164 & 0.475 & 0.991 \\
   $\phi^{(n);\mu;f_{\textrm{WC}_{0.5}}}_{2}$ &  & 0.061 & 0.054 & 0.060 & 0.072 & 0.116 & 0.342 & 0.114 & 0.273 & 0.836 & 0.187 & 0.498 & 0.991 \\
   $\phi^{(n);f_{\textrm{VM}_{10}}}_{2}$ &  & 0.820 & 0.820 & 0.817 & 0.828 & 0.855 & 0.933 & 0.846 & 0.905 & 0.994 & 0.858 & 0.964 & 0.999 \\
   $\phi^{(n);f_{\textrm{C}_{\rho}}}_{2}$ &  & 0.027 & 0.026 & 0.028 & 0.023 & 0.051 & 0.216 & 0.049 & 0.156 & 0.716 & 0.077 & 0.299 & 0.969 \\
   $\phi^{(n);f_{\textrm{WC}_{0.5}}}_{2}$ &  & 0.041 & 0.039 & 0.041 & 0.034 & 0.054 & 0.085 & 0.046 & 0.079 & 0.217 & 0.046 & 0.099 & 0.409 \\

   \hline \hline  Test & $k'$ & \multicolumn{12}{c|}{$g_0=f_{\textrm{VM}_{10}}$}\\
  \hline

   $\phi^{(n);\mu;f_{\textrm{VM}_{10}}}_{2}$ & 1 & 0.049 & 0.046 & 0.044 & 0.060 & 0.085 & 0.282 & 0.097 & 0.233 & 0.786 & 0.163 & 0.476 & 0.983 \\
   $\phi^{(n);\mu;f_{\textrm{C}_{\rho}}}_{2}$ &  & 0.011 & 0.005 & 0.008 & 0.010 & 0.016 & 0.105 & 0.016 & 0.086 & 0.564 & 0.046 & 0.218 & 0.939 \\
   $\phi^{(n);\mu;f_{\textrm{WC}_{0.5}}}_{2}$ &  & 0.013 & 0.006 & 0.012 & 0.013 & 0.021 & 0.119 & 0.022 & 0.099 & 0.601 & 0.057 & 0.243 & 0.949 \\
   $\phi^{(n);f_{\textrm{VM}_{10}}}_{2}$ &  & 0.050 & 0.047 & 0.040 & 0.045 & 0.045 & 0.034 & 0.045 & 0.040 & 0.039 & 0.047 & 0.036 & 0.044 \\
   $\phi^{(n);f_{\textrm{C}_{\rho}}}_{2}$ &  & 0 & 0 & 0 & 0 & 0 & 0 & 0 & 0 & 0 & 0 & 0 & 0 \\
   $\phi^{(n);f_{\textrm{WC}_{0.5}}}_{2}$ &  & 0 & 0 & 0 & 0 & 0 & 0 & 0 & 0 & 0 & 0 & 0 & 0 \\
\hline 
   $\phi^{(n);\mu;f_{\textrm{VM}_{10}}}_{2}$ & 2 & 0.049 & 0.046 & 0.044 & 0.083 & 0.183 & 0.694 & 0.222 & 0.586 & 0.998 & 0.417 & 0.911 & 1 \\
   $\phi^{(n);\mu;f_{\textrm{C}_{\rho}}}_{2}$ &  & 0.011 & 0.005 & 0.008 & 0.014 & 0.060 & 0.411 & 0.063 & 0.314 & 0.986 & 0.180 & 0.757 & 1 \\
   $\phi^{(n);\mu;f_{\textrm{WC}_{0.5}}}_{2}$ &  & 0.013 & 0.006 & 0.012 & 0.019 & 0.072 & 0.441 & 0.075 & 0.339 & 0.989 & 0.205 & 0.785 & 1 \\
   $\phi^{(n);f_{\textrm{VM}_{10}}}_{2}$ &  & 0.050 & 0.047 & 0.040 & 0.052 & 0.053 & 0.044 & 0.045 & 0.051 & 0.070 & 0.032 & 0.040 & 0.053 \\
   $\phi^{(n);f_{\textrm{C}_{\rho}}}_{2}$ &  & 0 & 0 & 0 & 0 & 0 & 0 & 0 & 0 & 0 & 0 & 0 & 0 \\
   $\phi^{(n);f_{\textrm{WC}_{0.5}}}_{2}$ &  & 0 & 0 & 0 & 0 & 0 & 0 & 0 & 0 & 0 & 0 & 0 & 0 \\
\hline 
   $\phi^{(n);\mu;f_{\textrm{VM}_{10}}}_{2}$ & 3 & 0.049 & 0.046 & 0.044 & 0.107 & 0.238 & 0.843 & 0.273 & 0.728 & 1 & 0.575 & 0.977 & 1 \\
   $\phi^{(n);\mu;f_{\textrm{C}_{\rho}}}_{2}$ &  & 0.011 & 0.005 & 0.008 & 0.024 & 0.089 & 0.616 & 0.100 & 0.485 & 0.998 & 0.291 & 0.908 & 1 \\
   $\phi^{(n);\mu;f_{\textrm{WC}_{0.5}}}_{2}$ &  & 0.013 & 0.006 & 0.012 & 0.031 & 0.101 & 0.648 & 0.120 & 0.520 & 0.999 & 0.320 & 0.925 & 1 \\
   $\phi^{(n);f_{\textrm{VM}_{10}}}_{2}$ &  & 0.050 & 0.047 & 0.040 & 0.061 & 0.058 & 0.138 & 0.050 & 0.098 & 0.412 & 0.059 & 0.138 & 0.543 \\
   $\phi^{(n);f_{\textrm{C}_{\rho}}}_{2}$ &  & 0 & 0 & 0 & 0 & 0 & 0 & 0 & 0 & 0 & 0 & 0 & 0 \\
   $\phi^{(n);f_{\textrm{WC}_{0.5}}}_{2}$ &  & 0 & 0 & 0 & 0 & 0 & 0 & 0 & 0 & 0 & 0 & 0 & 0 \\
   \hline
\end{tabular}
}

\end{table}

\begin{table}[!h]
\caption{Rejection rates, for a nominal significance level of
$\alpha=0.05$, of the $\phi^{(n);\mu;f_0}_{2}$ and
$\phi^{(n);f_0}_{2}$ tests when $f_0$ is posited to be
$f_{\textrm{VM}_{10}}$, $f_{\textrm{C}_{\rho}}$ (cardioid with any
valid value of $\rho$) or $f_{\textrm{WC}_{0.5}}$, calculated
using $1000$ samples of size $n$ simulated from the
$k'$-sine-skewed distribution with the specified base wrapped
normal density $g_0$ and values of $\lambda$ and
$k'$.}\label{estsim8b}\vspace{3pt} \centering \scriptsize{
\begin{tabular}{|ll||ccc|ccc|ccc|ccc|}
  \hline
   & $\lambda$ & \multicolumn{3}{c|}{0} & \multicolumn{3}{c|}{0.2} & \multicolumn{3}{c|}{0.4} & \multicolumn{3}{c|}{0.6}\\
   & $n$ & 30 & 100 & 500 & 30 & 100 & 500 & 30 & 100 & 500 & 30 & 100 & 500 \\

   \hline \hline  Test & $k'$ & \multicolumn{12}{c|}{$g_0=f_{\textrm{WN}_{0.5}}$}\\
  \hline
   $\phi^{(n);\mu;f_{\textrm{VM}_{10}}}_{2}$ & 1 & 0.138 & 0.133 & 0.165 & 0.161 & 0.254 & 0.547 & 0.246 & 0.479 & 0.964 & 0.395 & 0.747 & 1 \\
   $\phi^{(n);\mu;f_{\textrm{C}_{\rho}}}_{2}$ &  & 0.054 & 0.048 & 0.052 & 0.066 & 0.113 & 0.354 & 0.117 & 0.292 & 0.894 & 0.202 & 0.578 & 0.997 \\
   $\phi^{(n);\mu;f_{\textrm{WC}_{0.5}}}_{2}$ &  & 0.061 & 0.058 & 0.059 & 0.074 & 0.125 & 0.376 & 0.128 & 0.317 & 0.903 & 0.229 & 0.602 & 0.998 \\
   $\phi^{(n);f_{\textrm{VM}_{10}}}_{2}$ &  & 0.831 & 0.819 & 0.819 & 0.823 & 0.845 & 0.866 & 0.836 & 0.854 & 0.903 & 0.791 & 0.821 & 0.887 \\
   $\phi^{(n);f_{\textrm{C}_{\rho}}}_{2}$ &  & 0.023 & 0.031 & 0.034 & 0.027 & 0.034 & 0.106 & 0.016 & 0.044 & 0.154 & 0.011 & 0.023 & 0.070 \\
   $\phi^{(n);f_{\textrm{WC}_{0.5}}}_{2}$ &  & 0.032 & 0.039 & 0.048 & 0.035 & 0.046 & 0.133 & 0.034 & 0.062 & 0.200 & 0.016 & 0.031 & 0.120 \\
\hline 
   $\phi^{(n);\mu;f_{\textrm{VM}_{10}}}_{2}$ & 2 & 0.138 & 0.133 & 0.165 & 0.249 & 0.516 & 0.960 & 0.540 & 0.924 & 1 & 0.817 & 0.998 & 1 \\
   $\phi^{(n);\mu;f_{\textrm{C}_{\rho}}}_{2}$ &  & 0.054 & 0.048 & 0.052 & 0.116 & 0.289 & 0.894 & 0.327 & 0.826 & 1 & 0.656 & 0.995 & 1 \\
   $\phi^{(n);\mu;f_{\textrm{WC}_{0.5}}}_{2}$ &  & 0.061 & 0.058 & 0.059 & 0.132 & 0.317 & 0.905 & 0.350 & 0.835 & 1 & 0.676 & 0.996 & 1 \\
   $\phi^{(n);f_{\textrm{VM}_{10}}}_{2}$ &  & 0.831 & 0.819 & 0.819 & 0.846 & 0.927 & 1 & 0.923 & 0.990 & 1 & 0.967 & 1 & 1 \\
   $\phi^{(n);f_{\textrm{C}_{\rho}}}_{2}$ &  & 0.023 & 0.031 & 0.034 & 0.051 & 0.172 & 0.801 & 0.150 & 0.634 & 1 & 0.311 & 0.907 & 1 \\
   $\phi^{(n);f_{\textrm{WC}_{0.5}}}_{2}$ &  & 0.032 & 0.039 & 0.048 & 0.078 & 0.208 & 0.819 & 0.173 & 0.658 & 1 & 0.352 & 0.922 & 1 \\
\hline.118 & 0.365 & 0.114 & 0.284 & 0.889 & 0.187 & 0.553 & 0.999 \\
   $\phi^{(n);\mu;f_{\textrm{VM}_{10}}}_{2}$ & 3 & 0.138 & 0.133 & 0.165 & 0.155 & 0.251 & 0.545 & 0.238 & 0.469 & 0.958 & 0.395 & 0.751 & 1 \\
   $\phi^{(n);\mu;f_{\textrm{C}_{\rho}}}_{2}$ &  & 0.054 & 0.048 & 0.052 & 0.062 & 0.118 & 0.364 & 0.114 & 0.284 & 0.889 & 0.187 & 0.553 & 0.999 \\
   $\phi^{(n);\mu;f_{\textrm{WC}_{0.5}}}_{2}$ &  & 0.061 & 0.058 & 0.059 & 0.068 & 0.129 & 0.381 & 0.136 & 0.304 & 0.899 & 0.208 & 0.586 & 0.999 \\
   $\phi^{(n);f_{\textrm{VM}_{10}}}_{2}$ &  & 0.831 & 0.819 & 0.819 & 0.812 & 0.870 & 0.958 & 0.859 & 0.921 & 1 & 0.917 & 0.979 & 1 \\
   $\phi^{(n);f_{\textrm{C}_{\rho}}}_{2}$ &  & 0.023 & 0.031 & 0.034 & 0.036 & 0.073 & 0.318 & 0.058 & 0.221 & 0.884 & 0.125 & 0.486 & 1 \\
   $\phi^{(n);f_{\textrm{WC}_{0.5}}}_{2}$ &  & 0.032 & 0.039 & 0.048 & 0.035 & 0.062 & 0.140 & 0.047 & 0.105 & 0.423 & 0.061 & 0.196 & 0.764 \\
      \hline \hline  Test & $k'$ & \multicolumn{12}{c|}{$g_0=f_{\textrm{WN}_{0.9}}$}\\
  \hline

   $\phi^{(n);\mu;f_{\textrm{VM}_{10}}}_{2}$ & 1 & 0.096 & 0.105 & 0.110 & 0.117 & 0.199 & 0.562 & 0.187 & 0.481 & 0.981 & 0.342 & 0.792 & 1 \\
   $\phi^{(n);\mu;f_{\textrm{C}_{\rho}}}_{2}$ &  & 0.028 & 0.030 & 0.032 & 0.041 & 0.078 & 0.360 & 0.076 & 0.271 & 0.931 & 0.165 & 0.585 & 1 \\
   $\phi^{(n);\mu;f_{\textrm{WC}_{0.5}}}_{2}$ &  & 0.035 & 0.034 & 0.039 & 0.047 & 0.088 & 0.381 & 0.088 & 0.291 & 0.938 & 0.189 & 0.619 & 1 \\
   $\phi^{(n);f_{\textrm{VM}_{10}}}_{2}$ &  & 0.315 & 0.352 & 0.389 & 0.296 & 0.348 & 0.379 & 0.291 & 0.350 & 0.386 & 0.298 & 0.333 & 0.342 \\
   $\phi^{(n);f_{\textrm{C}_{\rho}}}_{2}$ &  & 0 & 0 & 0 & 0 & 0 & 0 & 0 & 0 & 0 & 0 & 0 & 0 \\
   $\phi^{(n);f_{\textrm{WC}_{0.5}}}_{2}$ &  & 0 & 0 & 0 & 0 & 0 & 0 & 0 & 0 & 0 & 0 & 0 & 0 \\
\hline 
   $\phi^{(n);\mu;f_{\textrm{VM}_{10}}}_{2}$ & 2 & 0.096 & 0.105 & 0.110 & 0.174 & 0.328 & 0.886 & 0.378 & 0.830 & 1 & 0.689 & 0.994 & 1 \\
   $\phi^{(n);\mu;f_{\textrm{C}_{\rho}}}_{2}$ &  & 0.028 & 0.030 & 0.032 & 0.061 & 0.176 & 0.751 & 0.198 & 0.645 & 1 & 0.464 & 0.963 & 1 \\
   $\phi^{(n);\mu;f_{\textrm{WC}_{0.5}}}_{2}$ &  & 0.035 & 0.034 & 0.039 & 0.068 & 0.189 & 0.765 & 0.216 & 0.668 & 1 & 0.495 & 0.972 & 1 \\
   $\phi^{(n);f_{\textrm{VM}_{10}}}_{2}$ &  & 0.315 & 0.352 & 0.389 & 0.304 & 0.398 & 0.541 & 0.309 & 0.446 & 0.783 & 0.296 & 0.470 & 0.817 \\
   $\phi^{(n);f_{\textrm{C}_{\rho}}}_{2}$ &  & 0 & 0 & 0 & 0 & 0 & 0 & 0 & 0 & 0 & 0 & 0 & 0 \\
   $\phi^{(n);f_{\textrm{WC}_{0.5}}}_{2}$ &  & 0 & 0 & 0 & 0 & 0 & 0 & 0 & 0 & 0 & 0 & 0 & 0 \\
\hline 
   $\phi^{(n);\mu;f_{\textrm{VM}_{10}}}_{2}$ & 3 & 0.096 & 0.105 & 0.110 & 0.177 & 0.355 & 0.908 & 0.415 & 0.848 & 1 & 0.720 & 0.996 & 1 \\
   $\phi^{(n);\mu;f_{\textrm{C}_{\rho}}}_{2}$ &  & 0.028 & 0.030 & 0.032 & 0.058 & 0.176 & 0.768 & 0.208 & 0.666 & 1 & 0.502 & 0.977 & 1 \\
   $\phi^{(n);\mu;f_{\textrm{WC}_{0.5}}}_{2}$ &  & 0.035 & 0.034 & 0.039 & 0.065 & 0.190 & 0.784 & 0.230 & 0.692 & 1 & 0.533 & 0.979 & 1 \\
   $\phi^{(n);f_{\textrm{VM}_{10}}}_{2}$ &  & 0.315 & 0.352 & 0.389 & 0.366 & 0.500 & 0.855 & 0.452 & 0.731 & 0.997 & 0.522 & 0.870 & 1 \\
   $\phi^{(n);f_{\textrm{C}_{\rho}}}_{2}$ &  & 0 & 0 & 0 & 0 & 0 & 0 & 0 & 0 & 0 & 0 & 0 & 0 \\
   $\phi^{(n);f_{\textrm{WC}_{0.5}}}_{2}$ &  & 0 & 0 & 0 & 0 & 0 & 0 & 0 & 0 & 0 & 0 & 0 & 0 \\
   \hline
\end{tabular}
}

\end{table}

\begin{table}[!h]
\caption{Rejection rates, for a nominal significance level of
$\alpha=0.05$, of the $\phi^{(n);\mu;f_0}_{2}$ and
$\phi^{(n);f_0}_{2}$ tests when $f_0$ is posited to be
$f_{\textrm{VM}_{10}}$, $f_{\textrm{C}_{\rho}}$ (cardioid with any
valid value of $\rho$) or $f_{\textrm{WC}_{0.5}}$, calculated
using $1000$ samples of size $n$ simulated from the
$k'$-sine-skewed distribution with the specified base density
$g_0$ and values of $\lambda$ and
$k'$.}\label{estsim9b}\vspace{3pt} \centering \scriptsize{
\begin{tabular}{|ll||ccc|ccc|ccc|ccc|}
  \hline
   & $\lambda$ & \multicolumn{3}{c|}{0} & \multicolumn{3}{c|}{0.2} & \multicolumn{3}{c|}{0.4} & \multicolumn{3}{c|}{0.6}\\
   & $n$ & 30 & 100 & 500 & 30 & 100 & 500 & 30 & 100 & 500 & 30 & 100 & 500 \\
        \hline \hline  Test & $k'$ & \multicolumn{12}{c|}{$g_0=f_{\textrm{C}_{0.45}}$}\\
  \hline

   $\phi^{(n);\mu;f_{\textrm{VM}_{10}}}_{2}$ & 1 & 0.141 & 0.142 & 0.139 & 0.162 & 0.206 & 0.487 & 0.246 & 0.425 & 0.917 & 0.327 & 0.671 & 1 \\
   $\phi^{(n);\mu;f_{\textrm{C}_{\rho}}}_{2}$ &  & 0.056 & 0.057 & 0.042 & 0.068 & 0.090 & 0.285 & 0.106 & 0.260 & 0.814 & 0.175 & 0.489 & 0.993 \\
   $\phi^{(n);\mu;f_{\textrm{WC}_{0.5}}}_{2}$ &  & 0.065 & 0.065 & 0.048 & 0.080 & 0.098 & 0.308 & 0.119 & 0.276 & 0.826 & 0.189 & 0.510 & 0.994 \\
   $\phi^{(n);f_{\textrm{VM}_{10}}}_{2}$ &  & 0.839 & 0.858 & 0.836 & 0.827 & 0.860 & 0.932 & 0.824 & 0.875 & 0.979 & 0.821 & 0.848 & 0.968 \\
   $\phi^{(n);f_{\textrm{C}_{\rho}}}_{2}$ &  & 0.054 & 0.061 & 0.044 & 0.041 & 0.064 & 0.246 & 0.042 & 0.101 & 0.485 & 0.025 & 0.055 & 0.338 \\
   $\phi^{(n);f_{\textrm{WC}_{0.5}}}_{2}$ &  & 0.068 & 0.082 & 0.079 & 0.062 & 0.101 & 0.341 & 0.055 & 0.167 & 0.621 & 0.041 & 0.094 & 0.482 \\
\hline 
   $\phi^{(n);\mu;f_{\textrm{VM}_{10}}}_{2}$ & 2 & 0.141 & 0.142 & 0.139 & 0.259 & 0.482 & 0.956 & 0.534 & 0.923 & 1 & 0.843 & 0.996 & 1 \\
   $\phi^{(n);\mu;f_{\textrm{C}_{\rho}}}_{2}$ &  & 0.056 & 0.057 & 0.042 & 0.120 & 0.289 & 0.905 & 0.336 & 0.830 & 1 & 0.661 & 0.991 & 1 \\
   $\phi^{(n);\mu;f_{\textrm{WC}_{0.5}}}_{2}$ &  & 0.065 & 0.065 & 0.048 & 0.134 & 0.310 & 0.911 & 0.359 & 0.848 & 1 & 0.688 & 0.991 & 1 \\
   $\phi^{(n);f_{\textrm{VM}_{10}}}_{2}$ &  & 0.839 & 0.858 & 0.836 & 0.874 & 0.930 & 0.998 & 0.931 & 0.996 & 1 & 0.965 & 0.999 & 1 \\
   $\phi^{(n);f_{\textrm{C}_{\rho}}}_{2}$ &  & 0.054 & 0.061 & 0.044 & 0.073 & 0.261 & 0.890 & 0.200 & 0.729 & 1 & 0.402 & 0.948 & 1 \\
   $\phi^{(n);f_{\textrm{WC}_{0.5}}}_{2}$ &  & 0.068 & 0.082 & 0.079 & 0.099 & 0.292 & 0.907 & 0.266 & 0.772 & 1 & 0.463 & 0.966 & 1 \\
\hline 
   $\phi^{(n);\mu;f_{\textrm{VM}_{10}}}_{2}$ & 3 & 0.141 & 0.142 & 0.139 & 0.161 & 0.223 & 0.494 & 0.245 & 0.415 & 0.910 & 0.324 & 0.677 & 0.995 \\
   $\phi^{(n);\mu;f_{\textrm{C}_{\rho}}}_{2}$ &  & 0.056 & 0.057 & 0.042 & 0.057 & 0.094 & 0.296 & 0.113 & 0.237 & 0.829 & 0.157 & 0.493 & 0.989 \\
   $\phi^{(n);\mu;f_{\textrm{WC}_{0.5}}}_{2}$ &  & 0.065 & 0.065 & 0.048 & 0.069 & 0.114 & 0.319 & 0.127 & 0.261 & 0.842 & 0.175 & 0.515 & 0.992 \\
   $\phi^{(n);f_{\textrm{VM}_{10}}}_{2}$ &  & 0.839 & 0.858 & 0.836 & 0.839 & 0.856 & 0.937 & 0.864 & 0.926 & 0.998 & 0.888 & 0.971 & 1 \\
   $\phi^{(n);f_{\textrm{C}_{\rho}}}_{2}$ &  & 0.054 & 0.061 & 0.044 & 0.051 & 0.092 & 0.287 & 0.070 & 0.217 & 0.816 & 0.105 & 0.435 & 0.987 \\
   $\phi^{(n);f_{\textrm{WC}_{0.5}}}_{2}$ &  & 0.068 & 0.082 & 0.079 & 0.074 & 0.084 & 0.157 & 0.077 & 0.121 & 0.361 & 0.081 & 0.188 & 0.648 \\
   \hline \hline  Test & $k'$ & \multicolumn{12}{c|}{$g_0=f_{\textrm{WC}_{0.5}}$}\\
  \hline
   $\phi^{(n);\mu;f_{\textrm{VM}_{10}}}_{2}$ & 1 & 0.144 & 0.119 & 0.133 & 0.143 & 0.194 & 0.388 & 0.190 & 0.354 & 0.815 & 0.260 & 0.555 & 0.988 \\
   $\phi^{(n);\mu;f_{\textrm{C}_{\rho}}}_{2}$ &  & 0.051 & 0.035 & 0.038 & 0.055 & 0.088 & 0.227 & 0.083 & 0.203 & 0.667 & 0.115 & 0.366 & 0.952 \\
   $\phi^{(n);\mu;f_{\textrm{WC}_{0.5}}}_{2}$ &  & 0.058 & 0.041 & 0.045 & 0.061 & 0.098 & 0.241 & 0.092 & 0.222 & 0.682 & 0.132 & 0.389 & 0.959 \\
   $\phi^{(n);f_{\textrm{VM}_{10}}}_{2}$ &  & 0.836 & 0.828 & 0.822 & 0.835 & 0.848 & 0.918 & 0.856 & 0.903 & 0.995 & 0.887 & 0.959 & 1 \\
   $\phi^{(n);f_{\textrm{C}_{\rho}}}_{2}$ &  & 0.036 & 0.045 & 0.042 & 0.038 & 0.058 & 0.207 & 0.051 & 0.140 & 0.664 & 0.084 & 0.313 & 0.966 \\
   $\phi^{(n);f_{\textrm{WC}_{0.5}}}_{2}$ &  & 0.045 & 0.054 & 0.043 & 0.046 & 0.064 & 0.162 & 0.059 & 0.122 & 0.507 & 0.083 & 0.264 & 0.894 \\
\hline 
   $\phi^{(n);\mu;f_{\textrm{VM}_{10}}}_{2}$ & 2 & 0.144 & 0.119 & 0.133 & 0.238 & 0.454 & 0.934 & 0.492 & 0.911 & 1 & 0.802 & 0.998 & 1 \\
   $\phi^{(n);\mu;f_{\textrm{C}_{\rho}}}_{2}$ &  & 0.051 & 0.035 & 0.038 & 0.096 & 0.271 & 0.836 & 0.306 & 0.787 & 1 & 0.592 & 0.980 & 1 \\
   $\phi^{(n);\mu;f_{\textrm{WC}_{0.5}}}_{2}$ &  & 0.058 & 0.041 & 0.045 & 0.114 & 0.294 & 0.852 & 0.322 & 0.807 & 1 & 0.629 & 0.985 & 1 \\
   $\phi^{(n);f_{\textrm{VM}_{10}}}_{2}$ &  & 0.836 & 0.828 & 0.822 & 0.823 & 0.897 & 0.976 & 0.870 & 0.949 & 1 & 0.925 & 0.981 & 1 \\
   $\phi^{(n);f_{\textrm{C}_{\rho}}}_{2}$ &  & 0.036 & 0.045 & 0.042 & 0.053 & 0.113 & 0.402 & 0.110 & 0.319 & 0.919 & 0.188 & 0.529 & 0.994 \\
   $\phi^{(n);f_{\textrm{WC}_{0.5}}}_{2}$ &  & 0.045 & 0.054 & 0.043 & 0.069 & 0.144 & 0.478 & 0.113 & 0.372 & 0.957 & 0.219 & 0.628 & 0.997 \\
\hline 
   $\phi^{(n);\mu;f_{\textrm{VM}_{10}}}_{2}$ & 3 & 0.144 & 0.119 & 0.133 & 0.150 & 0.232 & 0.500 & 0.222 & 0.459 & 0.944 & 0.346 & 0.708 & 1 \\
   $\phi^{(n);\mu;f_{\textrm{C}_{\rho}}}_{2}$ &  & 0.051 & 0.035 & 0.038 & 0.060 & 0.100 & 0.305 & 0.097 & 0.269 & 0.845 & 0.163 & 0.511 & 0.996 \\
   $\phi^{(n);\mu;f_{\textrm{WC}_{0.5}}}_{2}$ &  & 0.058 & 0.041 & 0.045 & 0.067 & 0.113 & 0.334 & 0.110 & 0.290 & 0.857 & 0.186 & 0.531 & 0.997 \\
   $\phi^{(n);f_{\textrm{VM}_{10}}}_{2}$ &  & 0.836 & 0.828 & 0.822 & 0.837 & 0.838 & 0.878 & 0.842 & 0.889 & 0.969 & 0.876 & 0.913 & 0.989 \\
   $\phi^{(n);f_{\textrm{C}_{\rho}}}_{2}$ &  & 0.036 & 0.045 & 0.042 & 0.035 & 0.064 & 0.137 & 0.055 & 0.103 & 0.412 & 0.061 & 0.202 & 0.717 \\
   $\phi^{(n);f_{\textrm{WC}_{0.5}}}_{2}$ &  & 0.045 & 0.054 & 0.043 & 0.046 & 0.061 & 0.057 & 0.056 & 0.060 & 0.088 & 0.047 & 0.060 & 0.127 \\
   \hline
\end{tabular}
}
\end{table}

\begin{table}[!h]
\caption{Rejection rates, for a nominal significance level of
$\alpha=0.05$, of the $\phi_{f_0;2}^{*(n)}$ test when $f_0$ is
posited to be $f_{\textrm{VM}_\kappa}$ (von Mises with any valid
value of $\kappa$), $f_{\textrm{C}_{0.45}}$ or
$f_{\textrm{WC}_{0.5}}$ and the $\phi^{*(n);\hat{\mu}^{(n)}}_{2}$
test ($\phi^{*(n);\mu}_{2}$ with $\mu$ estimated from the data),
calculated using $1000$ samples of size $n$ simulated from the
$k'$-sine-skewed distribution with the specified base von Mises
density $g_0$ and values of $\lambda$ and
$k'$.}\label{estsim10}\vspace{3pt} \centering \scriptsize{
\begin{tabular}{|ll||ccc|ccc|ccc|ccc|}
  \hline
   & $\lambda$ & \multicolumn{3}{c|}{0} & \multicolumn{3}{c|}{0.2} & \multicolumn{3}{c|}{0.4} & \multicolumn{3}{c|}{0.6}\\
   & $n$ & 30 & 100 & 500 & 30 & 100 & 500 & 30 & 100 & 500 & 30 & 100 & 500 \\
   \hline \hline  Test & $k'$ & \multicolumn{12}{c|}{$g_0=f_{\textrm{VM}_1}$}\\
  \hline
$\phi_{f_{\textrm{VM}_{\kappa}};2}^{*(n)}$ & 1 & 0.033 & 0.036 & 0.050 & 0.029 & 0.043 & 0.056 & 0.038 & 0.038 & 0.068 & 0.044 & 0.079 & 0.248 \\
  $\phi_{f_{\textrm{C}_{0.45}};2}^{*(n)}$ &  & 0.030 & 0.040 & 0.049 & 0.031 & 0.045 & 0.055 & 0.043 & 0.038 & 0.066 & 0.039 & 0.083 & 0.248 \\
  $\phi_{f_{\textrm{WC}_{0.5}};2}^{*(n)}$ &  & 0.031 & 0.040 & 0.048 & 0.027 & 0.044 & 0.053 & 0.037 & 0.046 & 0.064 & 0.043 & 0.074 & 0.239 \\
  $\phi^{*(n);\hat{\mu}^{(n)}}_{2}$ &  & 0.026 & 0.026 & 0.026 & 0.026 & 0.029 & 0.030 & 0.026 & 0.025 & 0.039 & 0.022 & 0.041 & 0.154 \\
  \hline $\phi_{f_{\textrm{VM}_{\kappa}};2}^{*(n)}$ & 2 & 0.033 & 0.036 & 0.050 & 0.039 & 0.183 & 0.777 & 0.139 & 0.550 & 0.999 & 0.235 & 0.747 & 1 \\
  $\phi_{f_{\textrm{C}_{0.45}};2}^{*(n)}$ &  & 0.030 & 0.040 & 0.049 & 0.047 & 0.179 & 0.755 & 0.141 & 0.512 & 0.995 & 0.227 & 0.704 & 0.999 \\
  $\phi_{f_{\textrm{WC}_{0.5}};2}^{*(n)}$ &  & 0.031 & 0.040 & 0.048 & 0.040 & 0.185 & 0.762 & 0.127 & 0.532 & 0.999 & 0.225 & 0.743 & 1 \\
  $\phi^{*(n);\hat{\mu}^{(n)}}_{2}$ &  & 0.026 & 0.026 & 0.026 & 0.044 & 0.157 & 0.703 & 0.133 & 0.512 & 0.999 & 0.275 & 0.784 & 1 \\
  \hline $\phi_{f_{\textrm{VM}_{\kappa}};2}^{*(n)}$ & 3 & 0.033 & 0.036 & 0.050 & 0.032 & 0.065 & 0.288 & 0.057 & 0.198 & 0.777 & 0.086 & 0.343 & 0.975 \\
  $\phi_{f_{\textrm{C}_{0.45}};2}^{*(n)}$ &  & 0.030 & 0.040 & 0.049 & 0.032 & 0.049 & 0.178 & 0.047 & 0.128 & 0.508 & 0.069 & 0.196 & 0.782 \\
  $\phi_{f_{\textrm{WC}_{0.5}};2}^{*(n)}$ &  & 0.031 & 0.040 & 0.048 & 0.030 & 0.050 & 0.161 & 0.048 & 0.125 & 0.481 & 0.063 & 0.183 & 0.785 \\
  $\phi^{*(n);\hat{\mu}^{(n)}}_{2}$ &  & 0.026 & 0.026 & 0.026 & 0.024 & 0.046 & 0.220 & 0.051 & 0.162 & 0.732 & 0.089 & 0.320 & 0.970 \\
   \hline \hline  Test & $k'$ & \multicolumn{12}{c|}{$g_0=f_{\textrm{VM}_{10}}$}\\
  \hline
$\phi_{f_{\textrm{VM}_{\kappa}};2}^{*(n)}$ & 1 & 0.028 & 0.040 & 0.040 & 0.029 & 0.037 & 0.034 & 0.029 & 0.030 & 0.041 & 0.027 & 0.038 & 0.053 \\
  $\phi_{f_{\textrm{C}_{0.45}};2}^{*(n)}$ &  & 0.028 & 0.039 & 0.039 & 0.028 & 0.037 & 0.034 & 0.028 & 0.031 & 0.041 & 0.027 & 0.037 & 0.053 \\
  $\phi_{f_{\textrm{WC}_{0.5}};2}^{*(n)}$ &  & 0.035 & 0.043 & 0.041 & 0.042 & 0.045 & 0.032 & 0.043 & 0.040 & 0.034 & 0.037 & 0.053 & 0.059 \\
  $\phi^{*(n);\hat{\mu}^{(n)}}_{2}$ &  & 0 & 0 & 0 & 0 & 0 & 0 & 0 & 0 & 0 & 0 & 0 & 0 \\
  \hline $\phi_{f_{\textrm{VM}_{\kappa}};2}^{*(n)}$ & 2 & 0.028 & 0.040 & 0.040 & 0.029 & 0.044 & 0.046 & 0.038 & 0.042 & 0.068 & 0.037 & 0.047 & 0.068 \\
  $\phi_{f_{\textrm{C}_{0.45}};2}^{*(n)}$ &  & 0.028 & 0.039 & 0.039 & 0.029 & 0.044 & 0.044 & 0.036 & 0.043 & 0.068 & 0.035 & 0.046 & 0.066 \\
  $\phi_{f_{\textrm{WC}_{0.5}};2}^{*(n)}$ &  & 0.035 & 0.043 & 0.041 & 0.045 & 0.053 & 0.047 & 0.049 & 0.048 & 0.062 & 0.046 & 0.056 & 0.053 \\
  $\phi^{*(n);\hat{\mu}^{(n)}}_{2}$ &  & 0 & 0 & 0 & 0 & 0 & 0 & 0 & 0 & 0 & 0 & 0 & 0 \\
  \hline $\phi_{f_{\textrm{VM}_{\kappa}};2}^{*(n)}$ & 3 & 0.028 & 0.040 & 0.040 & 0.034 & 0.057 & 0.146 & 0.033 & 0.087 & 0.397 & 0.031 & 0.091 & 0.524 \\
  $\phi_{f_{\textrm{C}_{0.45}};2}^{*(n)}$ &  & 0.028 & 0.039 & 0.039 & 0.033 & 0.057 & 0.143 & 0.033 & 0.085 & 0.395 & 0.030 & 0.090 & 0.521 \\
  $\phi_{f_{\textrm{WC}_{0.5}};2}^{*(n)}$ &  & 0.035 & 0.043 & 0.041 & 0.044 & 0.059 & 0.108 & 0.045 & 0.080 & 0.249 & 0.051 & 0.077 & 0.241 \\
  $\phi^{*(n);\hat{\mu}^{(n)}}_{2}$ &  & 0 & 0 & 0 & 0 & 0 & 0 & 0 & 0 & 0 & 0 & 0 & 0 \\
   \hline
\end{tabular}
}
\end{table}

\begin{table}[!h]
\caption{Rejection rates, for a nominal significance level of
$\alpha=0.05$, of the $\phi_{f_0;2}^{*(n)}$ test when $f_0$ is
posited to be $f_{\textrm{VM}_\kappa}$ (von Mises with any valid
value of $\kappa$), $f_{\textrm{C}_{0.45}}$ or
$f_{\textrm{WC}_{0.5}}$ and the $\phi^{*(n);\hat{\mu}^{(n)}}_{2}$
test ($\phi^{*(n);\mu}_{2}$ with $\mu$ estimated from the data),
calculated using $1000$ samples of size $n$ simulated from the
$k'$-sine-skewed distribution with the specified base wrapped
normal density $g_0$ and values of $\lambda$ and
$k'$.}\label{estsim11}\vspace{3pt} \centering \scriptsize{
\begin{tabular}{|ll||ccc|ccc|ccc|ccc|}
  \hline
   & $\lambda$ & \multicolumn{3}{c|}{0} & \multicolumn{3}{c|}{0.2} & \multicolumn{3}{c|}{0.4} & \multicolumn{3}{c|}{0.6}\\
   & $n$ & 30 & 100 & 500 & 30 & 100 & 500 & 30 & 100 & 500 & 30 & 100 & 500 \\
   \hline \hline  Test & $k'$ & \multicolumn{12}{c|}{$g_0=f_{\textrm{WN}_{0.5}}$}\\
  \hline

$\phi_{f_{\textrm{VM}_{\kappa}};2}^{*(n)}$ & 1 & 0.030 & 0.041 & 0.050 & 0.039 & 0.059 & 0.148 & 0.037 & 0.077 & 0.257 & 0.041 & 0.058 & 0.167 \\
  $\phi_{f_{\textrm{C}_{0.45}};2}^{*(n)}$ &  & 0.028 & 0.045 & 0.054 & 0.040 & 0.060 & 0.161 & 0.035 & 0.081 & 0.289 & 0.037 & 0.067 & 0.207 \\
  $\phi_{f_{\textrm{WC}_{0.5}};2}^{*(n)}$ &  & 0.027 & 0.045 & 0.055 & 0.039 & 0.052 & 0.155 & 0.037 & 0.083 & 0.277 & 0.038 & 0.062 & 0.190 \\
  $\phi^{*(n);\hat{\mu}^{(n)}}_{2}$ &  & 0.021 & 0.033 & 0.032 & 0.028 & 0.034 & 0.107 & 0.016 & 0.045 & 0.152 & 0.012 & 0.024 & 0.069 \\
  \hline $\phi_{f_{\textrm{VM}_{\kappa}};2}^{*(n)}$ & 2 & 0.030 & 0.041 & 0.050 & 0.084 & 0.249 & 0.866 & 0.186 & 0.728 & 1 & 0.355 & 0.930 & 1 \\
  $\phi_{f_{\textrm{C}_{0.45}};2}^{*(n)}$ &  & 0.028 & 0.045 & 0.054 & 0.077 & 0.253 & 0.870 & 0.186 & 0.715 & 1 & 0.363 & 0.927 & 1 \\
  $\phi_{f_{\textrm{WC}_{0.5}};2}^{*(n)}$ &  & 0.027 & 0.045 & 0.055 & 0.077 & 0.247 & 0.866 & 0.185 & 0.724 & 1 & 0.344 & 0.941 & 1 \\
  $\phi^{*(n);\hat{\mu}^{(n)}}_{2}$ &  & 0.021 & 0.033 & 0.032 & 0.056 & 0.174 & 0.796 & 0.153 & 0.643 & 1 & 0.324 & 0.908 & 1 \\
  \hline $\phi_{f_{\textrm{VM}_{\kappa}};2}^{*(n)}$ & 3 & 0.030 & 0.041 & 0.050 & 0.062 & 0.111 & 0.387 & 0.086 & 0.269 & 0.912 & 0.165 & 0.534 & 1 \\
  $\phi_{f_{\textrm{C}_{0.45}};2}^{*(n)}$ &  & 0.028 & 0.045 & 0.054 & 0.063 & 0.103 & 0.321 & 0.076 & 0.213 & 0.858 & 0.135 & 0.451 & 0.997 \\
  $\phi_{f_{\textrm{WC}_{0.5}};2}^{*(n)}$ &  & 0.027 & 0.045 & 0.055 & 0.054 & 0.097 & 0.305 & 0.072 & 0.201 & 0.829 & 0.125 & 0.434 & 0.994 \\
  $\phi^{*(n);\hat{\mu}^{(n)}}_{2}$ &  & 0.021 & 0.033 & 0.032 & 0.039 & 0.076 & 0.317 & 0.070 & 0.223 & 0.886 & 0.131 & 0.491 & 1 \\

   \hline \hline  Test & $k'$ & \multicolumn{12}{c|}{$g_0=f_{\textrm{WN}_{0.9}}$}\\
  \hline

$\phi_{f_{\textrm{VM}_{\kappa}};2}^{*(n)}$ & 1 & 0.024 & 0.044 & 0.060 & 0.033 & 0.044 & 0.055 & 0.032 & 0.039 & 0.063 & 0.033 & 0.036 & 0.065 \\
  $\phi_{f_{\textrm{C}_{0.45}};2}^{*(n)}$ &  & 0.024 & 0.046 & 0.053 & 0.031 & 0.042 & 0.054 & 0.026 & 0.038 & 0.061 & 0.032 & 0.035 & 0.063 \\
  $\phi_{f_{\textrm{WC}_{0.5}};2}^{*(n)}$ &  & 0.027 & 0.042 & 0.054 & 0.037 & 0.038 & 0.063 & 0.032 & 0.040 & 0.043 & 0.033 & 0.047 & 0.050 \\
  $\phi^{*(n);\hat{\mu}^{(n)}}_{2}$ &  & 0 & 0 & 0 & 0 & 0 & 0 & 0 & 0 & 0 & 0 & 0 & 0 \\
  \hline $\phi_{f_{\textrm{VM}_{\kappa}};2}^{*(n)}$ & 2 & 0.024 & 0.044 & 0.060 & 0.043 & 0.051 & 0.133 & 0.031 & 0.066 & 0.357 & 0.024 & 0.104 & 0.462 \\
  $\phi_{f_{\textrm{C}_{0.45}};2}^{*(n)}$ &  & 0.024 & 0.046 & 0.053 & 0.041 & 0.049 & 0.127 & 0.030 & 0.059 & 0.360 & 0.022 & 0.095 & 0.467 \\
  $\phi_{f_{\textrm{WC}_{0.5}};2}^{*(n)}$ &  & 0.027 & 0.042 & 0.054 & 0.036 & 0.041 & 0.061 & 0.031 & 0.043 & 0.072 & 0.033 & 0.038 & 0.045 \\
  $\phi^{*(n);\hat{\mu}^{(n)}}_{2}$ &  & 0 & 0 & 0 & 0 & 0 & 0 & 0 & 0 & 0 & 0 & 0 & 0 \\
  \hline $\phi_{f_{\textrm{VM}_{\kappa}};2}^{*(n)}$ & 3 & 0.024 & 0.044 & 0.060 & 0.049 & 0.123 & 0.486 & 0.077 & 0.297 & 0.958 & 0.123 & 0.497 & 0.997 \\
  $\phi_{f_{\textrm{C}_{0.45}};2}^{*(n)}$ &  & 0.024 & 0.046 & 0.053 & 0.047 & 0.113 & 0.481 & 0.076 & 0.287 & 0.952 & 0.117 & 0.485 & 0.996 \\
  $\phi_{f_{\textrm{WC}_{0.5}};2}^{*(n)}$ &  & 0.027 & 0.042 & 0.054 & 0.041 & 0.059 & 0.170 & 0.047 & 0.094 & 0.280 & 0.059 & 0.078 & 0.169 \\
  $\phi^{*(n);\hat{\mu}^{(n)}}_{2}$ &  & 0 & 0 & 0 & 0 & 0 & 0 & 0 & 0 & 0 & 0 & 0 & 0.028 \\
  \hline
\end{tabular}
}

\end{table}

\begin{table}[!h]
\caption{Rejection rates, for a nominal significance level of
$\alpha=0.05$, of the $\phi_{f_0;2}^{*(n)}$ test when $f_0$ is
posited to be $f_{\textrm{VM}_\kappa}$ (von Mises with any valid
value of $\kappa$), $f_{\textrm{C}_{0.45}}$ or
$f_{\textrm{WC}_{0.5}}$ and the $\phi^{*(n);\hat{\mu}^{(n)}}_{2}$
test ($\phi^{*(n);\mu}_{2}$ with $\mu$ estimated from the data),
calculated using $1000$ samples of size $n$ simulated from the
$k'$-sine-skewed distribution with the specified base density
$g_0$ and values of $\lambda$ and
$k'$.}\label{estsim12}\vspace{3pt} \centering \scriptsize{
\begin{tabular}{|ll||ccc|ccc|ccc|ccc|}
  \hline
   & $\lambda$ & \multicolumn{3}{c|}{0} & \multicolumn{3}{c|}{0.2} & \multicolumn{3}{c|}{0.4} & \multicolumn{3}{c|}{0.6}\\
   & $n$ & 30 & 100 & 500 & 30 & 100 & 500 & 30 & 100 & 500 & 30 & 100 & 500 \\
  \hline \hline  Test & $k'$ & \multicolumn{12}{c|}{$g_0=f_{\textrm{C}_{0.45}}$}\\
  \hline
$\phi_{f_{\textrm{VM}_{\kappa}};2}^{*(n)}$ & 1 & 0.049 & 0.057 & 0.038 & 0.045 & 0.067 & 0.269 & 0.055 & 0.148 & 0.577 & 0.048 & 0.118 & 0.488 \\
  $\phi_{f_{\textrm{C}_{0.45}};2}^{*(n)}$ &  & 0.050 & 0.058 & 0.038 & 0.050 & 0.074 & 0.280 & 0.056 & 0.155 & 0.624 & 0.047 & 0.140 & 0.599 \\
  $\phi_{f_{\textrm{WC}_{0.5}};2}^{*(n)}$ &  & 0.052 & 0.057 & 0.040 & 0.043 & 0.070 & 0.274 & 0.047 & 0.162 & 0.605 & 0.052 & 0.136 & 0.567 \\
  $\phi^{*(n);\hat{\mu}^{(n)}}_{2}$ &  & 0.053 & 0.059 & 0.040 & 0.041 & 0.068 & 0.247 & 0.043 & 0.100 & 0.478 & 0.028 & 0.058 & 0.333 \\
  \hline $\phi_{f_{\textrm{VM}_{\kappa}};2}^{*(n)}$ & 2 & 0.049 & 0.057 & 0.038 & 0.083 & 0.274 & 0.901 & 0.205 & 0.763 & 1 & 0.373 & 0.950 & 1 \\
  $\phi_{f_{\textrm{C}_{0.45}};2}^{*(n)}$ &  & 0.050 & 0.058 & 0.038 & 0.088 & 0.281 & 0.902 & 0.219 & 0.773 & 1 & 0.387 & 0.954 & 1 \\
  $\phi_{f_{\textrm{WC}_{0.5}};2}^{*(n)}$ &  & 0.052 & 0.057 & 0.040 & 0.086 & 0.273 & 0.903 & 0.201 & 0.772 & 1 & 0.369 & 0.962 & 1 \\
  $\phi^{*(n);\hat{\mu}^{(n)}}_{2}$ &  & 0.053 & 0.059 & 0.040 & 0.072 & 0.258 & 0.893 & 0.204 & 0.734 & 1 & 0.400 & 0.947 & 1 \\
  \hline $\phi_{f_{\textrm{VM}_{\kappa}};2}^{*(n)}$ & 3 & 0.049 & 0.057 & 0.038 & 0.050 & 0.085 & 0.302 & 0.074 & 0.215 & 0.802 & 0.106 & 0.408 & 0.981 \\
  $\phi_{f_{\textrm{C}_{0.45}};2}^{*(n)}$ &  & 0.050 & 0.058 & 0.038 & 0.047 & 0.089 & 0.289 & 0.072 & 0.202 & 0.816 & 0.093 & 0.402 & 0.987 \\
  $\phi_{f_{\textrm{WC}_{0.5}};2}^{*(n)}$ &  & 0.052 & 0.057 & 0.040 & 0.045 & 0.085 & 0.290 & 0.070 & 0.197 & 0.814 & 0.090 & 0.384 & 0.985 \\
  $\phi^{*(n);\hat{\mu}^{(n)}}_{2}$ &  & 0.053 & 0.059 & 0.040 & 0.054 & 0.096 & 0.290 & 0.079 & 0.223 & 0.814 & 0.122 & 0.438 & 0.987 \\
  \hline \hline  Test & $k'$ & \multicolumn{12}{c|}{$g_0=f_{\textrm{WC}_{0.5}}$}\\
  \hline
$\phi_{f_{\textrm{VM}_{\kappa}};2}^{*(n)}$ & 1 & 0.038 & 0.053 & 0.053 & 0.051 & 0.069 & 0.227 & 0.077 & 0.187 & 0.719 & 0.120 & 0.427 & 0.979 \\
  $\phi_{f_{\textrm{C}_{0.45}};2}^{*(n)}$ &  & 0.038 & 0.058 & 0.045 & 0.053 & 0.070 & 0.203 & 0.065 & 0.152 & 0.606 & 0.107 & 0.347 & 0.945 \\
  $\phi_{f_{\textrm{WC}_{0.5}};2}^{*(n)}$ &  & 0.040 & 0.053 & 0.046 & 0.046 & 0.066 & 0.167 & 0.060 & 0.133 & 0.534 & 0.091 & 0.301 & 0.905 \\
  $\phi^{*(n);\hat{\mu}^{(n)}}_{2}$ &  & 0.041 & 0.056 & 0.046 & 0.040 & 0.064 & 0.218 & 0.056 & 0.149 & 0.676 & 0.089 & 0.325 & 0.967 \\
  \hline $\phi_{f_{\textrm{VM}_{\kappa}};2}^{*(n)}$ & 2 & 0.038 & 0.053 & 0.053 & 0.060 & 0.123 & 0.417 & 0.106 & 0.310 & 0.910 & 0.160 & 0.467 & 0.990 \\
  $\phi_{f_{\textrm{C}_{0.45}};2}^{*(n)}$ &  & 0.038 & 0.058 & 0.045 & 0.058 & 0.129 & 0.431 & 0.103 & 0.303 & 0.902 & 0.169 & 0.469 & 0.986 \\
  $\phi_{f_{\textrm{WC}_{0.5}};2}^{*(n)}$ &  & 0.040 & 0.053 & 0.046 & 0.056 & 0.132 & 0.468 & 0.102 & 0.343 & 0.953 & 0.169 & 0.558 & 0.997 \\
  $\phi^{*(n);\hat{\mu}^{(n)}}_{2}$ &  & 0.041 & 0.056 & 0.046 & 0.055 & 0.123 & 0.432 & 0.109 & 0.350 & 0.932 & 0.206 & 0.574 & 0.995 \\
  \hline $\phi_{f_{\textrm{VM}_{\kappa}};2}^{*(n)}$ & 3 & 0.038 & 0.053 & 0.053 & 0.051 & 0.074 & 0.152 & 0.074 & 0.128 & 0.413 & 0.097 & 0.206 & 0.679 \\
  $\phi_{f_{\textrm{C}_{0.45}};2}^{*(n)}$ &  & 0.038 & 0.058 & 0.045 & 0.046 & 0.062 & 0.059 & 0.061 & 0.069 & 0.107 & 0.064 & 0.064 & 0.149 \\
  $\phi_{f_{\textrm{WC}_{0.5}};2}^{*(n)}$ &  & 0.040 & 0.053 & 0.046 & 0.044 & 0.066 & 0.053 & 0.055 & 0.064 & 0.100 & 0.047 & 0.067 & 0.174 \\
  $\phi^{*(n);\hat{\mu}^{(n)}}_{2}$ &  & 0.041 & 0.056 & 0.046 & 0.047 & 0.070 & 0.154 & 0.069 & 0.129 & 0.449 & 0.074 & 0.226 & 0.762 \\
   \hline
\end{tabular}
}
\end{table}

\begin{table}[!h]
\caption{Rejection rates, for a nominal significance level of
$\alpha=0.05$, of the $b^*_2$ based and $\bar{b}_2$ based tests
calculated using $1000$ samples of size $n$ simulated from: the
Kato and Jones (2010) distribution with parameters $\mu=0$,
$r=0.5$ and the values of $\nu$ and $\kappa$ specified
($\mbox{KJ}_{10}$); the three--parameter asymmetric submodel given
in the Equation (7) of Kato and Jones (2015) with parameters
$\mu=0$, $r=0.5$ and the values of $\gamma$ and
$\bar{\beta}_2=\nu\gamma(1-\gamma)$ specified
($\mbox{KJ}_{15}$).}\label{estsim13} \vspace{3pt}\centering
\scriptsize{
\begin{tabular}{|l||ccc|ccc|ccc|ccc|}
  \hline
    $\nu$ & \multicolumn{3}{c|}{0} & \multicolumn{3}{c|}{0.2} & \multicolumn{3}{c|}{0.4} & \multicolumn{3}{c|}{0.6}\\
    $n$ & 30 & 100 & 500 & 30 & 100 & 500 & 30 & 100 & 500 & 30 & 100 & 500 \\
   \hline \hline  Test  & \multicolumn{12}{c|}{$\mbox{KJ}_{10}; \kappa=0.5$}\\
  \hline

  $b^*_2$  & 0.053 & 0.051 & 0.058 & 0.194 & 0.538 & 0.995 & 0.524 & 0.962 & 1 & 0.747 & 0.997 & 1 \\
$\bar{b}_2$ & 0.044 & 0.057 & 0.050 & 0.047 & 0.057 & 0.040 & 0.053 & 0.059 & 0.047 & 0.044 & 0.061 & 0.067 \\
   \hline \hline  Test  & \multicolumn{12}{c|}{$\mbox{KJ}_{10}; \kappa=0.9$}\\
  \hline

  $b^*_2$  & 0.048 & 0.038 & 0.059 & 0.271 & 0.724 & 1 & 0.707 & 0.997 & 1 & 0.916 & 1 & 1 \\
$\bar{b}_2$ & 0.049 & 0.042 & 0.053 & 0.053 & 0.048 & 0.062 & 0.052 & 0.067 & 0.085 & 0.064 & 0.070 & 0.142 \\
    \hline \hline  Test  & \multicolumn{12}{c|}{$\mbox{KJ}_{15};\gamma=0.5$}\\
  \hline

  $b^*_2$  & 0.054 & 0.043 & 0.047 & 0.070 & 0.103 & 0.399 & 0.127 & 0.307 & 0.923 & 0.199 & 0.569 & 0.999 \\
$\bar{b}_2$ & 0.038 & 0.053 & 0.053 & 0.069 & 0.113 & 0.364 & 0.103 & 0.296 & 0.895 & 0.175 & 0.567 & 1 \\

   \hline \hline  Test  & \multicolumn{12}{c|}{$\mbox{KJ}_{15};\gamma=0.9$}\\
  \hline

  $b^*_2$  & 0.064 & 0.049 & 0.036 & 0.051 & 0.067 & 0.165 & 0.066 & 0.154 & 0.496 & 0.123 & 0.263 & 0.791 \\
$\bar{b}_2$ & 0.025 & 0.045 & 0.062 & 0.032 & 0.058 & 0.173 & 0.048 & 0.157 & 0.516 & 0.061 & 0.279 & 0.867 \\
 \hline
\end{tabular}
}
\end{table}

\clearpage

\bibliographystyle{apa-Jose}

\bibliography{references}

\end{document}